\documentclass[12pt,preprint,iop,apj]{emulateapj}  

\usepackage{natbib}

\newcommand{\cn}{c_{\mathrm{n}}}
\newcommand{\nuin}{\nu_{\mathrm{in}}}
\newcommand{\nuni}{\nu_{\mathrm{ni}}}
\newcommand{\cie}{c_{\mathrm{i}}}
\newcommand{\ca}{c_{\mathrm{A}}}
\newcommand{\rhon}{\rho_{\rm n}}
\newcommand{\rhoi}{\rho_{\rm i}}

\newcommand{\pie}{p_{\rm i}}
\newcommand{\vi}{{\bf v}_{\rm i}}
\newcommand{\pn}{p_{\rm n}}
\newcommand{\vn}{{\bf v}_{\rm n}}

\newcommand{\Pie}{P_{\rm i}}
\newcommand{\Pn}{P_{\rm n}}

\newcommand{\ain}{\alpha_{\rm in}}

\newcommand{\deln}{\Delta_{\mathrm{n}}}
\newcommand{\deli}{\Delta_{\mathrm{i}}}

\begin{document}

\title{Magnetoacoustic waves in a partially ionized two-fluid plasma}

\shorttitle{Magnetoacoustic waves in a two-fluid plasma}

\author{Roberto Soler$^{1}$, Marc Carbonell$^{2}$, \&  Jose Luis Ballester$^{1}$}
\affil{$^{1}$Departament de  F\'{\i}sica,  Universitat de les Illes Balears, E-07122 Palma de Mallorca, Spain}
\affil{$^{2}$Departament de Matem\`{a}tiques i Inform\`{a}tica, Universitat de les Illes Balears, E-07122 Palma de Mallorca, Spain}

\email{roberto.soler@uib.es, marc.carbonell@uib.es, joseluis.ballester@uib.es}    

\begin{abstract}

Compressible disturbances propagate in a plasma in the form of magnetoacoustic waves driven by both gas pressure and magnetic forces. In partially ionized  plasmas the  dynamics of ionized and neutral species are coupled due to ion-neutral collisions. As a consequence, magnetoacoustic waves propagating through a partially ionized medium are affected by  the ion-neutral coupling. The degree to which the behavior of the classic waves is modified  depends on the physical properties of the various species and on the relative value of the wave frequency compared to  the ion-neutral collision frequency.  Here, we perform a comprehensive theoretical investigation of magnetoacoustic wave propagation in a partially ionized plasma using the two-fluid formalism. We consider an extensive range of values for the collision frequency, ionization ratio, and plasma $\beta$, so that the results are applicable to a wide variety of astrophysical plasmas. We determine the modification  of the wave frequencies  and study the frictional damping due to ion-neutral collisions. Approximate analytic expressions to the frequencies are given in the limit case of strongly coupled ions and neutrals, while numerically obtained dispersion diagrams are provided for arbitrary collision frequencies. In addition, we discuss the presence of cutoffs in the dispersion diagrams that constrain wave propagation for certain combinations of parameters. A specific application to propagation of compressible waves in the solar chromosphere is given.
\end{abstract}


\section{INTRODUCTION}

The theoretical study of magnetohydrodynamic (MHD) waves in partially ionized plasmas has received increasing attention in the recent years. Energy dissipation due to ion-neutral collisions may be important in many astrophysical plasmas, e.g.,  in the low solar atmosphere, because of its connection to plasma heating \citep[e.g.,][]{khomenko2012,martinez-sykora2012,russell2013}. Some recent examples of theoretical works that focused on the investigation of MHD waves in partially ionized plasmas are, e.g., \cite{pecseli2000,depontieu2001,kumar2003,khodachenko2004,khodachenko2006,leake2005,zaqarashvili2011a,zaqarashvili2013,soler2012} among many others.

In the present paper, we continue the research started in \citet{soler2013}, hereafter Paper~I, about wave propagation in a partially ionized two-fluid plasma.  The two-fluid formalism used here assumes that ions and electrons form together an ion-electron fluid, while neutrals form a separate fluid. The ion-electron fluid and the neutral fluid interact by means of ion-neutral collisions. Paper~I was devoted to the study of  Alfv\'en waves. Here we tackle the investigation of  magnetoacoustic waves. The dynamics of magnetoacoustic waves is more involved than that of Alfv\'en waves even in the case of a fully ionized plasma. The reason is that Alfv\'en waves are incompressible and are only driven by the magnetic tension force. On the contrary, magnetoacoustic waves produce compression of the plasma and are driven by both magnetic and gas pressure forces \citep[see, e.g.,][]{goossens2003}. As a consequence, the behavior of magnetoacoustic waves is  determined by the relative values of the sound and Alfv\'en velocities in the plasma.  When the medium is partially ionized, the situation is even more complex because, on the one hand, magnetoacoustic waves also produce compression in the neutral fluid due to  collisions with ions and, on the other hand, the neutral fluid is able to support its own acoustic waves that may couple with the plasma magnetoacoustic waves. Thus, the study of propagation of magnetoacoustic disturbances in a partially ionized medium is interesting from the physical point of view and challenging from the theoretical point of view.

This work can be  related to the previous papers by \citet{zaqarashvili2011a} and \citet{mouschovias2011}. \citet{zaqarashvili2011a} derived the dispersion relation for magnetoacoustic waves in a partially ionized two-fluid plasma. They focused their study on the derivation of the two-fluid equations from the more general three-fluid equations and in comparing the two-fluid results with those in the single-fluid approximation. They concluded that the single-fluid approximation breaks down when the value of the ion-neutral collision frequency is on the same order of magnitude or lower than the wave frequency. Here we proceed as in  \citet{zaqarashvili2011a} and use the more general two-fluid theory instead of the single-fluid approximation. \citet{mouschovias2011} also considered the two-fluid theory and performed a very detailed investigation of wave propagation and instabilities in interstellar molecular clouds. 

To realistically represent the plasma in molecular clouds,  \citet{mouschovias2011} included in addition to ion-neutral collisions other effects as, e.g., ionization and recombination  and self-gravity. The main purpose of the present work is to obtain physical insight on the effect of ion-neutral collisions on waves without taking into account other effects that could hinder the specific role of collisions. For this reason, our approach is closer to \citet{zaqarashvili2011a} than to \citet{mouschovias2011}. As stated before, \citet{zaqarashvili2011a} put their emphasis in the derivation of the  equations and the dispersion relation. As examples, \citet{zaqarashvili2011a} computed some solutions, but they restricted themselves to a specific set of parameters and did not perform an in-depth parameter study. We follow the spirit of Paper~I and do not focus on a specific astrophysical plasma. Instead, we  take the relevant physical quantities as, e.g., the collision frequency, the ionization fraction, the Alfv\'en and sound velocities, etc., as free parameters whose values can be conveniently chosen. An inconvenience of the present approach is that the space of parameters  to be explored is very big. The advantage is that this approach makes possible to apply the results of this work  to a wide variety of  situations. Compared to  \citet{zaqarashvili2011a}, we cover a wider parameter range. In particular, we consider lower ionization levels and weaker magnetic fields than those used by  \citet{zaqarashvili2011a}.  The output of this investigation will be also useful to understand the recent results by \citet{soler2013aa} concerning the behavior of slow  waves propagating along a partially ionized magnetic flux tube. 

This paper is organized as follows. Section~\ref{sec:basic} contains the description of the equilibrium state and the basic equations. In Section~\ref{sec:lim} we explore the limit cases of uncoupled fluids and strongly coupled fluids and derive approximate expressions to the wave frequencies in those cases. Later, we investigate the general case in Section~\ref{sec:gen} by plotting dispersion diagrams when the collision frequency is varied between the uncoupled and strongly coupled regimes. We perform a specific application to magnetoacoustic waves propagating in the low solar atmosphere in Section~\ref{sec:app}.  Finally, Section~\ref{sec:discussion} contains some  concluding remarks.

\section{EQUILIBRIUM AND BASIC EQUATIONS}
\label{sec:basic}

The equilibrium configuration considered here is a uniform and unbounded partially ionized hydrogen plasma composed of ions, electrons, and neutrals. As in Paper~I, the plasma dynamics is studied using the two-fluid formalism \citep[see also, e.g.,][]{zaqarashvili2011a}. Ions and electrons are considered together as forming a single fluid, while neutrals are treated as a separate fluid. The ion-electron fluid and the neutral fluid interact by means of particle collisions, which transfer momentum between the species. It is implicitly assumed in the present version of the two-fluid theory that ion-electron and ion-ion collisions occur always more frequently than ion-neutral collisions, so that ions and electrons display fluid behavior regardless of the presence of neutrals.

The medium is permeated by a straight and constant magnetic field, $\bf B$. We use Cartesian coordinates and conveniently choose the reference frame so that the equilibrium magnetic field is orientated along the $z$-direction, namely ${\bf B} = B \hat{z}$, with $B$ constant. We ignore the effect of gravity. The omission of gravity is approximately valid as long as we consider wavelengths much shorter than the gravitational stratification scale height. We also assume that there are no flows in the equilibrium. We study linear adiabatic perturbations superimposed on the equilibrium state. The equations governing the behavior of the perturbations are the same as in Paper~I, namely
\begin{eqnarray}
\rhoi \frac{\partial \vi}{\partial t} &=& - \nabla \pie + \frac{1}{\mu} \left( \nabla \times {\bf b} \right) \times {\bf B} - \ain \left( \vi - \vn \right), \label{eq:momlinion} \\
\rhon \frac{\partial \vn}{\partial t} &=& - \nabla \pn  - \ain \left( \vn - \vi \right), \\
\frac{\partial {\bf b}}{\partial t} &=& \nabla \times \left( \vi \times {\bf B} \right), \\
\frac{\partial \pie}{\partial t} &=& -\gamma \Pie \nabla \cdot \vi, \\
\frac{\partial \pn}{\partial t} &=& -\gamma \Pn \nabla \cdot \vn, \label{eq:presslin}
\end{eqnarray}
where $\bf \vi$, $\pie$, $\Pie$, and $\rhoi$ are the velocity perturbation, gas pressure perturbation, equilibrium gas pressure, and equilibrium density of the ion-electron fluid, $\bf \vn$, $\pn$, $\Pn$, and $\rhon$ are the respective quantities for the neutral fluid, $\bf b$ is the magnetic field perturbation, $\mu$ is the magnetic permeability, $\gamma$ is the adiabatic index, and $\ain$ is the ion-neutral friction coefficient.

The friction coefficient, $\ain$, plays a very important role. It determines the strength of the ion-neutral friction force and, therefore, the importance of ion-neutral coupling. \citet{brag} gives the expression of $\ain$ in a hydrogen plasma. As in Paper~I, we take $\ain$ as a free parameter and do not use a specific expression of $\ain$. We do so to conveniently control the strength of the ion-neutral friction force. Our aim is to keep the investigation as general as possible for the results of the present article to be applicable in a wide range of astrophysical situations. Instead of using $\ain$, in the remaining of this article we write the equations using the ion-neutral, $\nuin$, and neutral-ion, $\nuni$ collision frequencies, which have a more obvious physical meaning. They are defined as
\begin{equation}
\nuin = \frac{\ain}{\rhoi}, \qquad \nuni = \frac{\ain}{\rhon}.
\end{equation}
Hence, both frequencies are related by $\rhoi \nuin = \rhon \nuni$.

\subsection{Normal mode analysis}

Here we perform a normal mode analysis. The temporal dependence of the perturbations is put proportional to $\exp\left( - i \omega t\right)$, where $\omega$ is the angular frequency. In addition, we perform a Fourier analysis of the perturbations in space. The spatial dependence of perturbations is put proportional to $\exp \left(i k_x x + ik_y y + i k_z z \right)$, where $k_x$, $k_y$, and $k_z$ are the components of the wavevector in the $x$-, $y$-, and $z$-directions, respectively.

Alfv\'en waves and magnetoacoustic waves are two distinct classes of MHD waves in a uniform plasma \citep[see, e.g.,][]{cramer2001,goossens2003}. Alfv\'en waves are incompressible and propagate vorticity perturbations. Magnetoacoustic waves are compressible and do not produce  vorticity perturbations. Alfv\'en waves were investigated in Paper~I. Here we focus on magnetoacoustic waves. Compressibility is an appropriate quantity to describe magnetoacoustic waves  \citep[see, e.g.,][]{lighthill1960}. By working with compressibility perturbations we are able to decouple magnetoacoustic waves from Alfv\'en waves without imposing any restriction on the direction of the wavevector.  We define $\deli$ and $\deln$ as the compressibility perturbations of the ion-electron fluid and the neutral fluid, respectively,
\begin{eqnarray}
\deli &\equiv& \nabla \cdot \vi = i k_x v_{x,\rm i} + i k_y v_{y,\rm i} + i k_z v_{z,\rm i}, \\
\deln &\equiv& \nabla \cdot \vn = i k_x v_{x,\rm n} + i k_y v_{y,\rm n} + i k_z v_{z,\rm n}
\end{eqnarray}
We combine Equations~(\ref{eq:momlinion})--(\ref{eq:presslin}) and after  some algebraic manipulations we obtain the two following coupled equations involving $\deli$ and $\deln$ only, namely
\begin{eqnarray}
&&\left( \omega^4 - \omega^2 k^2 \left( \cie^2 + \ca^2 \right) + k^2 k_z^2 \ca^2 \cie^2 \right) \deli = \nonumber \\
&& - i \nuin \omega^3 \left( \deli-\deln \right) \nonumber \\
&& + \frac{i\nuin}{\omega+i\left( \nuni+\nuin \right)} k^2 k_z^2 \ca^2  \left( \cie^2 \deli - \cn^2 \deln \right) , \label{eq:deli} \\
&&\left( \omega^2 - k^2 \cn^2 \right)\deln = - i \nuni \omega  \left( \deln-\deli \right), \label{eq:deln}
 \end{eqnarray}
where $k^2 = k_x^2 + k_y^2 + k_z^2$ and $\ca$, $\cie$, and $\cn$ are the Alfv\'en velocity, the ion-electron sound velocity, and the neutral sound velocity defined as
\begin{equation}
\ca^2 = \frac{B^2}{\mu \rhoi}, \qquad \cie^2 = \frac{\gamma\Pie}{\rhoi}, \qquad \cn^2 = \frac{\gamma\Pn}{\rhon}.
\end{equation}
Note that the Alfv\'en velocity is here defined using the density of the ionized fluid only. Equations~(\ref{eq:deli}) and (\ref{eq:deln}) are the governing equations for linear compressibility perturbations. Therefore, they govern the behavior of magnetoacoustic waves. 

The magnetoacoustic  character of the waves is determined by the relative values of the  Alfv\'en and sound velocities. In this work we use the following definition of the ion-electron $\beta_{\rm i}$, 
\begin{equation}
\beta_{\rm i} = \frac{\cie^2}{\ca^2} = \frac{\gamma \Pie}{B^2 / \mu}. \label{eq:beta}
\end{equation}
 Thus, using the present definition of $\beta_{\rm i}$ we can quantify the relative importance of the Alfv\'en velocity, $\ca$, and ion-electron sound velocity, $\cie$. No restriction on the value of the  neutral sound velocity, $\cn$, has been imposed so far. However, to simplify matters and decrease the number of parameters involved, we proceed as in \citet{zaqarashvili2011a} and \citet{soler2013aa} and assume that there is a strong thermal coupling between the ion-electron fluid and the neutral fluid. As a consequence, the equilibrium temperature, $T$, is the same for both fluids. We use the ideal gas law to relate the equilibrium gas pressure, density, and temperature as
\begin{equation}
\Pie = 2 \rhoi R\, T, \qquad \Pn = \rhon R\, T, \label{eq:gaslaw}
\end{equation}
where $R$ is the ideal gas constant. The factor 2 in the expression for ion-electrons is present because the partial pressures of both electrons and ions have to be summed up. In a hydrogen plasma, the partial pressures of electrons and ions are the same. From Equation~(\ref{eq:gaslaw}), we deduce that the requirement that all the species have the same temperature results in imposing a relation between the sound velocities as,
\begin{equation}
\cie^2 = 2\cn^2.  \label{eq:cicn}
\end{equation}
Hence, using Equations~(\ref{eq:beta}) and (\ref{eq:cicn}) we see that fixing the value of $\beta_{\rm i}$ is enough to establish the relation between the  Alfv\'en velocity and the two sound velocities.

For the subsequent analysis, we decompose the wavevector, ${\bf k}$, in its components parallel and perpendicular to the equilibrium magnetic field as ${\bf k} = k_z \hat{z} + k_\perp \bf{1}_\perp$, where $k_\perp = \sqrt{k_x^2 + k_y^2}$ and $\bf{1}_\perp$ is the unity vector in the perpendicular direction to $\bf B$. We define $\theta$ as the angle that forms the wavevector, $\bf k$, with the equilibrium magnetic field, $\bf B$. Then, we can write $k_z = k \cos\theta$ and $k_\perp = k\sin\theta$. For propagation perpendicular to the magnetic field, $\theta = \pi/2$, whereas for propagation parallel to the magnetic field, $\theta = 0$. Intermediate values of $\theta$ represent oblique propagation.

\section{APPROXIMATIONS FOR LIMIT VALUES OF THE COLLISION FREQUENCIES}
\label{sec:lim}

Before tackling the general case for arbitrary values of the collision frequencies,  it is instructive to consider two paradigmatic limit cases, namely the  uncoupled case in which the two fluids do not interact, and the strongly coupled case in which the two fluids behave as a single fluid.

\subsection{Uncoupled case}

\label{sec:collisionless}

Here, we remove the effect of ion-neutral collisions. We represent this situation by setting $\nuin = \nuni = 0$.  Equations~(\ref{eq:deli}) and (\ref{eq:deln}) become
\begin{eqnarray}
\left( \omega^4 - \omega^2 k^2 \left( \cie^2 + \ca^2 \right) + k^4 \ca^2 \cie^2  \cos^2\theta \right) \deli &=& 0, \label{eq:deli2} \\
\left( \omega^2 - k^2 \cn^2 \right)\deln &=& 0. \label{eq:deln2}
 \end{eqnarray}
Now, the equations governing compressibility perturbations in the two fluids are decoupled. The waves in the two fluids can be studied separately.

We start by considering waves in the ion-electron fluid. For nonzero $\deli$, the solutions to Equation~(\ref{eq:deli2}) must satisfy
\begin{equation}
 \omega^4 - \omega^2 k^2 \left( \cie^2 + \ca^2 \right) + k^4 \ca^2 \cie^2  \cos^2\theta  = 0. \label{eq:reldisperma}
\end{equation}
Equation~(\ref{eq:reldisperma}) is the well-known dispersion relation of magnetoacoustic waves in a fully ionized plasma \citep[see, e.g.,][]{lighthill1960}. The solutions to Equation~(\ref{eq:reldisperma}) are
\begin{equation}
\omega^2 = k^2 \frac{\ca^2+\cie^2}{2}  \pm k^2 \frac{\ca^2+\cie^2}{2} \left[ 1 - \frac{4\ca^2\cie^2 \cos^2\theta}{\left(\ca^2 + \cie^2\right)^2} \right]^{1/2}, \label{eq:magnetoideal}
\end{equation}
where the $+$ sign is for the fast wave and the $-$ sign is for the slow wave. The magnetoacoustic character of these two distinct wave modes depends on the relative values of $\ca$ and $\cie$ \citep[see, e.g.,][]{goossens2003}. 

In the low-$\beta_{\rm i}$ case, $\ca^2 \gg \cie^2$ and the solutions simplify to
\begin{equation}
\omega^2 \approx k^2 \left( \ca^2 + \cie^2 \right) \approx k^2 \ca^2, \label{eq:fast}
\end{equation}
for the fast wave and
\begin{equation}
\omega^2 \approx k^2 \frac{\ca^2\cie^2}{\ca^2 + \cie^2} \cos^2\theta \approx k^2 \cie^2 \cos^2\theta, \label{eq:slow}
\end{equation}
for the slow wave. When $\ca^2 \gg \cie^2$ the fast wave is a magnetic wave that propagates isotropically at the Alfv\'en velocity, $\ca$, while the slow wave is essentially a sound wave that is guided by the magnetic field and travels at the ion-electron sound velocity, $\cie$. Conversely, in the high-$\beta_{\rm i}$ case $\cie^2 \gg \ca^2$ and the solutions simplify to
\begin{equation}
\omega^2 \approx k^2 \left( \ca^2 + \cie^2 \right) \approx k^2 \cie^2, \label{eq:fasth}
\end{equation}
for the fast wave and
\begin{equation}
\omega^2 \approx k^2 \frac{\ca^2\cie^2}{\ca^2 + \cie^2} \cos^2\theta \approx k^2 \ca^2 \cos^2\theta, \label{eq:slowh}
\end{equation}
for the slow wave. Now the fast wave behaves as an isotropic sound wave, while the slow wave is a magnetic wave guided by the magnetic field.

We turn to the waves in the neutral fluid. The solutions to Equation~(\ref{eq:deln2}) for nonzero $\deln$ are
\begin{equation}
\omega^2 = k^2 \cn^2. \label{eq:reldispersound}
\end{equation}
Equation~(\ref{eq:reldispersound}) represent acoustic (or sound) waves  in a gas. These waves propagate isotropically at the neutral sound velocity, $\cn$, and are unaffected by the magnetic field. 

As expected, in the absence of ion-neutral collisions we consistently recover the classic magnetoacoustic waves in the ion-electron fluid and the classic acoustic waves in the neutral fluid. Therefore, three distinct waves are present in the uncoupled, collisionless case. These waves do not interact and are undamped in the absence of collisions. The waves would be damped if nonideal processes as, e.g., viscosity, resistivity, or thermal conduction, are taken into account. In the remaining of this article, we use the adjective `classic' to refer to the waves found in the uncoupled case.

\subsection{Strongly coupled case}

\label{sec:highly}

Conversely to the uncoupled case, the strongly coupled  limit represents the situation in which ion-electrons and neutrals  behave as a single fluid. To study this case, we take the limits $\nuin \to \infty$ and $\nuni \to \infty$ in Equations~(\ref{eq:deli}) and (\ref{eq:deln}). We realize that, if $\omega \neq 0$, it is necessary that $\deli = \deln$ for the equations to remain finite. This is equivalent to assume that the two fluids move as a whole. Then,  when $\nuin \to \infty$ and $\nuni \to \infty$, Equations~(\ref{eq:deli}) and (\ref{eq:deln}) reduce to a single equation, namely
\begin{eqnarray}
 &&\left( \omega^4 - \omega^2 k^2 \frac{\ca^2 + \cie^2 + \chi \cn^2}{1+\chi} \right. \nonumber \\ 
 &+& \left. k^4 \frac{\ca^2 \left( \cie^2 +\chi \cn^2\right)}{\left( 1+\chi \right)^2} \cos^2\theta \right) \Delta_{\rm i,n} = 0,  \label{eq:high}
\end{eqnarray}
where we use $\Delta_{\rm i,n}$ to represent either $\deli$ or $\deln$ and the ionization fraction of the plasma, $\chi$, is defined as 
\begin{equation}
\chi = \frac{\rhon}{\rhoi}.
\end{equation}
For nonzero $\Delta_{\rm i,n}$, the solutions to Equation~(\ref{eq:high}) must satisfy
\begin{equation}
\omega^4 - \omega^2 k^2 \frac{\ca^2 + \cie^2 + \chi \cn^2}{1+\chi} + k^4 \frac{\ca^2 \left( \cie^2 +\chi \cn^2\right)}{\left( 1+\chi \right)^2} \cos^2\theta  = 0.  \label{eq:reldisperhigh}
\end{equation} 
This is the wave dispersion relation in the strongly coupled case. Its solutions are
\begin{eqnarray}
\omega^2 &=& k^2 \frac{\ca^2+\cie^2+\chi\cn^2}{2(1+\chi)} \nonumber \\
 &\pm & k^2\frac{\ca^2+\cie^2+\chi\cn^2}{2(1+\chi)} \left[ 1 - \frac{4\ca^2(\cie^2+\chi \cn^2)\cos^2\theta}{\left(\ca^2 + \cie^2+\chi \cn^2\right)^2} \right]^{1/2}, \label{eq:magnetolim}
\end{eqnarray}
where the $+$ sign is for the modified fast wave and the $-$ sign is for the modified slow wave. We use the adjective `modified' to stress that these waves are the counterparts of the classic fast and slow modes but modified by the presence of the neutral fluid. 

The first important difference between the uncoupled and strongly coupled cases is in the number of solutions. In the uncoupled case there are three distinct waves, namely the classic slow and fast magnetoacoustic waves and the neutral acoustic wave, but in the strongly coupled  limit we  only find the modified version of the slow and fast magnetoacoustic modes. The modified counterpart of the classic neutral acoustic mode is apparently absent. The question then arises, what happened to the neutral acoustic mode? To answer this question, we study in the following paragraphs the character of the modified fast and slow modes depending on the relative values of the Alfv\'en and sound velocities. 

A remark should be made before tackling the approximate study of the solutions in Equation~(\ref{eq:magnetolim}). We recall that the neutral and ion-electron sound velocities are related by $\cie^2 = 2\cn^2$. This condition means that $\cie^2$ and $\cn^2$ are of the same order of magnitude. However, this is not the end of the story. We notice that the value of the ionization fraction, $\chi$, also plays a role in determining the solutions to  Equation~(\ref{eq:magnetolim}). Specifically, $\chi$ appears multiplying $\cn^2$, which means that $\chi$ can increase or decrease the effective value of the neutral sound velocity. Thus, we need to  compare $\chi \cn^2$ with $\ca^2$ and $\cie^2$ in order to determine the nature of the solutions.

The approximate study of the solutions given in Equation~(\ref{eq:magnetolim}) is done considering eight typical cases: 
\begin{enumerate}
\item Case $\ca^2 \gg \cie^2 \gg \chi\cn^2$. This case corresponds to an almost fully ionized ($\chi \ll 1$) low-$\beta_{\rm i}$ plasma. The solutions given in Equation~(\ref{eq:magnetolim}) simplify to
\begin{equation}
\omega^2 \approx  k^2 \ca^2, \label{eq:modfast1}
\end{equation}
for the modified fast wave and
\begin{equation}
\omega^2 \approx  k^2 \cie^2 \cos^2\theta,\label{eq:modslow1}
\end{equation}
for the modified slow wave.  We recover the fast and slow waves of the uncoupled case (Equations~(\ref{eq:fast}) and (\ref{eq:slow})). No trace of the neutral acoustic mode remains in this limit because the amount of neutrals is negligible.

\item Case  $\ca^2 \gg \cie^2 \sim \chi\cn^2$.  This case corresponds to a low-$\beta_{\rm i}$ plasma with $\chi \sim 1$. The solutions are approximated as
\begin{equation}
\omega^2  \approx k^2 \frac{\ca^2}{1+\chi}, \label{eq:modfast2}
\end{equation}
for the modified fast wave and
\begin{equation}
\omega^2 \approx k^2 \frac{\cie^2+\chi\cn^2}{1+\chi} \cos^2\theta,\label{eq:modslow2}
\end{equation}
for the modified slow wave. By comparing Equations~(\ref{eq:fast}) and (\ref{eq:modfast2}), we see that the square of the fast mode frequency is reduced by the factor $\left(\chi+1\right)^{-1}$ compared to the value in the uncoupled case. As a consequence, the effective Alfv\'en  velocity in the strongly coupled  limit is $\ca/\sqrt{1+\chi}$.   This is equivalent to replace the ion density, $\rhoi$, by the sum of the ion and neutral densities, $\rhoi+\rhon$, in the definition of the Alfv\'en velocity. The same result was obtained by \citet{kumar2003} and \citet{soler2013aa} for fast magnetoacoustic waves in the case $\cn^2 = 0$. This is also the same result obtained in Paper~I and \citet{soler2012} for Alfv\'en waves. On the other hand, the modification of the slow mode frequency is more complicated (compare Equations~(\ref{eq:slow}) and (\ref{eq:modslow2})). The expression of the slow mode frequency involves both $\cie$ and $\cn$. Consequently, an effective sound velocity, $c_{\rm eff}$, can be defined as 
\begin{equation}
c_{\rm eff}^2= \frac{\cie^2 + \chi \cn^2 }{ 1+\chi } = \frac{\rhoi \cie^2 + \rhon \cn^2 }{ \rhoi+\rhon }. \label{eq:ceff}
\end{equation}
The effective sound velocity is the weighted average of the sound velocities of ion-electrons and neutrals, and was first obtained by \citet{soler2013aa} for slow magnetoacoustic waves propagating along magnetic flux tubes. The fact that the expression of the effective sound velocity (Equation~(\ref{eq:ceff})) involves both neutral and ion-electron sound velocities suggests that, in this case, the modified slow mode is the descendant of both the {\bf classic} slow mode and the neutral acoustic mode. 

\item Case  $\ca^2 \gg \chi\cn^2 \gg \cie^2 $.  This case corresponds to a weakly ionized ($\chi \gg 1$) low-$\beta_{\rm i}$ plasma, but the magnetic field is strong enough for $\ca^2$ to be much larger than $\chi \cn^2$. The solutions given in Equation~(\ref{eq:magnetolim}) become
\begin{equation}
\omega^2  \approx k^2 \frac{\ca^2}{\chi}, \label{eq:modfast3}
\end{equation}
for the modified fast wave and
\begin{equation}
\omega^2 \approx k^2 \cn^2 \cos^2\theta,\label{eq:modslow3}
\end{equation}
for the modified slow wave. The effective Alfv\'en  velocity  is now $\ca/\sqrt{\chi}$.   This is equivalent to replace the ion density, $\rhoi$, by the neutral density, $\rhon$, in the definition of the Alfv\'en velocity. On the other hand, the slow wave is the descendant of the neutral acoustic wave that becomes guided by the magnetic field. Therefore, the frequencies of both fast and slow waves are determined by the density of the neutral fluid alone, which is indirectly affected by the magnetic field.

\item Case $\chi \cn^2 \gg \ca^2 \gg \cie^2$. In this case we consider a very weakly ionized ($\chi \to \infty$) low-$\beta_{\rm i}$ plasma.  The approximate frequencies turn to be
\begin{equation}
\omega^2 \approx k^2 \cn^2, \label{eq:modfasth4}
\end{equation}
for the modified fast wave and
\begin{equation}
\omega^2 \approx  k^2 \frac{\ca^2}{\chi} \cos^2\theta,\label{eq:modslowh4}
\end{equation}
for the modified slow wave. The modified fast wave can straightforwardly be related to the isotropic acoustic mode of the neutral fluid (Equation~(\ref{eq:reldispersound})), while the modified slow wave is a guided magnetic wave with the effective Alfv\'en velocity depending on the neutral density alone. As  in case (3), neither the ion-electron sound velocity nor the ion density  play a role. The fact that the plasma is weakly ionized changes completely the physical nature of the solutions compared to the classic magnetoacoustic waves in a low-$\beta_{\rm i}$ fully ionized plasma. 

\item Case $\cie^2  \gg \ca^2 \gg \chi\cn^2$. This corresponds to an almost fully ionized ($\chi \ll 1$) high-$\beta_{\rm i}$ plasma. The solutions given in Equation~(\ref{eq:magnetolim}) simplify to
\begin{equation}
\omega^2 \approx  k^2 \cie^2, \label{eq:modfast25}
\end{equation}
for the modified fast wave and
\begin{equation}
\omega^2 \approx k^2 \ca^2 \cos^2\theta,\label{eq:modslow25}
\end{equation}
for the modified slow wave. We recover Equations~(\ref{eq:fasth}) and (\ref{eq:slowh}) obtained in the high-$\beta_{\rm i}$ collisionless case. No trace of the neutral acoustic mode remains in this limit.

\item Case $\cie^2  \gg \chi\cn^2 \gg \ca^2 $. This corresponds again to an almost fully ionized ($\chi \ll 1$) high-$\beta_{\rm i}$ plasma, but now the magnetic field is so weak that $ \chi\cn^2 \gg \ca^2 $. The solutions given in Equation~(\ref{eq:magnetolim}) simplify to
\begin{equation}
\omega^2 \approx  k^2 \cie^2, \label{eq:modfast26}
\end{equation}
for the modified fast wave and
\begin{equation}
\omega^2 \approx k^2 \ca^2 \cos^2\theta,\label{eq:modslow26}
\end{equation}
for the modified slow wave. As in the previous case (5), we revert to Equations~(\ref{eq:fasth}) and (\ref{eq:slowh}) obtained in the uncoupled case. Cases (5) and (6) point out that the relative values of $\ca^2$ and $\chi\cn^2$ are not important as long as $\cie^2$ remains much larger than both of them. The same approximations to the frequencies are found in cases (5) and (6). The properties of the neutral fluid are not important in  both cases (5) and (6).

\item Case $\chi\cn^2 \sim \cie^2  \gg \ca^2$. This is a  high-$\beta_{\rm i}$ plasma with $\chi \sim 1$. The solutions given in Equation~(\ref{eq:magnetolim}) simplify to
\begin{equation}
\omega^2 \approx k^2 \frac{\cie^2 + \chi \cn^2}{1+\chi} = k^2 c_{\rm eff}^2, \label{eq:modfast27}
\end{equation}
for the modified fast wave and
\begin{equation}
\omega^2 \approx k^2 \frac{\ca^2}{1+\chi} \cos^2\theta,\label{eq:modslow27}
\end{equation}
for the modified slow wave. Equations~(\ref{eq:modfast27}) and (\ref{eq:modslow27}) are the modified version of Equations~(\ref{eq:fasth}) and (\ref{eq:slowh}) obtained in the high-$\beta_{\rm i}$ uncoupled case, where  $c_{\rm eff}$ replaces $\cie$ and $\ca/\sqrt{1+\chi}$ replaces $\ca$. Here we find that the modified fast mode is the descendant of both the classic fast mode and the neutral acoustic mode, while the modified slow mode is a guided magnetic wave whose frequency depends on both ion and neutral densities.

\item Case $\chi\cn^2 \gg \cie^2  \gg \ca^2$. This corresponds to a weakly ionized ($\chi \gg 1$) high-$\beta_{\rm i}$ plasma. The solutions given in Equation~(\ref{eq:magnetolim}) simplify to
\begin{equation}
\omega^2 \approx k^2 \cn^2, \label{eq:modfast228}
\end{equation}
for the modified fast wave and
\begin{equation}
\omega^2 \approx k^2 \frac{\ca^2}{\chi} \cos^2\theta,\label{eq:modslow228}
\end{equation}
for the modified slow wave. The approximation found here are the same as in case (4). The properties of the neutral fluid completely determine the behavior of the waves. Neutrals feel the magnetic field indirectly through the collisions with ions.

\end{enumerate}

The results of the above approximate study are summarized in Table~\ref{tab:summ}. Based on this study, we can now answer the previous question about the apparent absence of the neutral acoustic mode in a strongly coupled plasma. We conclude that the classic neutral acoustic mode  and the classic ion-electron magnetoacoustic modes heavily interact when ion-neutral collisions are at work. The two resulting modes in the strongly coupled  regime (modified fast mode and modified slow mode) have, in general, mixed properties and are affected by the physical conditions in the two fluids. The degree to which the properties of the classic waves are present in the resulting waves depends on the relative values of the  Alfv\'en and sound velocities and on the ionization fraction of the plasma. 

In addition, the wave frequencies in the strongly coupled limit are real as in the uncoupled case. This means that the waves are undamped in the limit of high collision frequencies as well. The damping of magnetoacoustic waves due to ion-neutral collisions takes place for intermediate collision frequencies.  As stated before, the waves would be damped if additional damping mechanisms are considered, but this falls beyond the aim of the present article.  Both the ion-neutral damping and the coupling between modes are investigated in Section~\ref{sec:gen} by studying the modification of the wave frequencies when a progressive variation of the collision frequencies between the uncoupled limit and strongly coupled  limit is considered.

\section{NUMERICAL STUDY FOR ARBITRARY COLLISION FREQUENCIES}
\label{sec:gen}

From here on, we consider arbitrary values of $\nuin$ and $\nuni$. With no approximation, we combine the coupled Equations~(\ref{eq:deli}) and (\ref{eq:deln}) to obtain separate equations for $\deli$ and $\deln$ as
\begin{eqnarray}
\mathcal{D}\left( \omega\right) \deli &=& 0, \\
i\nuni\omega \frac{\mathcal{D} \left( \omega\right)}{D_{\rm n} \left( \omega\right) }\deln &=& 0.
 \end{eqnarray}
with
\begin{equation}
\mathcal{D}\left( \omega \right) = D_{\rm i} \left( \omega\right) D_{\rm n}\left( \omega\right) + D^2_{\rm c}\left( \omega\right), \label{eq:relma}
\end{equation}
where $D_{\rm i}\left( \omega\right)$, $D_{\rm n}\left( \omega\right)$, and $D_{\rm c}^2\left( \omega\right)$ are defined as 
\begin{eqnarray}
D_{\rm i}\left( \omega\right) &=& \omega^3 \left( \omega + i \nuin \right) - \omega^2 k^2 \left( \ca^2 + \cie^2  \right) \nonumber \\
 &&+ \frac{\omega + i  \nuni}{\omega + i (\nuin+\nuni)} k^4 \ca^2 \cie^2 \cos^2\theta, \\
D_{\rm n}\left( \omega\right) &=& \omega \left( \omega + i \nuni \right) - k^2 \cn^2, \\
 D_{\rm c}^2\left( \omega\right) &=& \frac{\omega \nuin \nuni}{\omega + i (\nuin+\nuni)} \left[ \omega^3 \left( \omega + i (\nuin+\nuni) \right) \right. \nonumber \\
 && \left. - k^4  \ca^2 \cn^2 \cos^2\theta \right].
\end{eqnarray}
The waves in the ion-electron fluid and in the neutral fluid are described by the same dispersion relation. The dispersion relation is 
\begin{equation}
\mathcal{D}\left( \omega\right) = 0, \label{eq:reldispertot}
\end{equation}
and is equivalent to the dispersion relation  given by \citet{zaqarashvili2011a} in their Equation~(57), although here we use a different notation. Expressed in polynomial form, the dispersion relation is a 7th-order polynomial in $\omega$, so that there are seven solutions \citep[see][]{zaqarashvili2011a}. The frequencies  are complex, namely $\omega = \omega_{\rm R} + i\omega_{\rm I}$, where $\omega_{\rm R} $ and $\omega_{\rm I}$ are the real and imaginary parts. The imaginary part of the frequency is negative and accounts for the exponential damping rate of the perturbations due to ion-neutral collisions. 

Before exploring the solutions to the dispersion relation, it is necessary to make a comment on the nature of the solutions. We distinguish  between two different kind of solutions depending of whether $\omega_{\rm R}$ is zero or nonzero. (1) The first kind of solutions are those with $\omega_{\rm R} = 0$, i.e., solutions that represent purely decaying disturbances. These solutions are not waves in the strict sense because they do not propagate.  The solutions with $\omega_{\rm R} = 0$ were called `vortex modes' by \citet{zaqarashvili2011a} but, in our view, the name `vortex modes' does not reflect the true nature of these solutions.  The reason is that vortex modes in a partially ionized plasma are intrinsically related to Alfv\'en waves (see Paper~I), not to magnetoacoustic waves.  Magnetoacoustic modes do not produce vorticity perturbations \citep[see, e.g.,][]{goossens2003}, hence the purely imaginary solutions of the magnetoacoustic dispersion relation cannot represent fluid vorticity. Instead, we propose to physically interpret the  solutions with $\omega_{\rm R} = 0$  as `entropy modes' \citep[see, e.g.,][]{goedbloed2004,murawski2011}. Entropy modes represent perturbations of density and, therefore, compression of the plasma. Compression is an intrinsic property of magnetoacoustic modes. Entropy modes have zero frequency in the absence of collisions. Due to ion-neutral collisions, perturbations in density are damped and the entropy mode frequency acquires an imaginary part, while its real part  remains zero. (2) The second kind of solutions are those with $\omega_{\rm R} \neq 0$. These solutions represent  propagating waves and appear in pairs, namely $\omega_1 = \omega_{\rm R} + i \omega_{\rm I}$ and $\omega_2 = -\omega_{\rm R} + i \omega_{\rm I}$. Both $\omega_1$ and $\omega_2$ represent the same magnetoacoustic wave, but the real parts of their frequencies have opposite signs \citep[see][]{carbonell2004}.  The different signs of the real parts  account for forward ($\omega_{\rm R} > 0$) and backward ($\omega_{\rm R} < 0$) propagation with respect to the direction of the wavevector, $\bf k$.  Since the equilibrium is static, the two directions of propagation are physically equivalent. As shown later, the number of entropy modes and propagating waves is not constant. Depending on the physical parameters considered, there can be conversion between these two kind of solutions.

Unlike the limit cases studied in Section~\ref{sec:lim}, no simple analytic solutions of the dispersion relation  can be obtained when $\nuin$ and $\nuni$ are arbitrary. Instead, we solve the dispersion relation numerically. The numerically obtained dispersion diagrams are plotted  as functions of the averaged collision frequency, $\bar{\nu}$, defined as
\begin{equation}
\bar{\nu} = \frac{\rhoi \nuin + \rhon\nuni}{\rhoi+\rhon} = \frac{2}{\rhoi+\rhon} \alpha_{\rm in}.
\end{equation}
 
 Because of the complex behavior displayed by the solutions and the large number of parameters involved, we  organize the presentation of the results and their discussion as follows.  First, we investigate the specific situation in which the wave propagation is strictly perpendicular to the magnetic field direction. This paradigmatic case is helpful  to get  physical insight of the behavior of the various solutions and to  illustrate some general results that repeatedly appear in this investigation. Slow modes are absent from the discussion for perpendicular propagation because they are unable to propagate across the magnetic field. Subsequently, we study the other limit case, i.e., wave propagation parallel to the magnetic field, so that slow modes are added to the discussion. Finally, the general case of oblique propagation is taken into account.

\subsection{Perpendicular propagation}

We start by studying the case of perpendicular propagation to the magnetic field.  Hence we set $\theta = \pi/2$.  The dispersion relation (Equation~(\ref{eq:reldispertot})) reduces to a 5th-order polynomial because the slow magnetoacoustic wave is absent.  Conversely, the fast magnetoacoustic mode and the neutral acoustic mode can propagate across the field and so they remain when $\theta = \pi/2$.

\begin{figure*}
	\centering
	\includegraphics[width=.49\columnwidth]{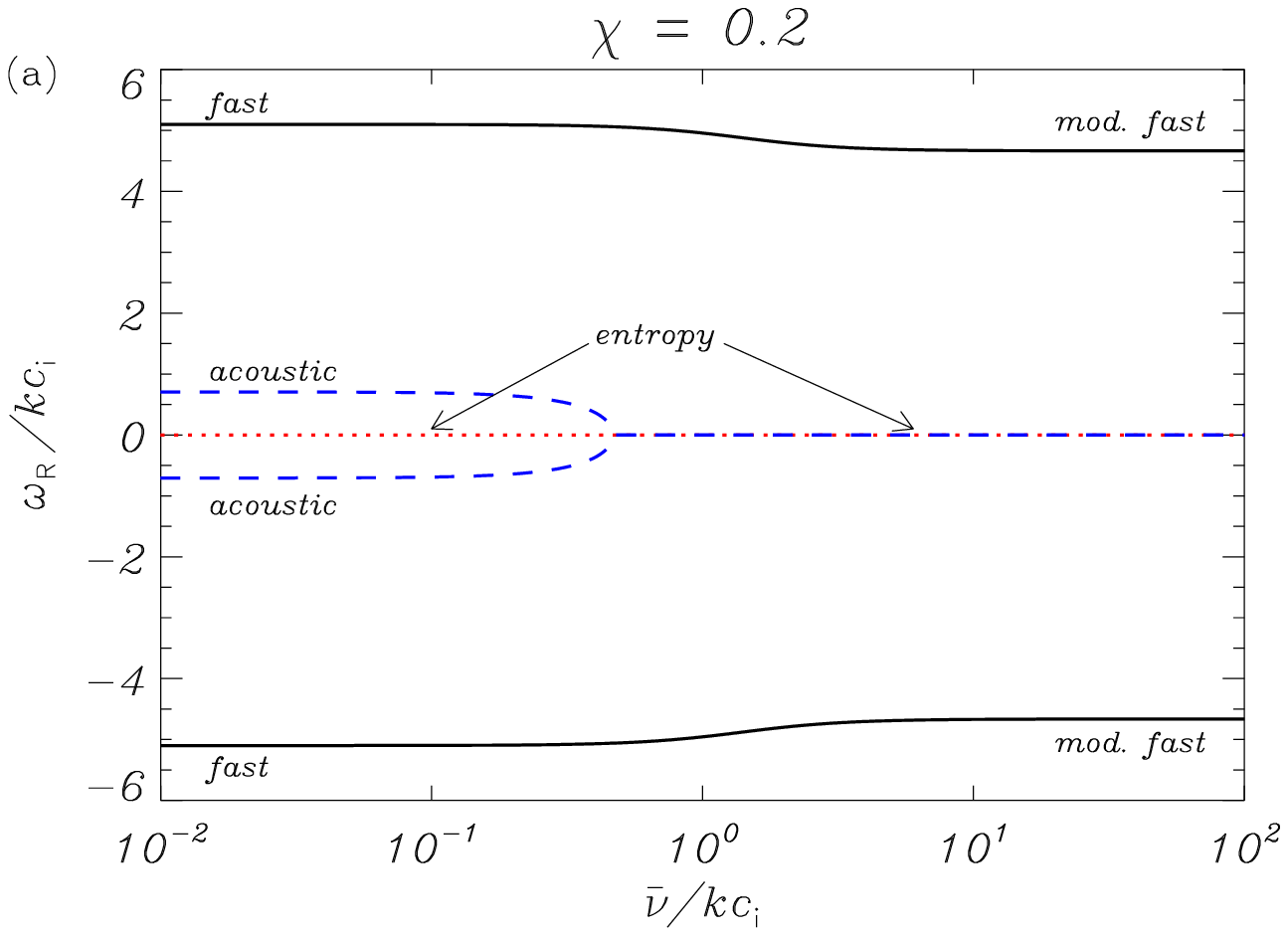}
	\includegraphics[width=.49\columnwidth]{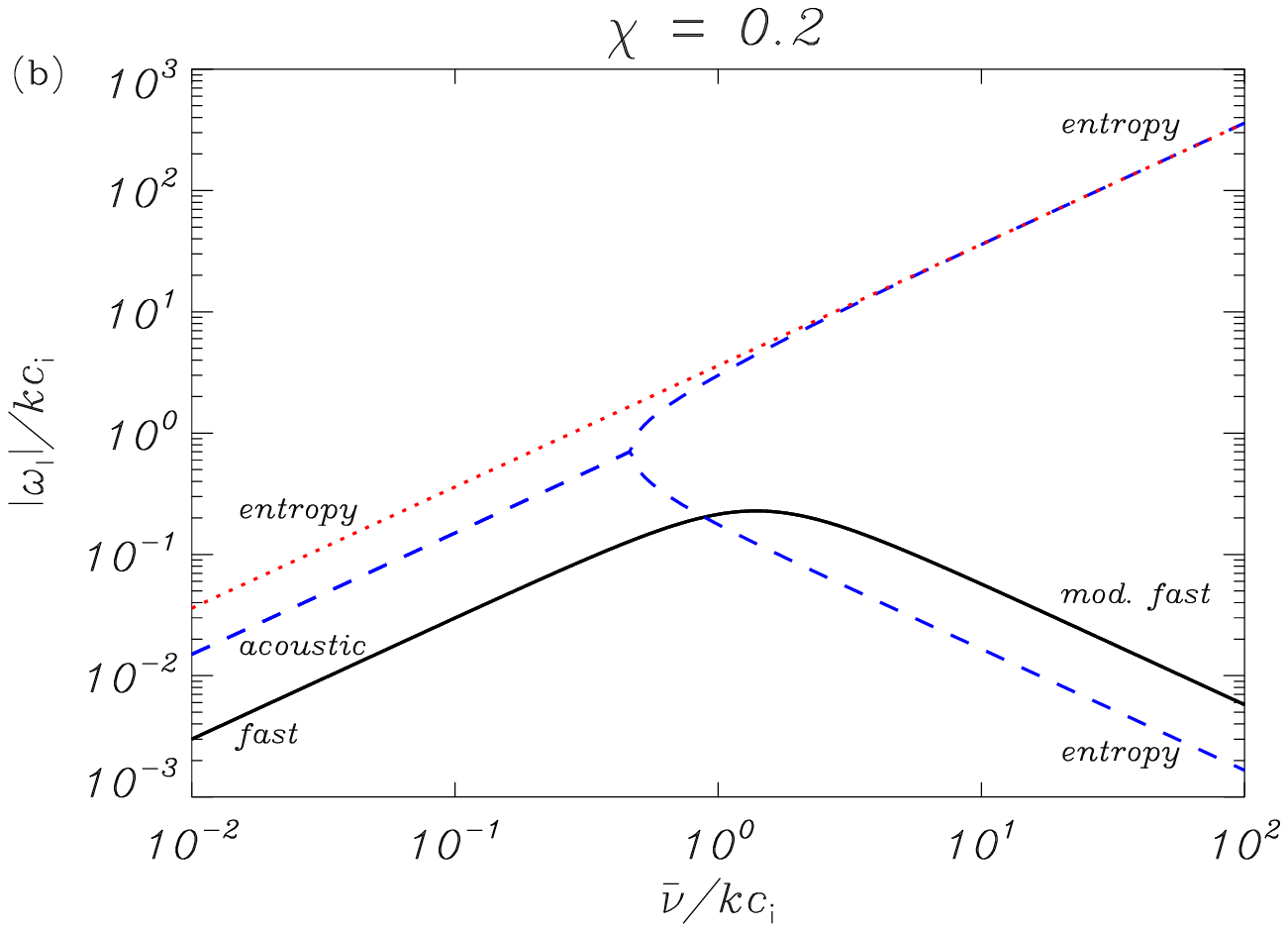}
	\includegraphics[width=.49\columnwidth]{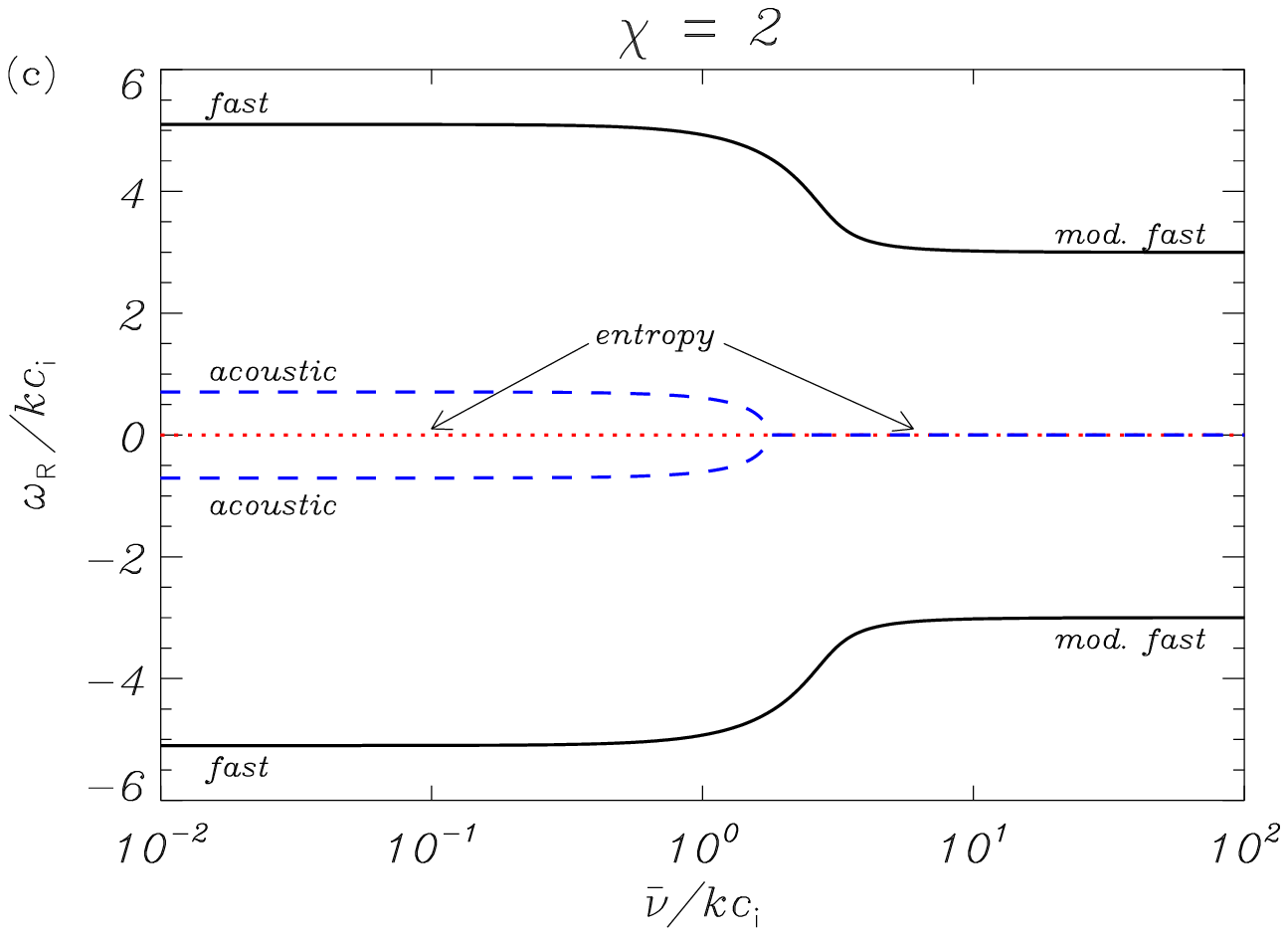}
	\includegraphics[width=.49\columnwidth]{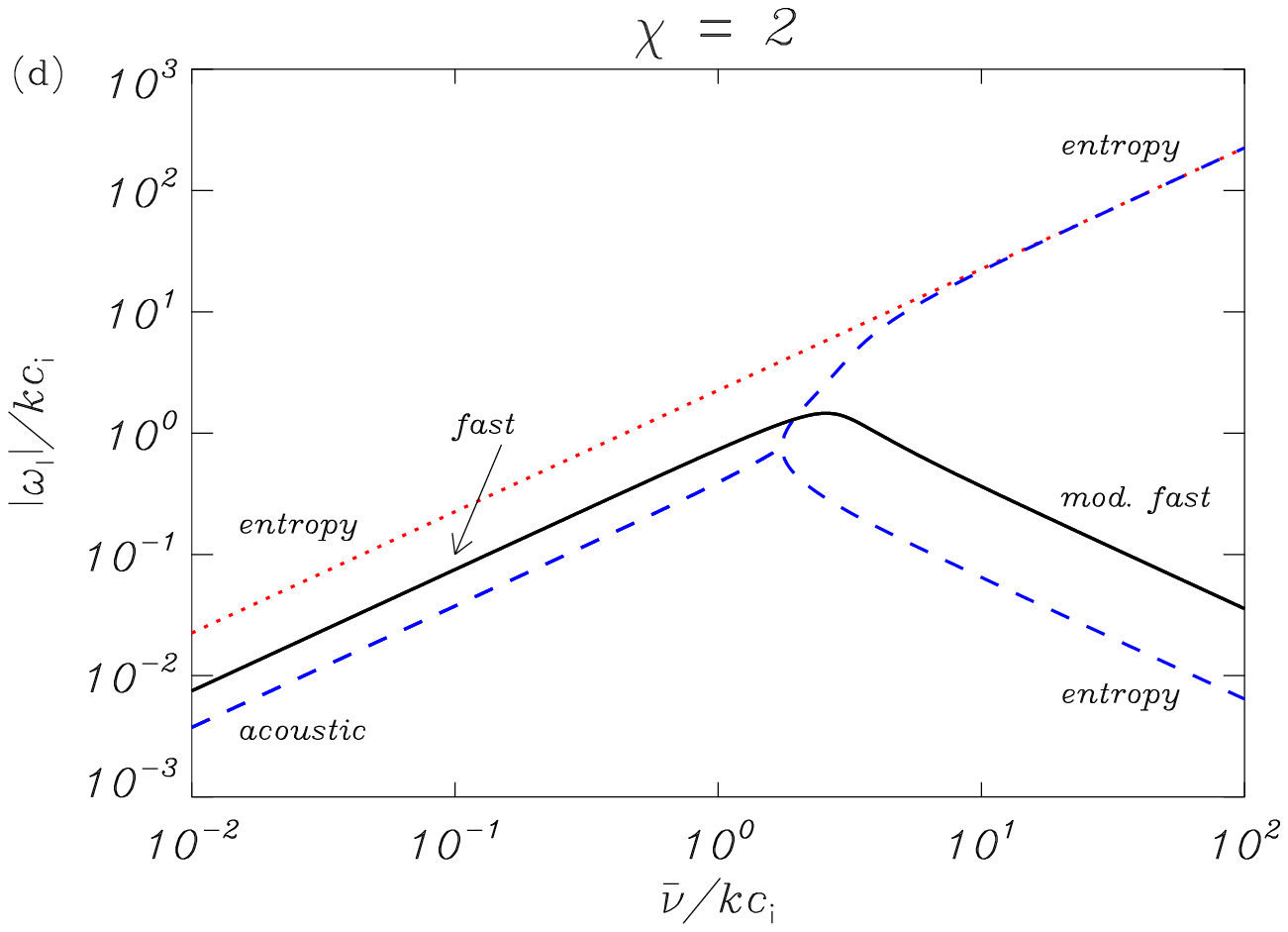}
	\includegraphics[width=.49\columnwidth]{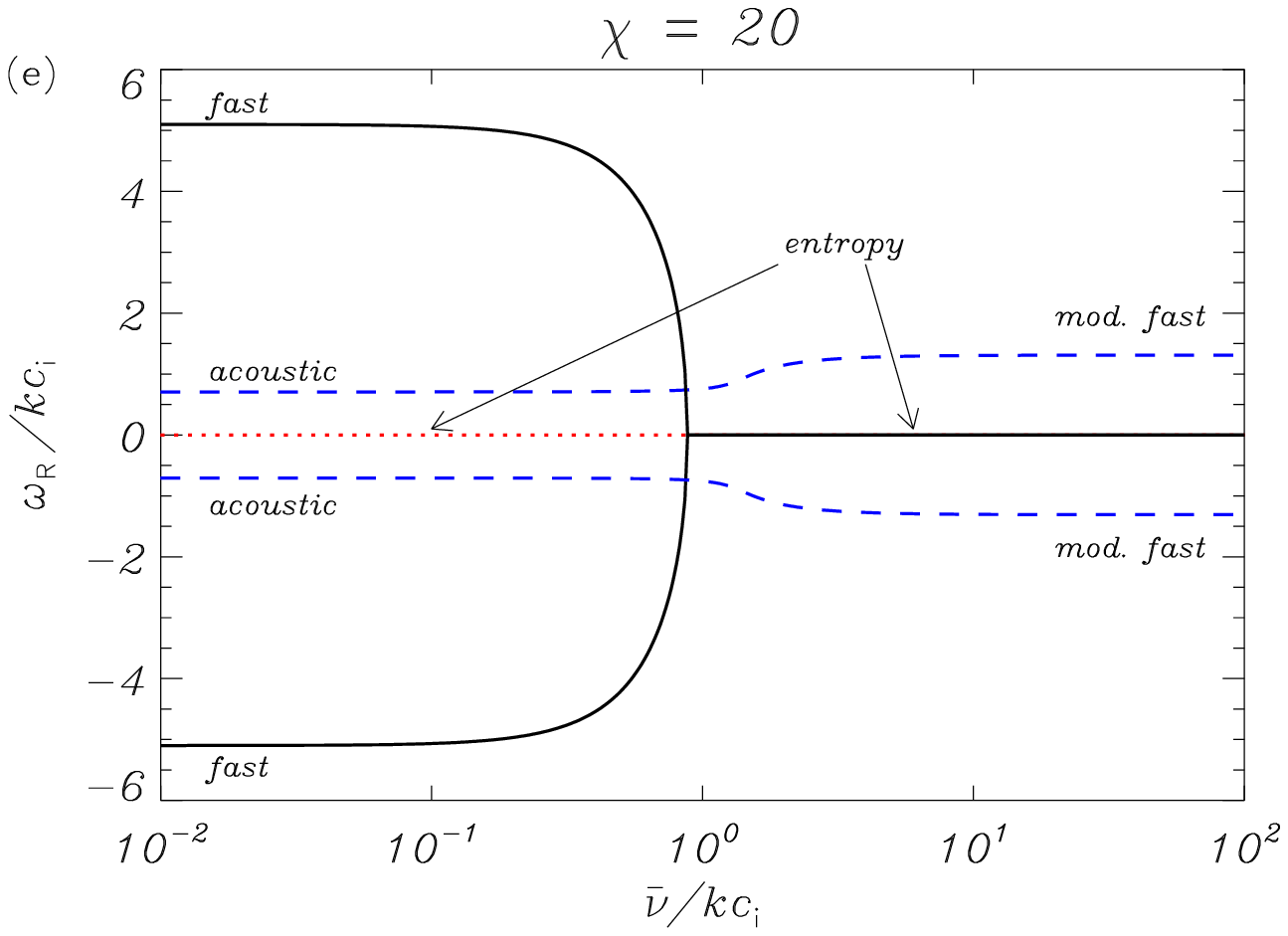}
	\includegraphics[width=.49\columnwidth]{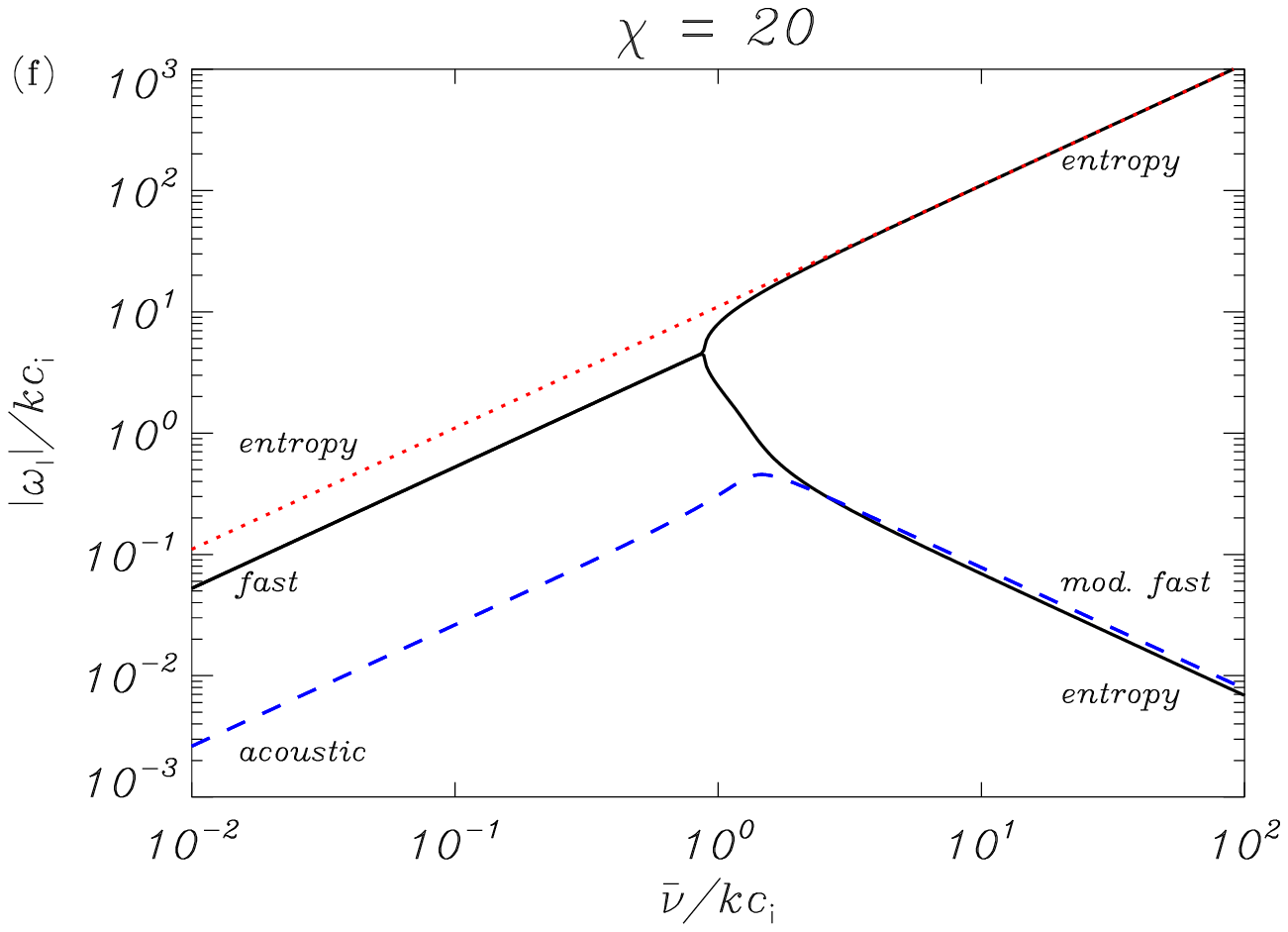}
	\caption{Real (left) and imaginary (right) parts of the frequency of the various waves versus the averaged collision frequency (in logarithmic scale) for propagation perpendicular to the magnetic field, i.e., $\theta=\pi/2$, with $\beta_{\rm i} = 0.04$. Panels (a)--(b) are for $\chi=0.2$, panels (c)--(b) for $\chi = 2$, and panels (c)--(b) for $\chi = 20$. All frequencies are expressed in units of $k\cie$. Note that the absolute value of $\omega_{\rm I}$ is plotted.}
	\label{fig:perp}
\end{figure*}

\begin{figure*}
	\centering
	\includegraphics[width=.49\columnwidth]{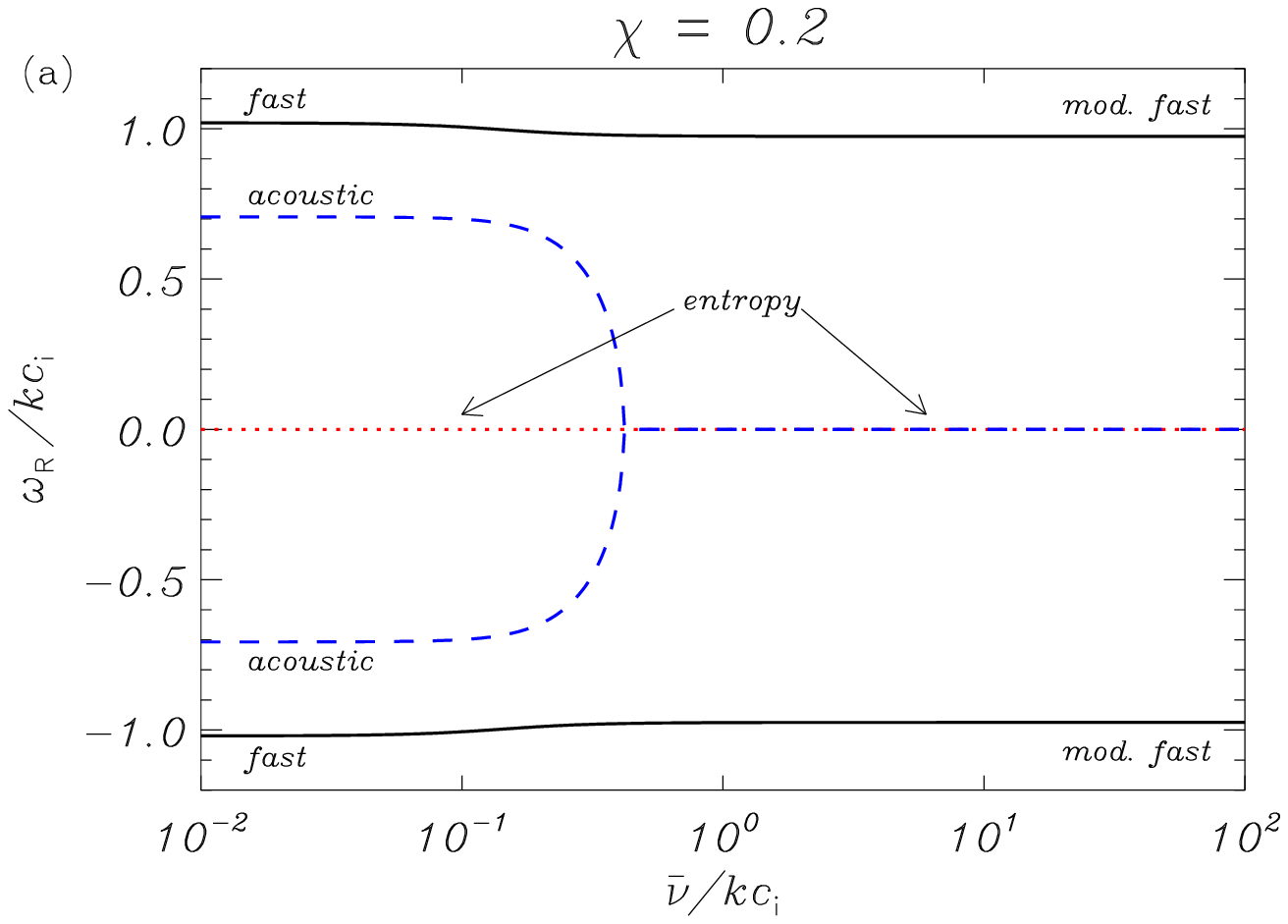}
	\includegraphics[width=.49\columnwidth]{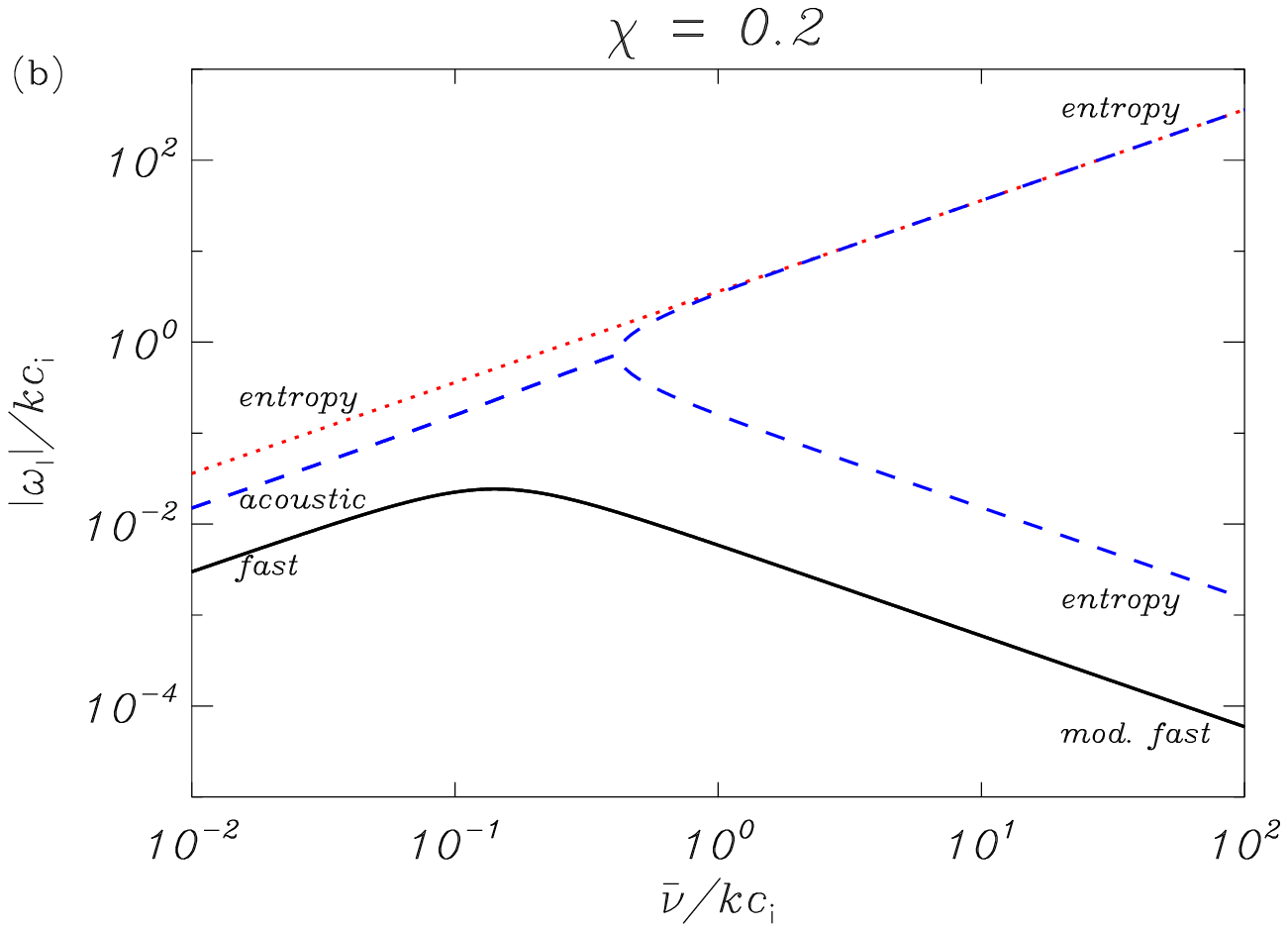}
	\includegraphics[width=.49\columnwidth]{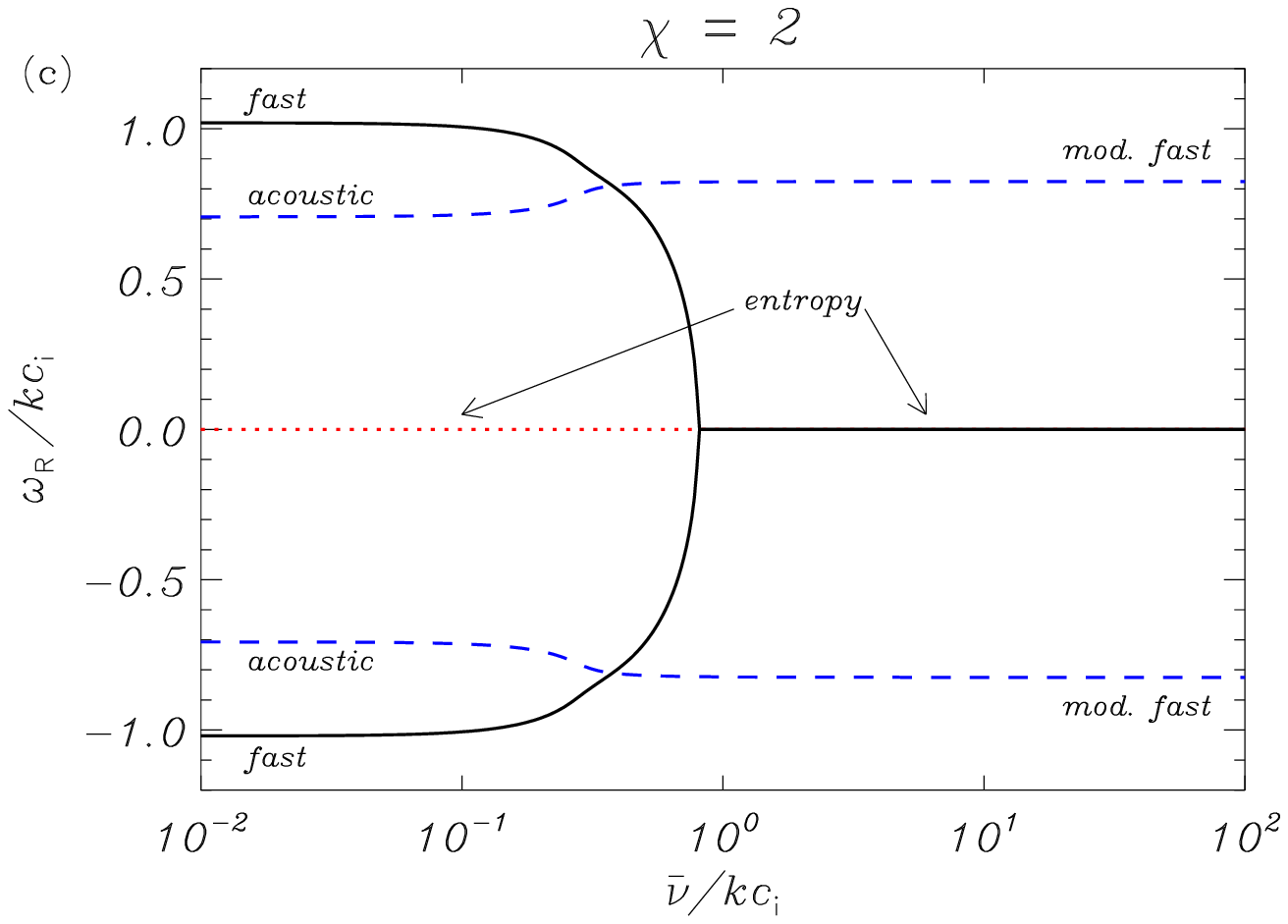}
	\includegraphics[width=.49\columnwidth]{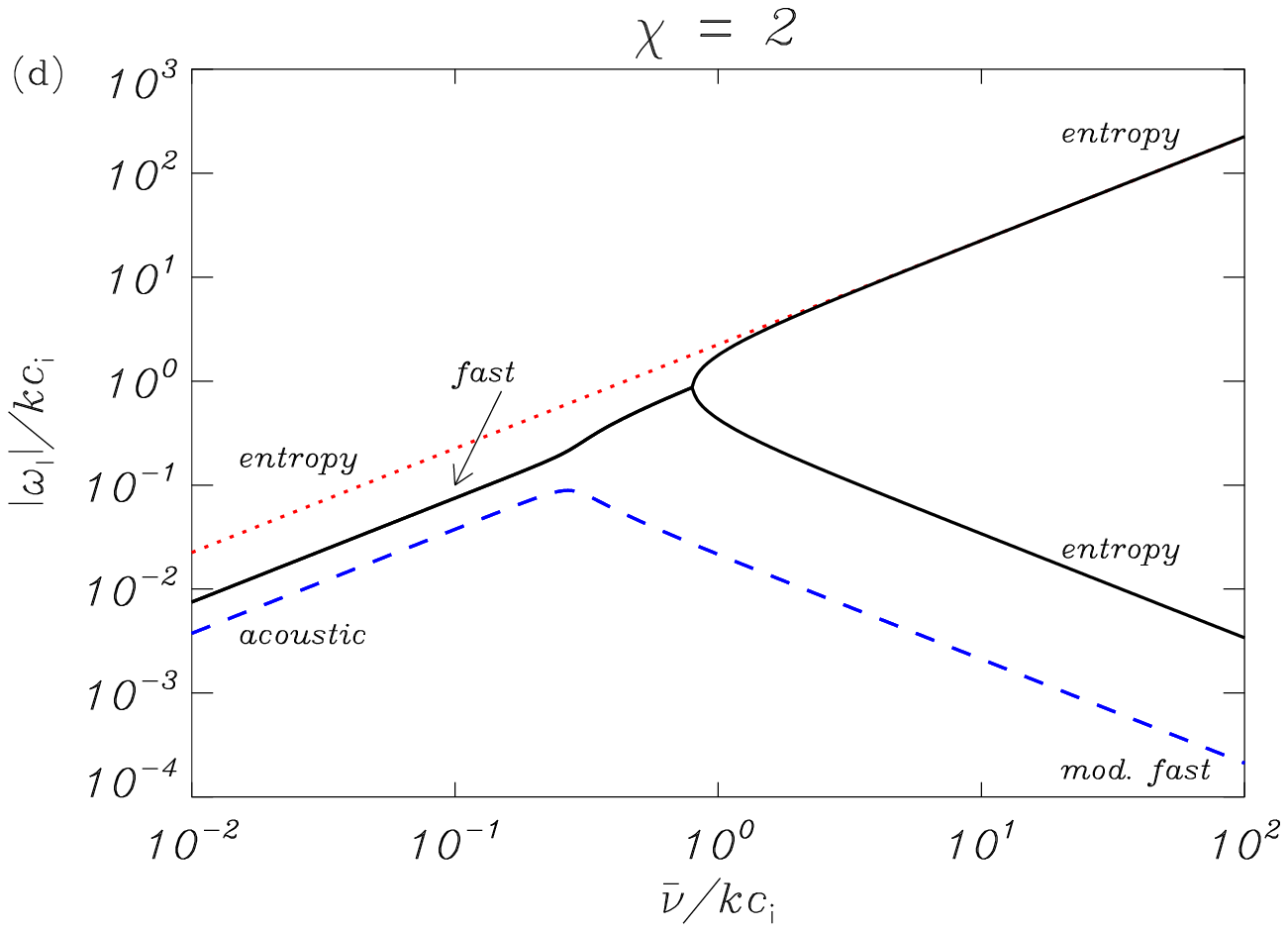}
	\includegraphics[width=.49\columnwidth]{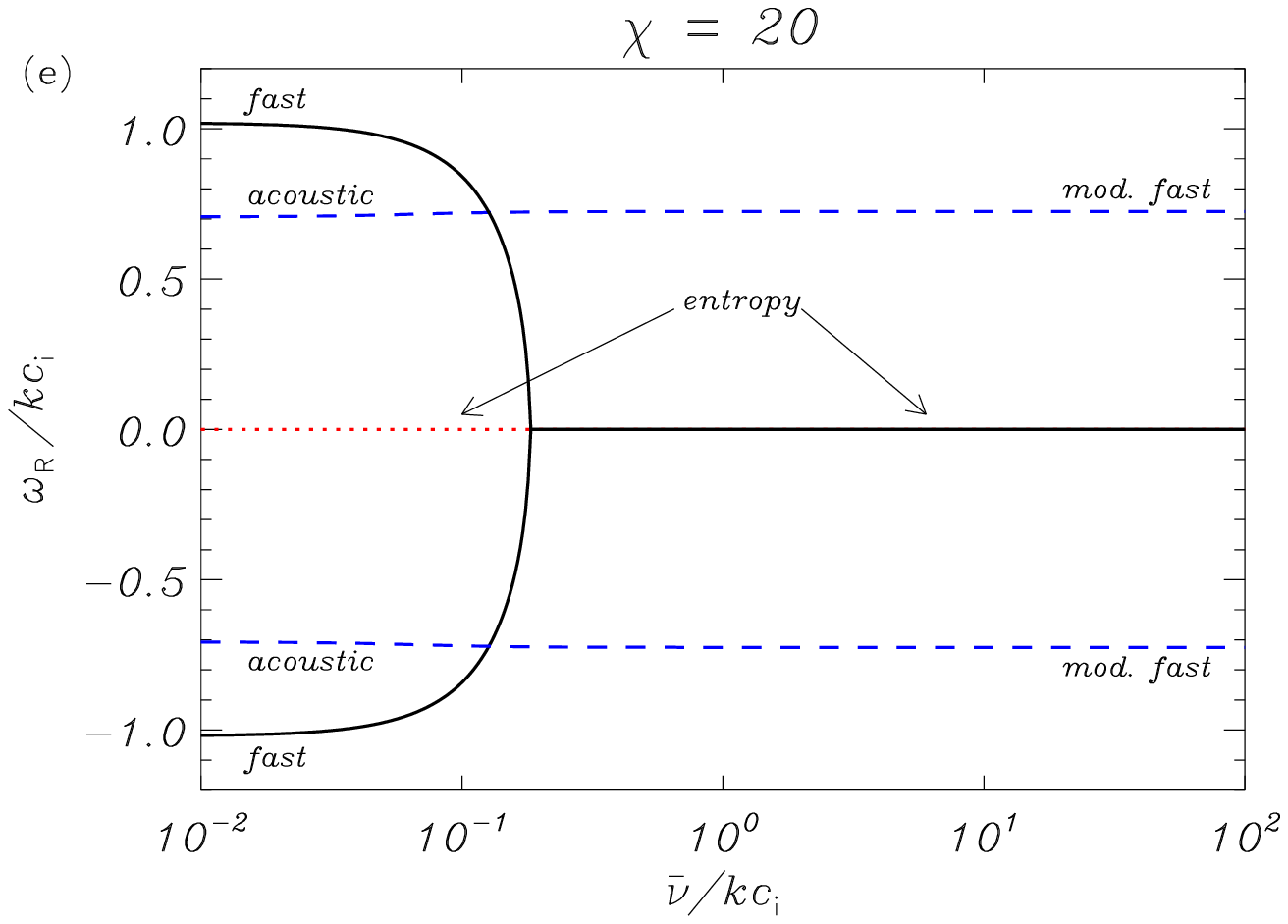}
	\includegraphics[width=.49\columnwidth]{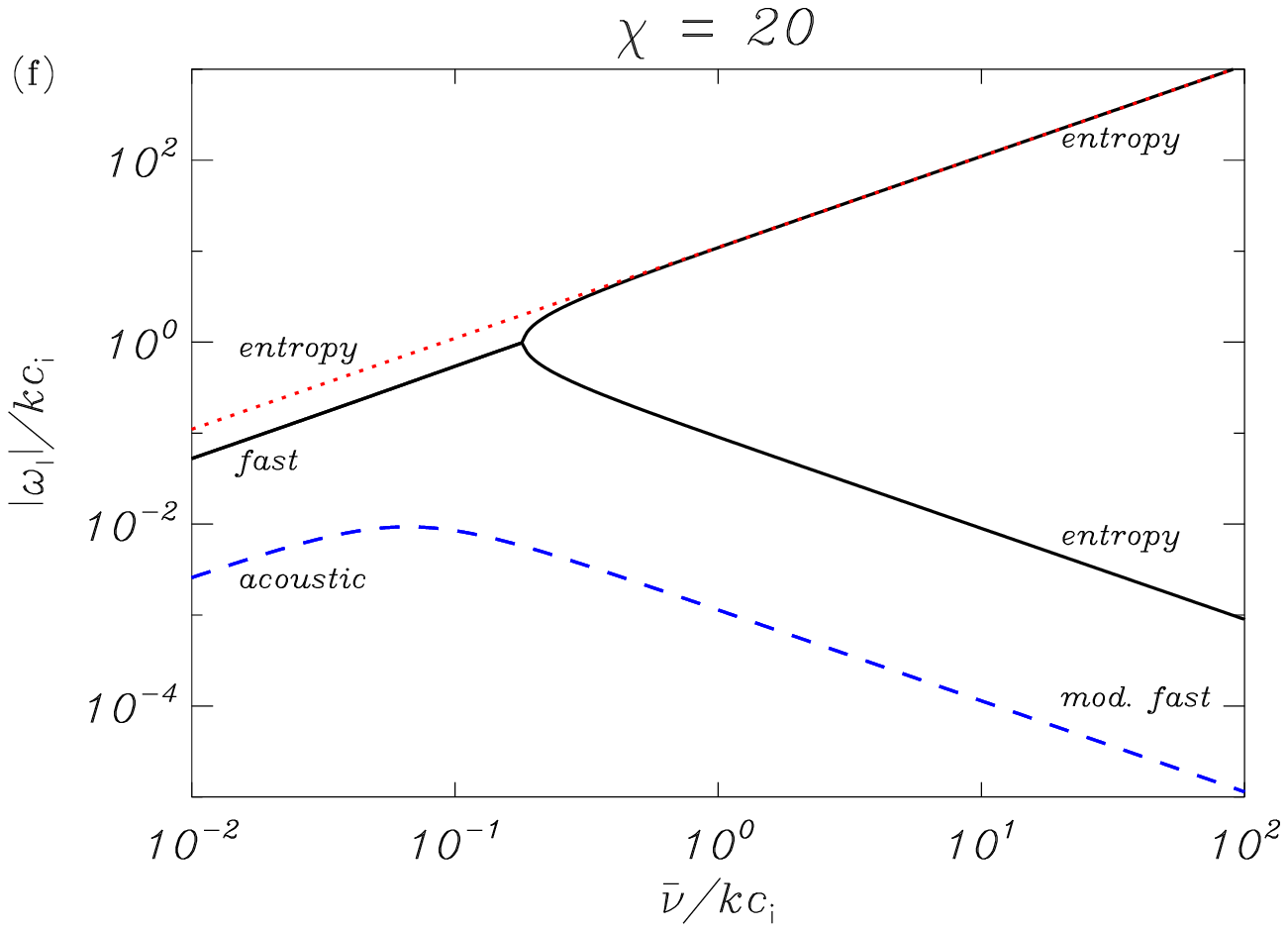}
	\caption{Same as Figure~\ref{fig:perp} but for $\beta_{\rm i}= 25$.}
	\label{fig:perph}
\end{figure*}

We compute the real and imaginary parts of the frequency of the various solutions of the dispersion relation as functions of $\bar{\nu}$. Frequencies are expressed in units of $k\cie$ so that the results can be applied to the astrophysical plasma of interest by providing appropriate numeric values of $k$ and $\cie$. The averaged collision frequency is varied four orders of magnitude between $\bar{\nu} / k\cie = 0.01 $ and  $\bar{\nu} / k\cie = 100$. These two  values are chosen to represent the uncoupled and strongly coupled limits, respectively. Beyond these two values of $\bar{\nu} / k\cie $  the behavior of the solutions remains essentially unaltered.  Figures~\ref{fig:perp} and \ref{fig:perph} show the results for $\beta_{\rm i} = 0.04$ (low-$\beta_{\rm i}$ case) and $\beta_{\rm i} = 25$ (high-$\beta_{\rm i}$ case), respectively.  Three values of $\chi$ are considered in each case, namely $\chi=0.2$, 2, and 20, which cover the cases of largely ionized, moderately ionized, and weakly ionized plasmas, respectively. 

First, we analyze the real part of the frequency. The comparison of the numerically obtained $\omega_{\rm R}$  with the analytic approximations of Section~\ref{sec:lim}  helps us identifying the various propagating waves in the limit values of $\bar{\nu}$. The modes are labeled and plotted with different line styles in Figures~\ref{fig:perp} and \ref{fig:perph}.  When $\bar{\nu}/k\cie =0.01$ we find the classic fast magnetoacoustic (solid black line) and neutral acoustic (dashed blue line) modes, while for  $\bar{\nu}/k\cie = 100$ only the modified fast wave is present as a propagating solution, in agreement with Section~\ref{sec:lim}. As expected, the modes do not display the properties of the classic waves when the averaged collision frequency takes intermediate values. To understand how the solutions change due to the effect of collisions we continuously follow the various solutions when $\bar{\nu}/k\cie$ increases from {\bf  $\bar{\nu}/k\cie = 0.01$ to $\bar{\nu}/k\cie = 100$}. We find that one of the two solutions present  at the low collision frequency limit has a cutoff when $\bar{\nu}/k\cie$ reaches a certain value. At the cutoff, the forward ($\omega_{\rm R} > 0$) and backward ($\omega_{\rm R} < 0$) waves of this specific solution merge and their  $\omega_{\rm R}$ becomes zero. On the contrary,  the other solution  does not have a cutoff and eventually becomes the modified fast wave  at the high collision frequency limit.

\begin{figure*}
	\centering
	\includegraphics[width=.69\columnwidth]{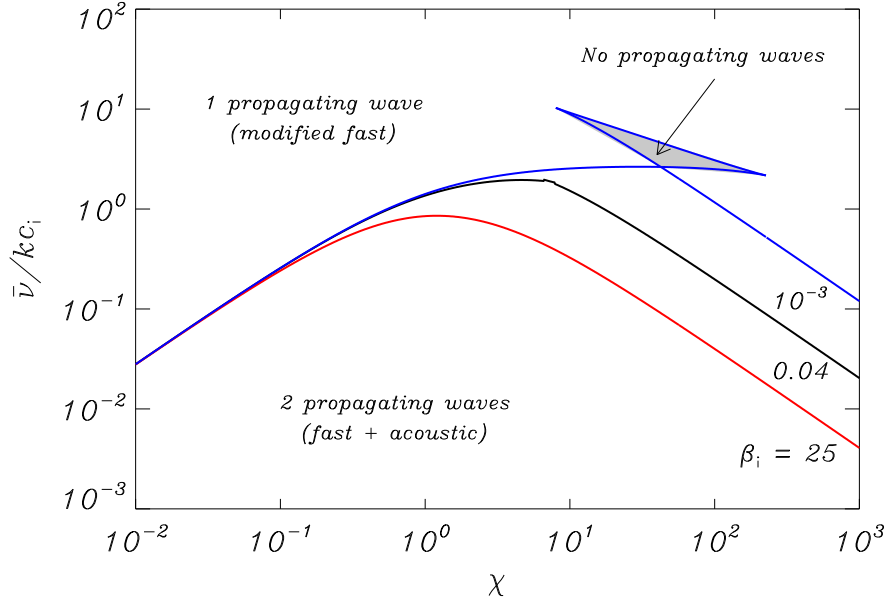}
	\caption{Number and nature of the propagating solutions of the dispersion relation in the $\chi$--$\bar{\nu}/k\cie$ plane  for $\theta = \pi/2$ and $\beta_{\rm i} = 10^{-3}$, 0.04, and 25. Each line denotes the boundary between the various regions for a given value of $\beta_{\rm i}$. The shaded area denotes a region in the case $\beta_{\rm i} = 10^{-3}$ where propagation is forbidden. This forbidden region is also distinguished as a tiny feature in the line for $\beta_{\rm i} = 0.04$.}
	\label{fig:discperp}
\end{figure*}

The location of the cutoff is important because it determines the number of distinct waves that are able to propagate for a certain combination of parameters. The effect of the various parameters on the location of the cutoff can be studied as in Paper~I by taking advantage of the fact that the dispersion relation is a polynomial in $\omega$. So, we can compute the polynomial discriminant of the dispersion relation. By definition, the discriminant is zero when the dispersion relation has a multiple solution. Since the cutoff takes place when a forward and a backward propagating wave merge, the dispersion relation has necessarily a double root at the cutoff. The discriminant is a function of $\chi$, $\bar{\nu}/k\cie$, and $\beta_{\rm i}$. Hence, the zeros of the discriminant inform us about the relation between these parameters at the cutoff. We omit here the expression of the discriminant because it is  rather cumbersome and can be straightforwardly obtained from the dispersion relation using standard algebraic methods. The number of propagating solutions for a given set of parameters is represented in Figure~\ref{fig:discperp} in the $\chi$--$\bar{\nu}/k\cie$ plane for three different values of $\beta_{\rm i}$. The zero of the discriminant is plotted with a  line. Below this line two propagating waves are possible, namely the classic fast and acoustic modes, while above the line only the modified  fast wave is present as a propagating wave. Importantly, we find that a third region,  where  propagating waves are not possible, appears when $\beta_{\rm i} \ll 1$   (see the shaded area in Figure~\ref{fig:discperp} when $\beta_{\rm i} = 10^{-3}$). This forbidden region takes place for large ionization ratio (weakly ionized plasmas), very low $\beta_{\rm i}$ (strong magnetic fields) and relatively high collision frequency (strongly coupled plasmas). This could be the case of the plasma in intense magnetic tubes of the low solar atmosphere.

The physical reason for the existence of wave cutoffs is qualitatively the same as discussed in Paper~I for the case of Alfv\'en waves \citep[see also, e.g.,][]{kulsrud1969,pudritz1990,kamaya1998,mouschovias2011}. There is competition between the ion-neutral friction force and the wave restoring forces. Friction becomes the dominant force and overcomes the remaining forces at the cutoff. Hence, waves are unable to propagate when the strength of friction greatly exceeds those of the restoring forces. The cutoff found here corresponds to the situation in which ions and neutrals are sufficiently coupled for the neutral fluid to be significantly affected by the magnetic field. As a consequence, the neutral fluid is not  able to support its own pure acoustic modes, which start to behave as slow-like magnetoacoustic waves. For $\theta = \pi/2$ slow modes cannot propagate and so these solutions are cut off.

We turn to the imaginary part of the frequency. For presentation purposes, we plot the absolute value of $\omega_{\rm I}$ in logarithmic scale. The line styles used in the graphics of $\omega_{\rm I}$ are the same as in the plots of $\omega_{\rm R}$ so that the various solutions can be easily identified.  The inspection of $\omega_{\rm I}$ in Figures~\ref{fig:perp} and \ref{fig:perph} informs us of the existence of a third purely imaginary mode (red dotted line). This mode is the descendant of the  undamped entropy mode, which becomes damped by collisions when $\bar{\nu}/k\cie$ increases. The damping rate of this nonpropagating mode monotonically increases with $\bar{\nu}/k\cie$. Regarding the propagating waves, we see that the solution that has the cutoff in $\omega_{\rm R}$ displays a bifurcation in $\omega_{\rm I}$ at the same location. Two purely imaginary branches emerge at the bifurcation. We identify these new branches as entropy modes. On the one hand, the upper branch is heavily damped and follows the behavior of the original entropy mode plotted with a red dotted line. On the other hand, the lower branch is less damped and becomes an undamped entropy mode when $\bar{\nu}/k\cie \to \infty$. Finally, the remaining solution whose $\omega_{\rm I}$ does not have a bifurcation corresponds to the propagating wave that  becomes the modified fast wave when $\bar{\nu}/k\cie \gg 1$.  The damping rate of this wave is nonmonotonic. Its $\left|\omega_{\rm I}\right|$ is  maximal, i.e., the damping is most efficient, when $\bar{\nu} \approx \omega_{\rm R}$. This is the same result as for Alfv\'en waves (see Paper~I).

\subsection{Parallel propagation}

\begin{figure*}
	\centering
	\includegraphics[width=.49\columnwidth]{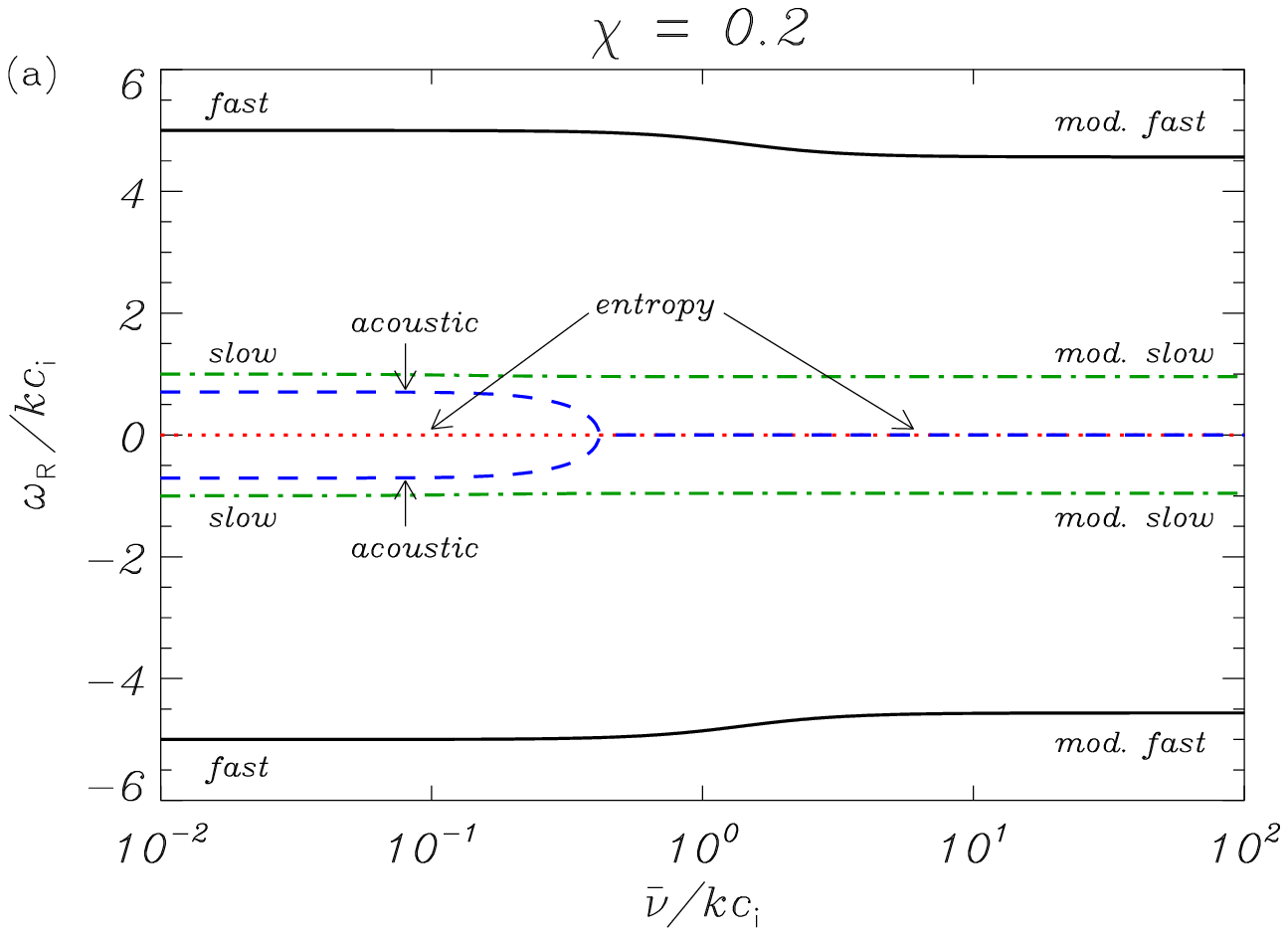}
	\includegraphics[width=.49\columnwidth]{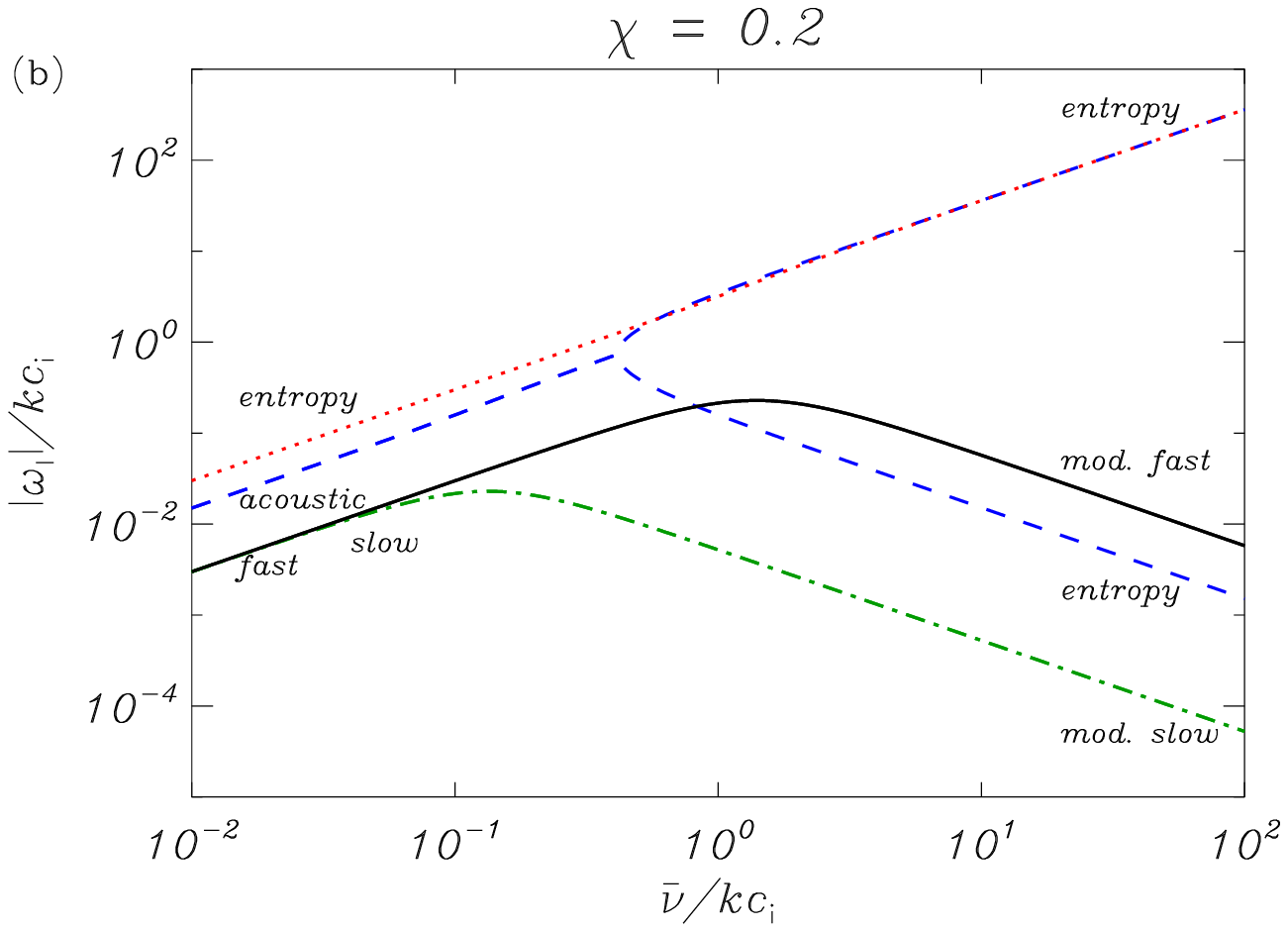}
	\includegraphics[width=.49\columnwidth]{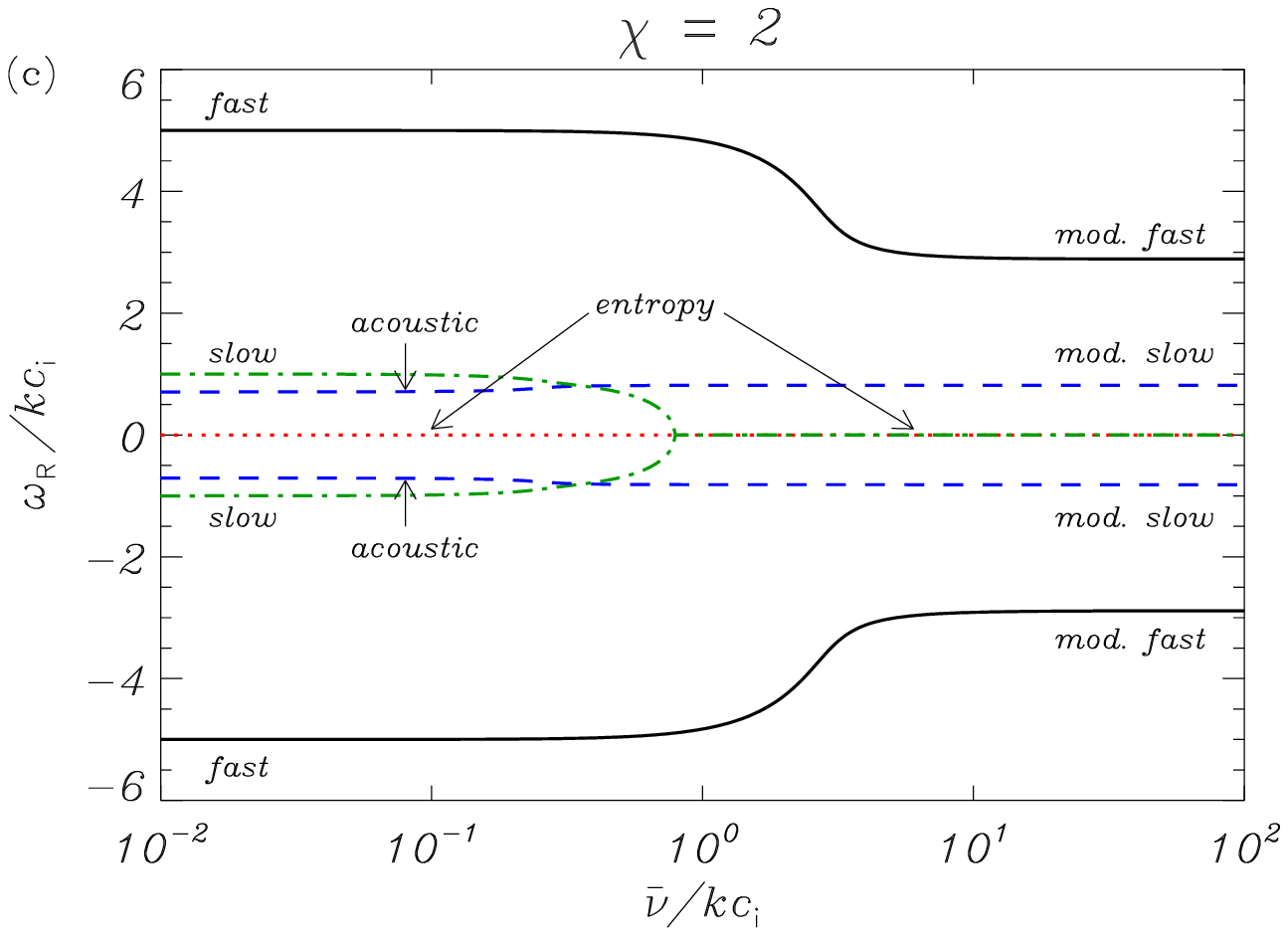}
	\includegraphics[width=.49\columnwidth]{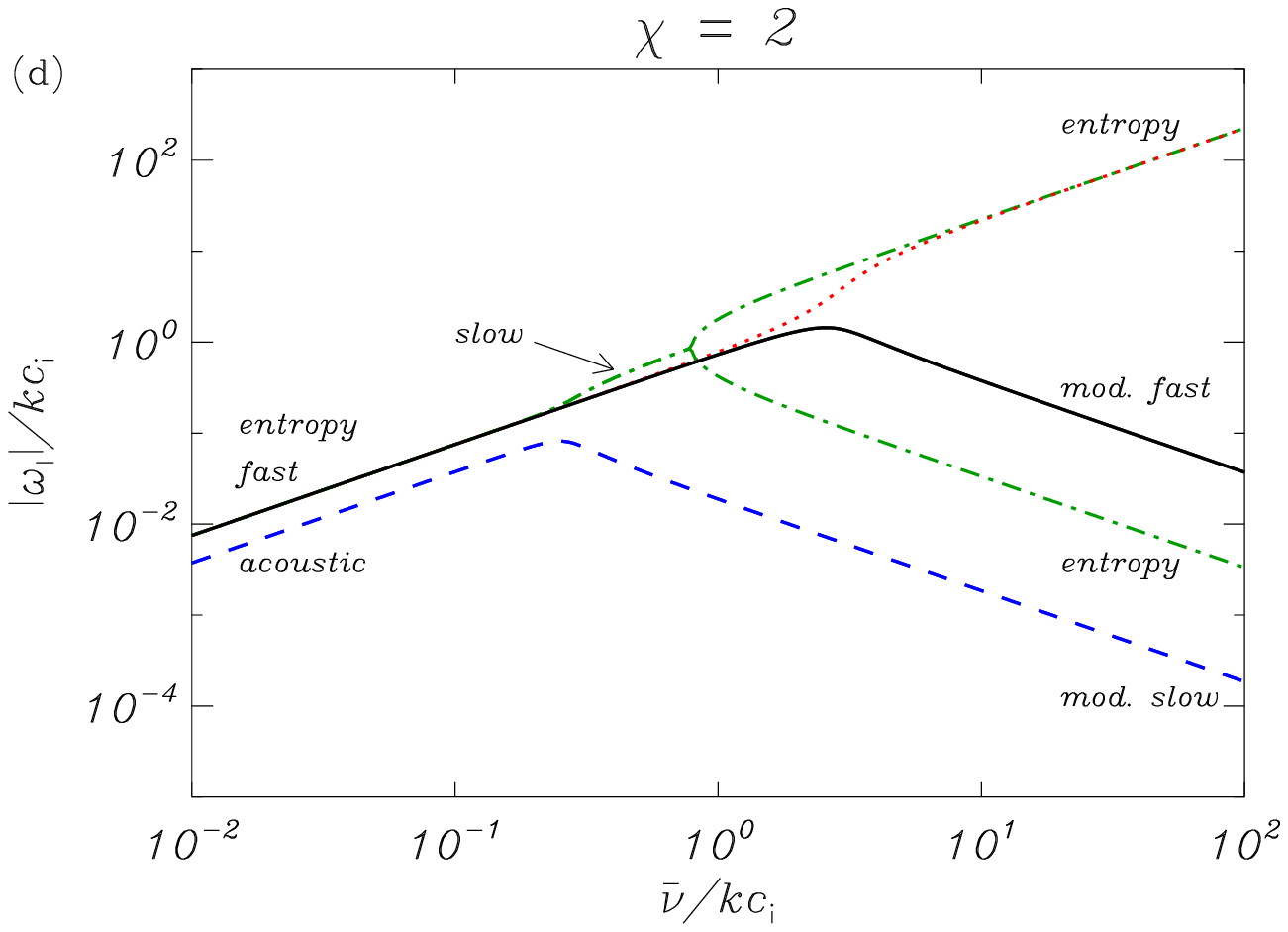}
	\includegraphics[width=.49\columnwidth]{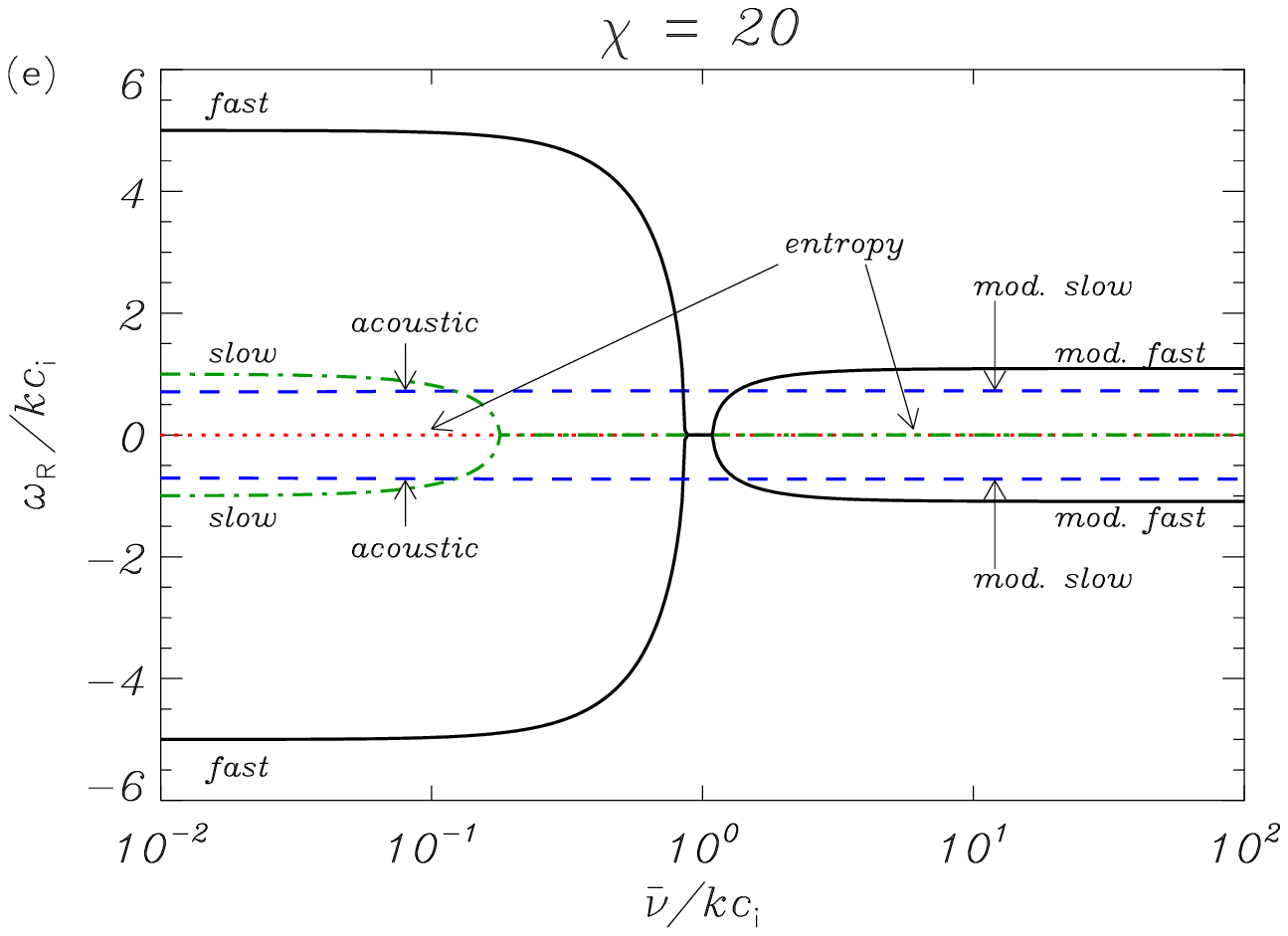}
	\includegraphics[width=.49\columnwidth]{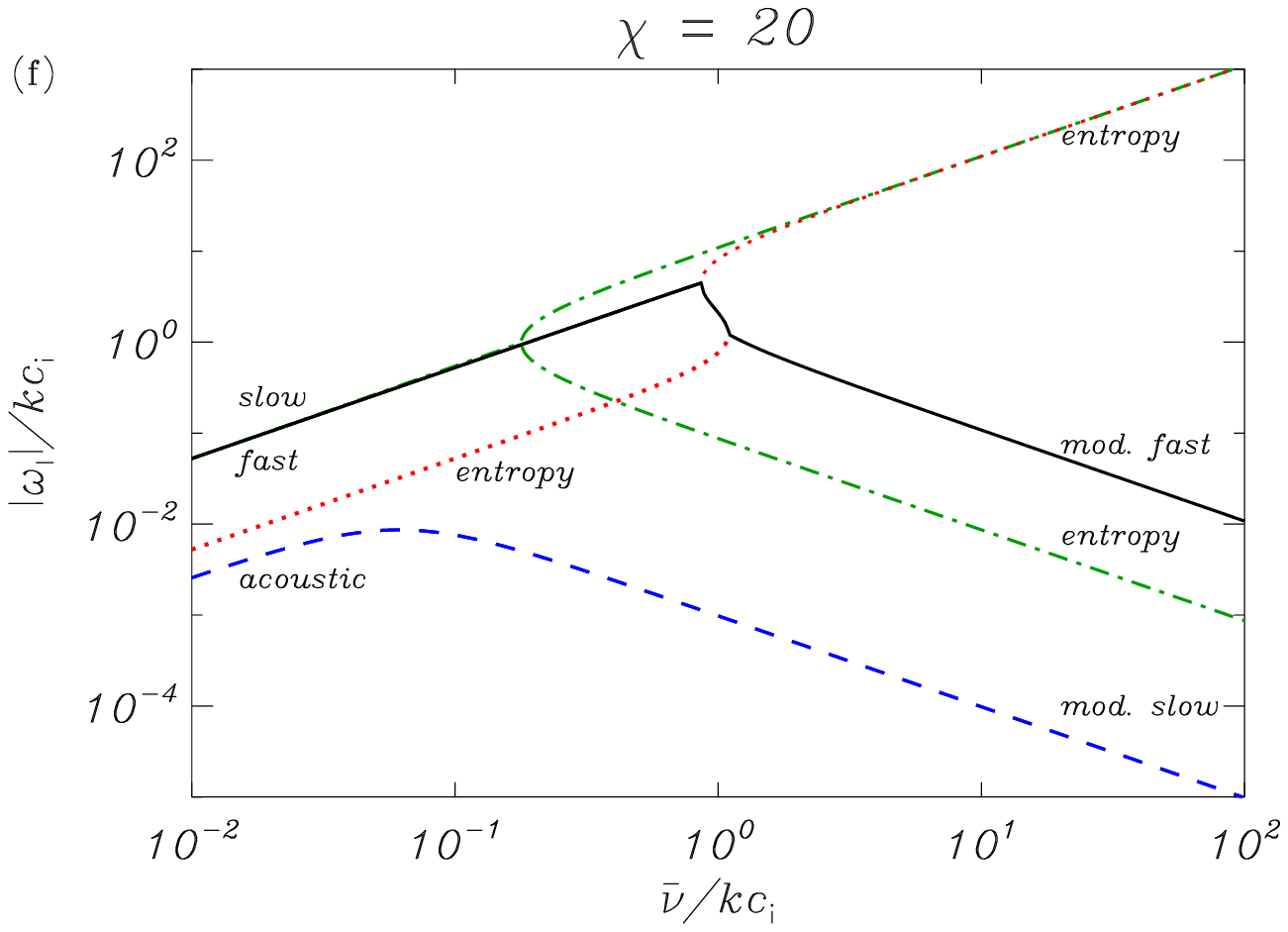}
	\caption{Same as Figure~\ref{fig:perp} but for propagation parallel to the magnetic field, i.e, $\theta = 0$. We use $\beta_{\rm i}=0.04$.}
	\label{fig:para}
\end{figure*}

\begin{figure*}
	\centering
	\includegraphics[width=.49\columnwidth]{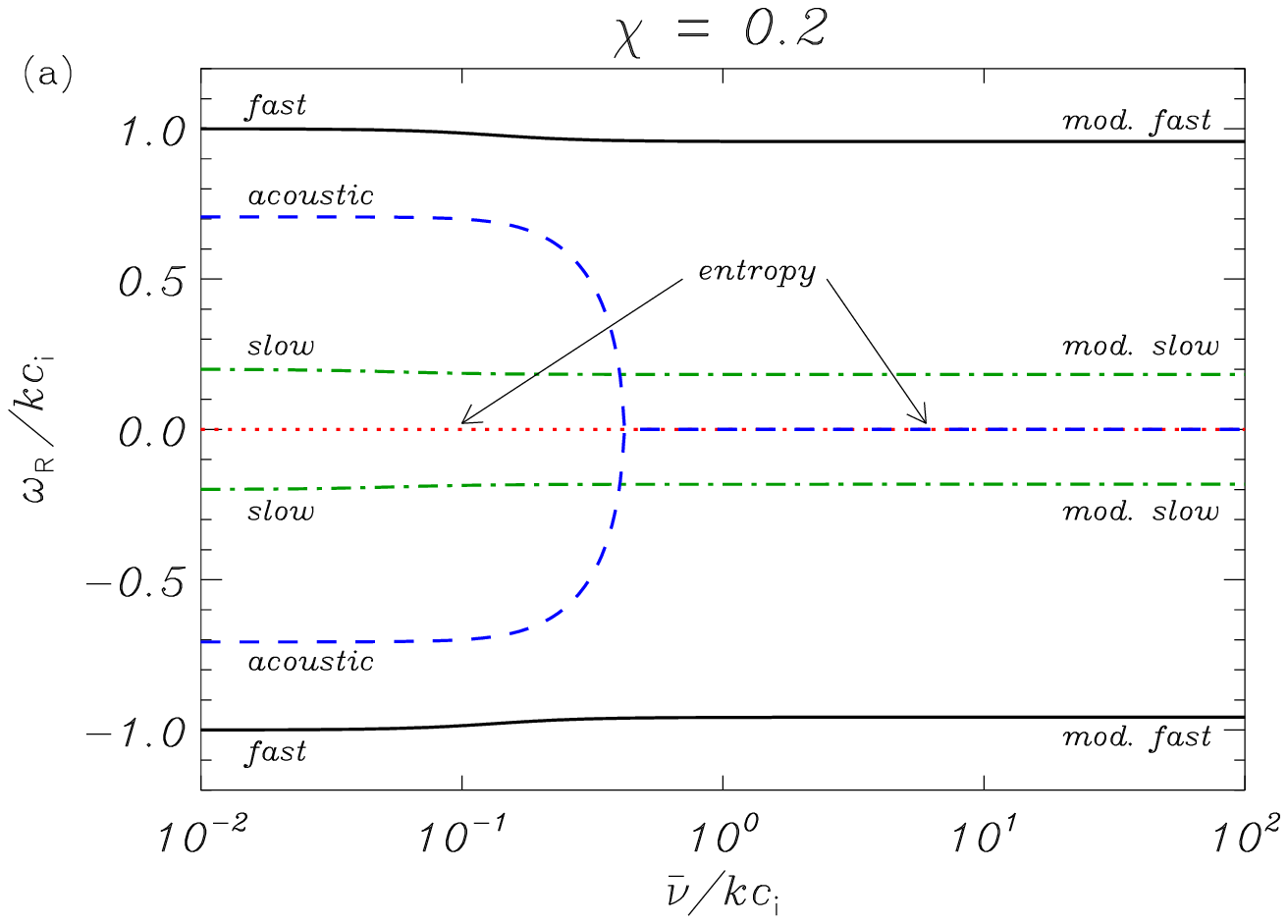}
	\includegraphics[width=.49\columnwidth]{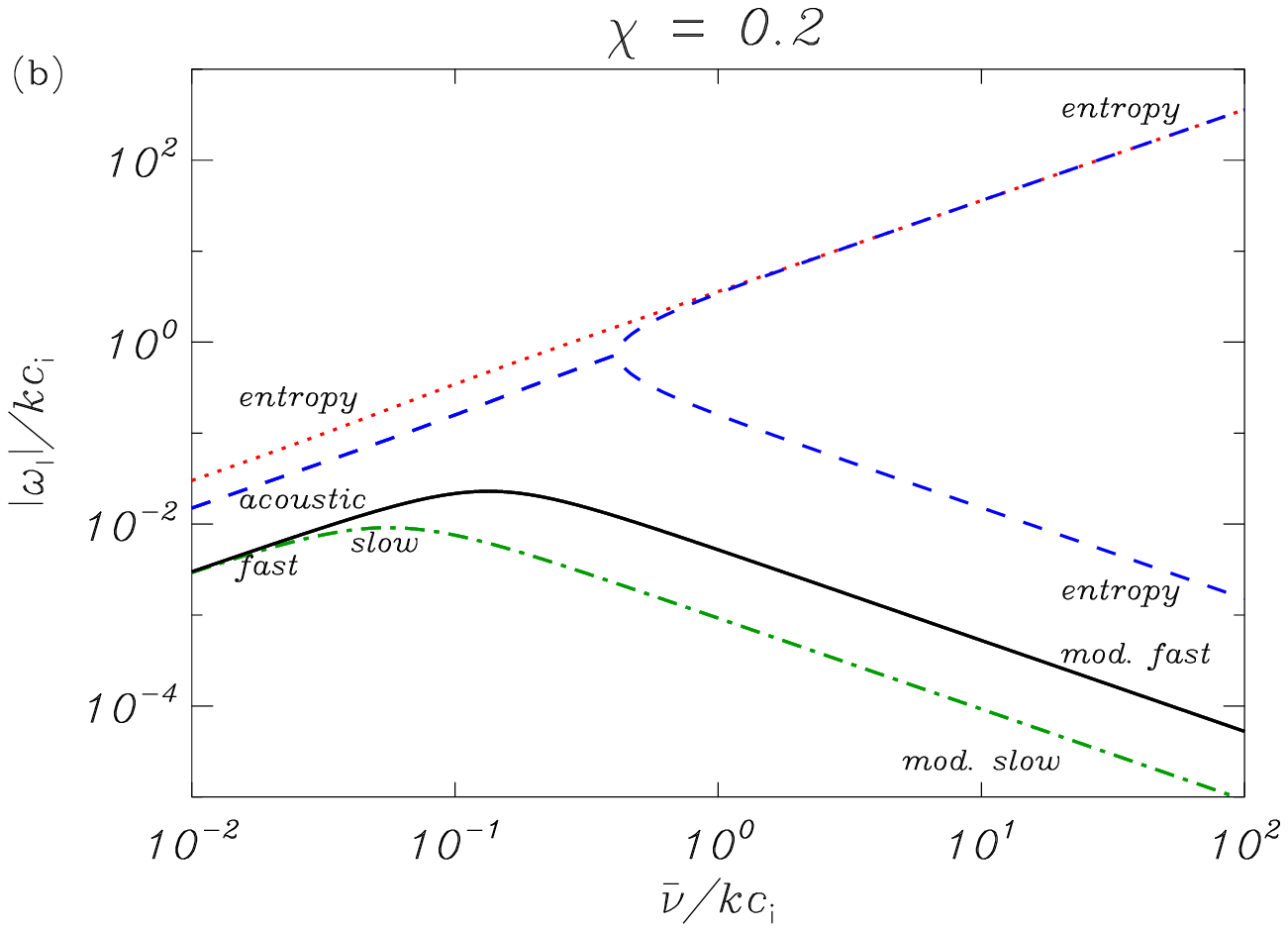}
	\includegraphics[width=.49\columnwidth]{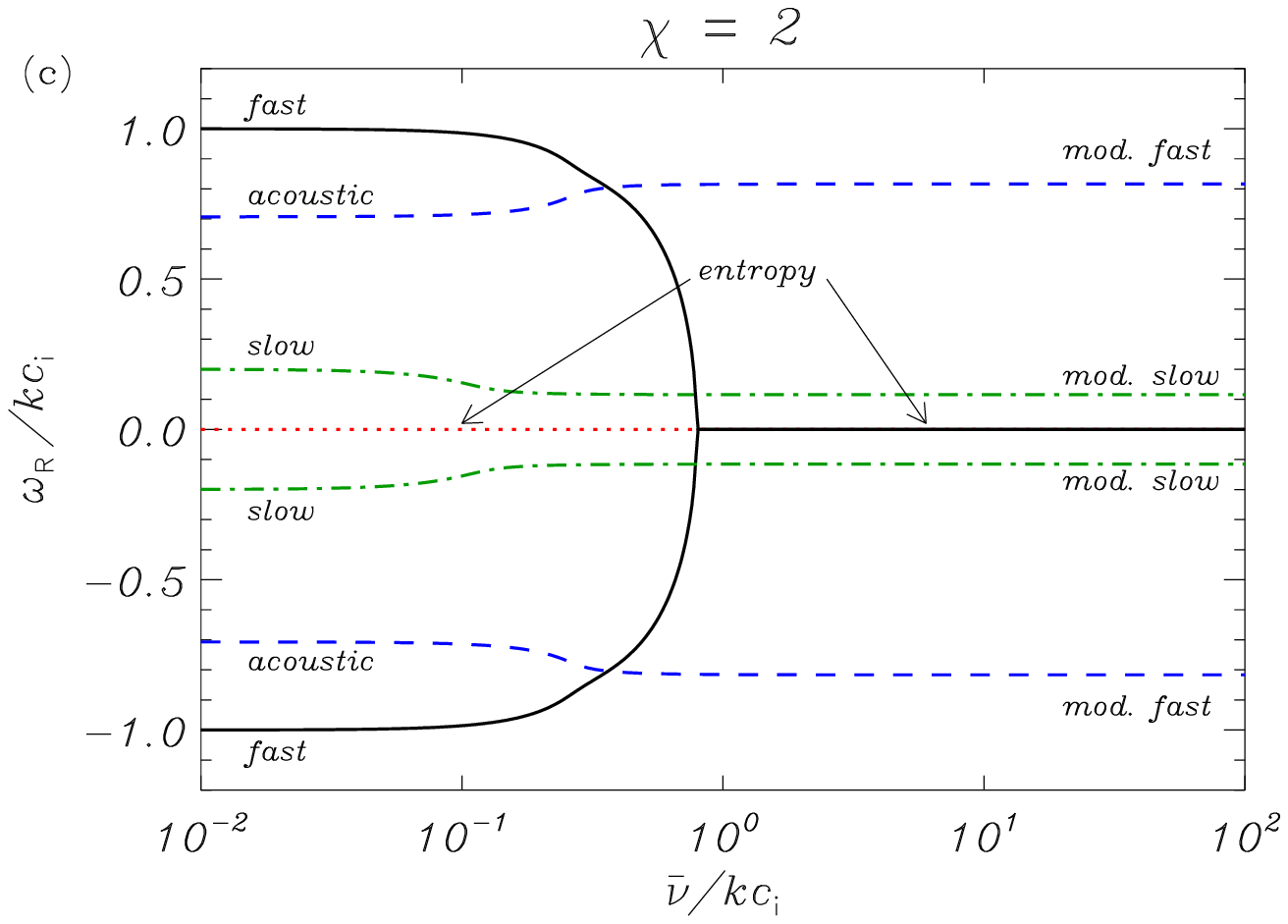}
	\includegraphics[width=.49\columnwidth]{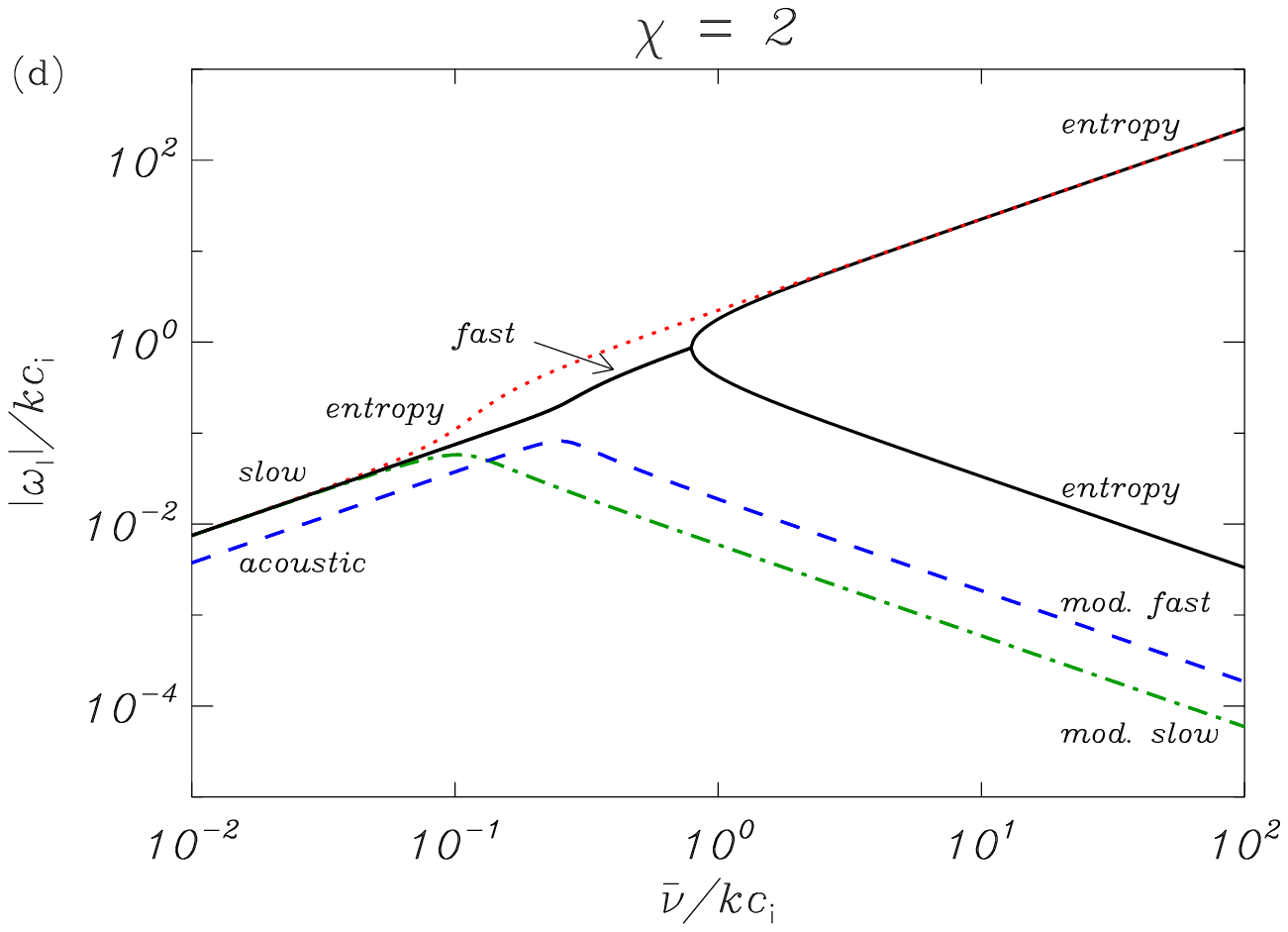}
	\includegraphics[width=.49\columnwidth]{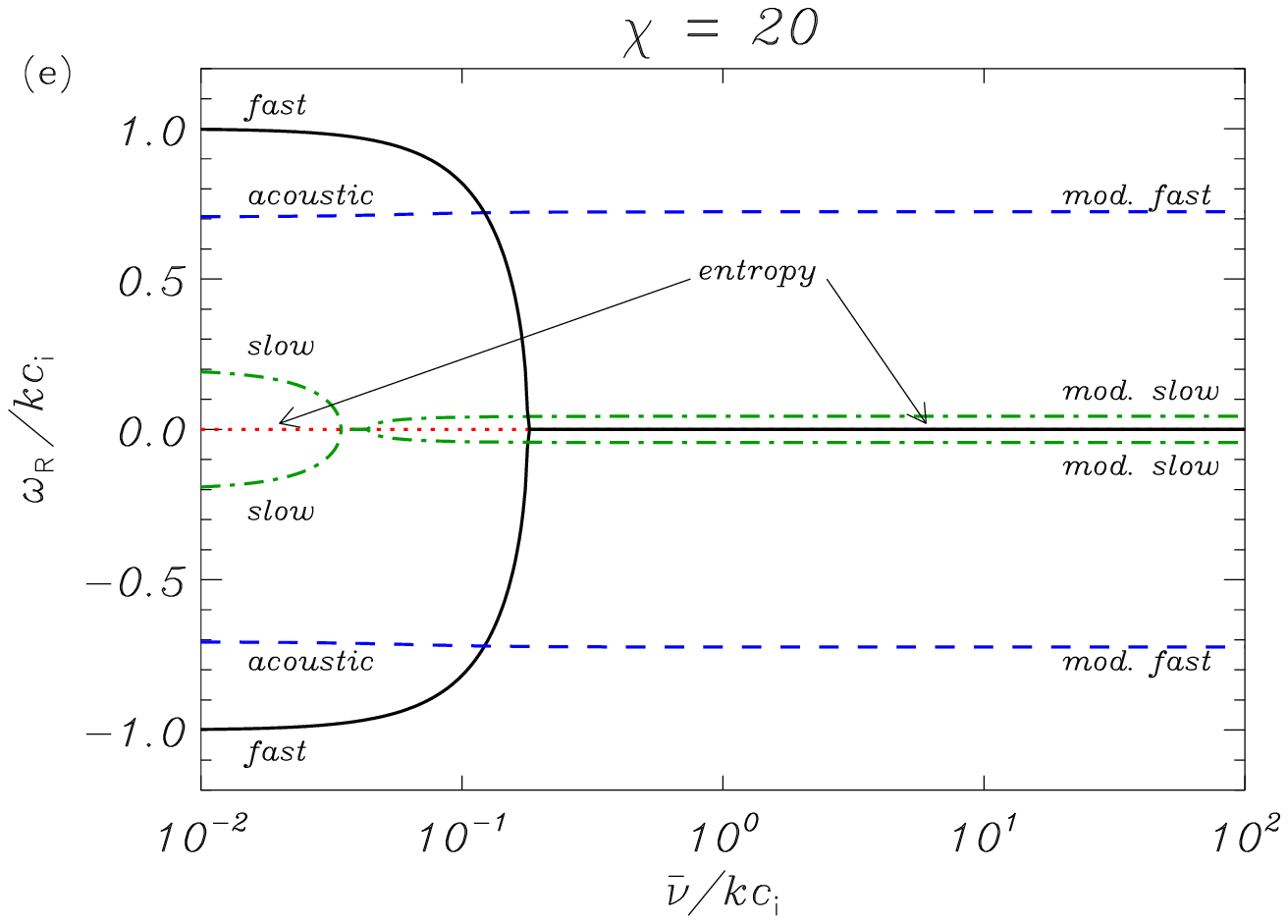}
	\includegraphics[width=.49\columnwidth]{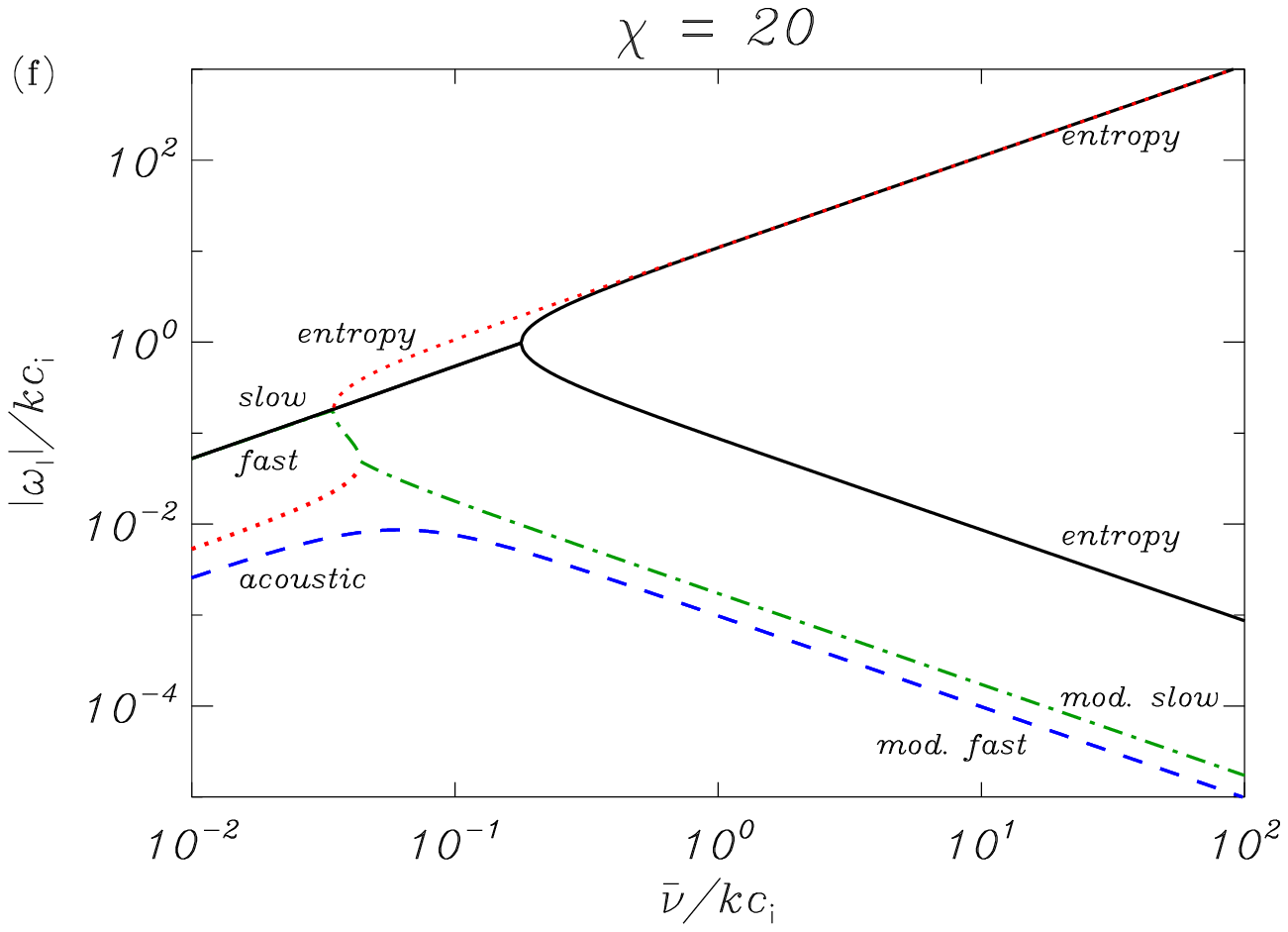}
	\caption{Same as Figure~\ref{fig:para} but for $\beta_{\rm i}=25$.}
	\label{fig:parah}
\end{figure*}

We move to the case of parallel propagation to the magnetic field. We set $\theta = 0$ and perform the same computations as in the previous section. These results are displayed in  Figures~\ref{fig:para} and \ref{fig:parah}, where we have considered the same parameters as before.   As expected, the slow magnetoacoustic wave is now present (dash-dotted green line). As in the perpendicular propagation case, we analyze Figures~\ref{fig:para} and \ref{fig:parah} by observing how the modes present  at the low collision frequency limit evolve as we increase the collision frequency toward the high collision frequency limit.

In the low-$\beta_{\rm i}$ case (Figure~\ref{fig:para}), the slow magnetoacoustic wave and the neutral acoustic wave strongly interact. There is a change of character between these two waves depending on the value of $\chi$. When $\chi\ll 1$ , i.e., largely ionized plasma, the neutral acoustic mode has a cutoff and the classic slow mode becomes the modified slow mode in the limit of large $\bar{\nu}/k\cie$. On the contrary, the situation is reversed when $\chi \gtrsim 1$, so that the solution that is cut off is the classic slow mode while the neutral acoustic mode is the solution that becomes the  modified slow mode. These results are consistent with the approximate study of Section~\ref{sec:highly}. On the other hand, the behavior of the fast magnetoacoustic wave in the low-$\beta_{\rm i}$ case is very similar to that of the Alfv\'en waves studied in Paper~I. It is well-known that in a low-$\beta_{\rm i}$ plasma fast waves behave as Alfv\'en waves for parallel propagation to the magnetic field \citep[see, e.g.,][]{goossens2003}. As in the case of Alfv\'en waves, fast magnetoacoustic waves are nonpropagating in an interval of $\bar{\nu}/k\cie$ when $\chi > 8$ (see extensive details in Paper~I for the case of Alfv\'en waves). This nonpropagating interval can be seen in Figure~\ref{fig:para} in the results for $\chi = 20$. In this interval the fast wave becomes a purely imaginary solution. The boundaries of the nonpropagating interval  can be obtained from Equation~(20) of Paper~I. Using the present notation, the boundaries of the nonpropagating interval are
\begin{equation}
\frac{\bar{\nu}}{k\cie} = \frac{2\chi }{\beta_{\rm i}^{1/2}\left(\chi+1\right)} \left[ \frac{\chi^2 + 20\chi - 8}{8\left( \chi+1 \right)^3} \pm \frac{\chi^{1/2}\left( \chi-8 \right)^{3/2}}{8\left( \chi+1 \right)^3} \right]^{1/2}, \label{eq:nonprop}
\end{equation}
where the $-$ and $+$ signs correspond to the left and right boundaries of the nonpropagating range.

In the high-$\beta_{\rm i}$ case (Figure~\ref{fig:parah}), the slow magnetoacoustic wave displays little interaction with the other two solutions. In this case, the behavior of the fast magnetoacoustic wave and the neutral acoustic wave is similar to that found when $\theta=\pi/2$, while the slow wave behaves as an Alfv\'en wave (Paper~I). Now, it is the slow wave and not the fast wave that has a  nonpropagating interval of $\bar{\nu}/k\cie$. Equation~(\ref{eq:nonprop}) holds for the slow wave in the high-$\beta_{\rm i}$ case too. Since Equation~(\ref{eq:nonprop}) depends upon $\beta_{\rm i}^{-1/2}$, the nonpropagating interval is now shifted towards smaller $\bar{\nu}/k\cie$ than in the low-$\beta_{\rm i}$ case (compare  the results for $\chi = 20$ in Figure~\ref{fig:para} and \ref{fig:parah}).

\begin{figure*}
	\centering
	\includegraphics[width=.69\columnwidth]{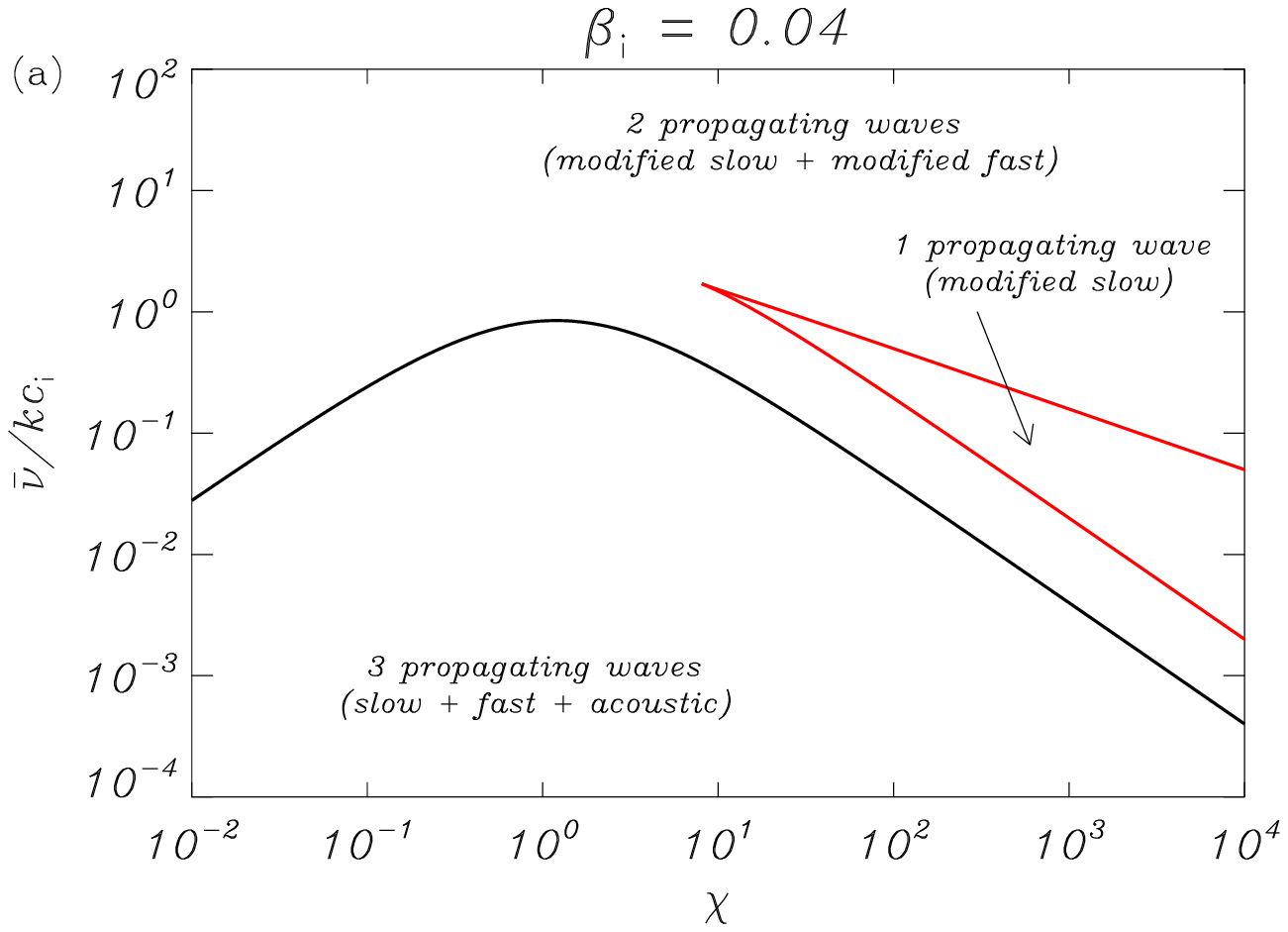}
		\includegraphics[width=.69\columnwidth]{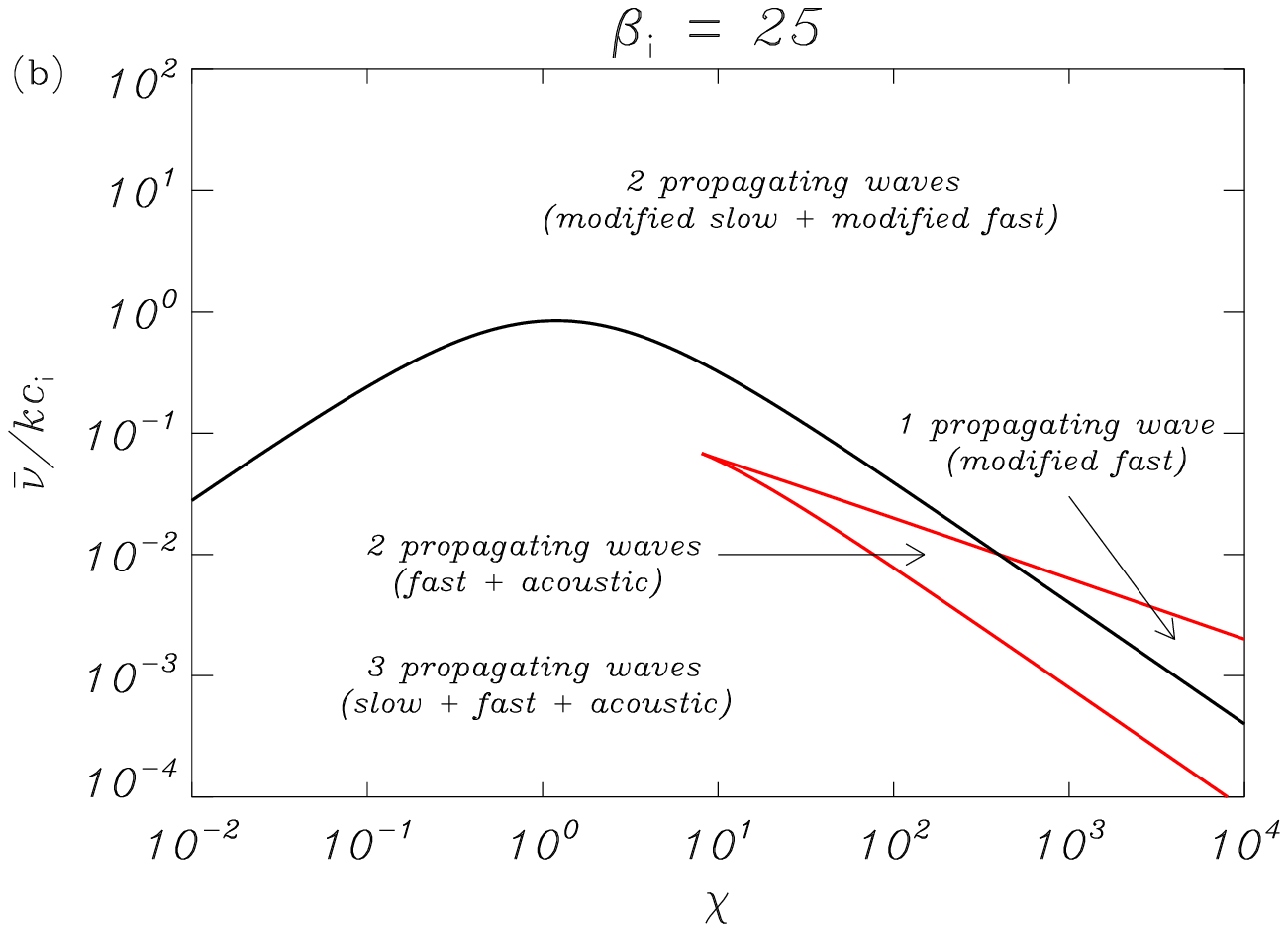}
	\caption{Number and nature of the propagating solutions of the dispersion relation in the $\chi$--$\bar{\nu}/k\cie$ plane  for $\theta = 0$ and (a) $\beta_{\rm i} = 0.04$ and (b) $\beta_{\rm i} = 25$. The red lines correspond to the interval given in Equation~(\ref{eq:nonprop}).}
	\label{fig:discpara}
\end{figure*}

We study in more detail the presence of cutoffs and nonpropagating intervals by using again the discriminant of the dispersion relation. This is shown in Figure~\ref{fig:discpara} for $\beta_{\rm i} = 0.04$ and $\beta_{\rm i} = 25$. As in the case of perpendicular propagation (see Figure~\ref{fig:discperp})  we find two  zones in the $\chi$--$\bar{\nu}/k\cie$ plane separated by a cutoff (black solid line). These separate zones correspond to the region where the classic slow, fast, and acoustic waves live and the region where the modified slow and fast modes exist. In addition, we find the presence of the nonpropagating interval previously discussed  (area between  the red lines). In the low-$\beta_{\rm i}$ case (Figure~\ref{fig:discpara}a) the nonpropagating interval is located above the boundary between the two regions. Within the nonpropagating interval, the modified slow wave is the only propagating solution. On the contrary, in the high-$\beta_{\rm i}$ case (Figure~\ref{fig:discpara}b) the nonpropagating interval intersects the boundary between the two regions. Hence the number of possible scenarios increases, as schematically indicated in Figure~\ref{fig:discpara}b.

\begin{figure*}
	\centering
	\includegraphics[width=.69\columnwidth]{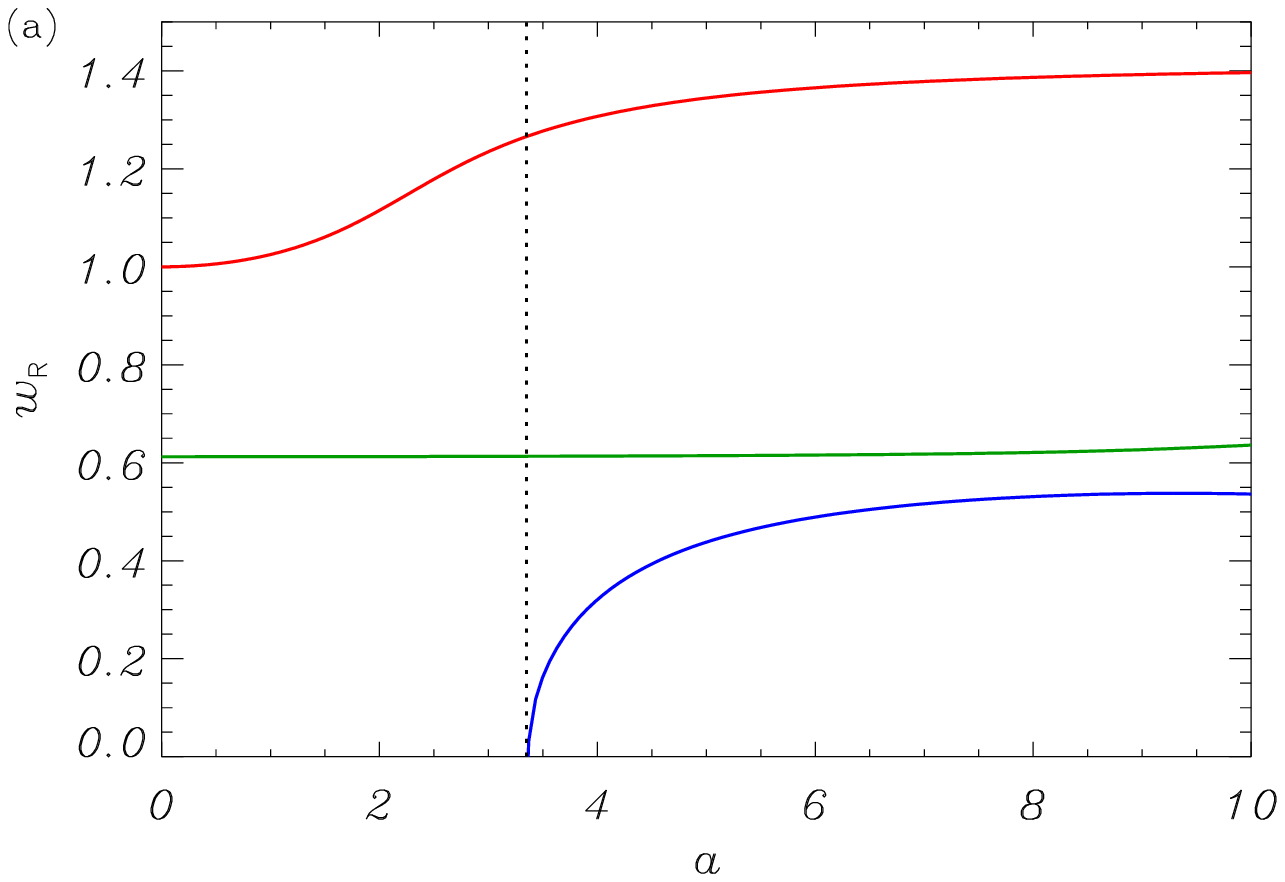}
		\includegraphics[width=.69\columnwidth]{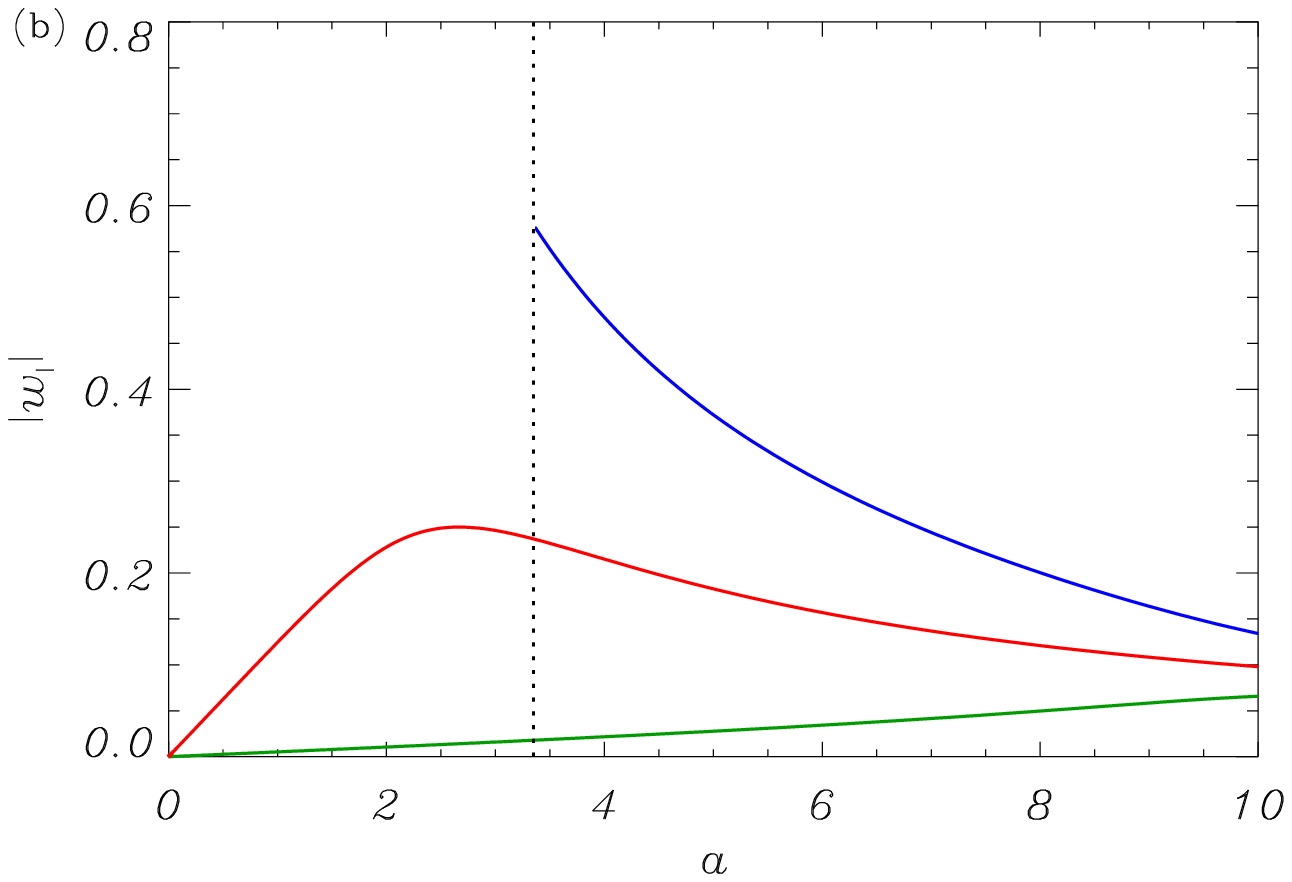}
	\caption{Reproduction of Figure~3 of \citet{zaqarashvili2011a} with the results of the present work. (a) Real and (b) imaginary parts of $w$ as functions of $a$. The line colors have the same meaning as in Figure~3 of \citet{zaqarashvili2011a}. The vertical dotted line denote the location of the neutral acoustic wave cutoff. Purely imaginary solutions are omitted in these plots.}
	\label{fig:tem}
\end{figure*}

It is appropriate to compare our results  with those plotted in Figure~3 of \citet{zaqarashvili2011a} in order to check whether the present work is consistent with that previous study. In our notation, the parameters used by \citet{zaqarashvili2011a} are $\chi = 1$, $\beta_{\rm i} = 0.25$, and $\theta = 0$. These parameters are close to those used in Figure~\ref{fig:para}(c)--(d). To perform a proper comparison we should use the same notation as \citet{zaqarashvili2011a}. They used two dimensionless quantities, $w$ and $a$, that correspond to the wave frequency and to the inverse of the collision frequency, respectively. These two quantities are expressed in our notation as 
\begin{eqnarray}
w &=& \frac{\sqrt{\beta_{\rm i}\left( 1+\chi \right)}}{\cos\theta} \frac{\omega}{k \cie}, \\ 
a &=& \frac{2\cos\theta}{\sqrt{\beta_{\rm i} \left( 1+\chi \right)}} \frac{k \cie}{\bar{\nu}} .
\end{eqnarray}
 We solve the dispersion relation using the above parameters and reproduce in Figure~\ref{fig:tem} the results shown in Figure~3 of \citet{zaqarashvili2011a}. A perfect agreement between our results and those of \citet{zaqarashvili2011a} is obtained. With the help of the discriminant of the dispersion relation, we compute that the cutoff of the neutral acoustic wave takes place at $\bar{\nu}/k\cie \approx 0.84$, which corresponds to $a \approx 3.4$ in the notation of \citet{zaqarashvili2011a}. Since $\chi < 8$, the fast mode does not have a nonpropagating interval.  Thus, our results confirm and fully agree with the specific example studied by \citet{zaqarashvili2011a}.

\subsection{Oblique propagation}

\begin{figure*}
	\centering
	\includegraphics[width=.49\columnwidth]{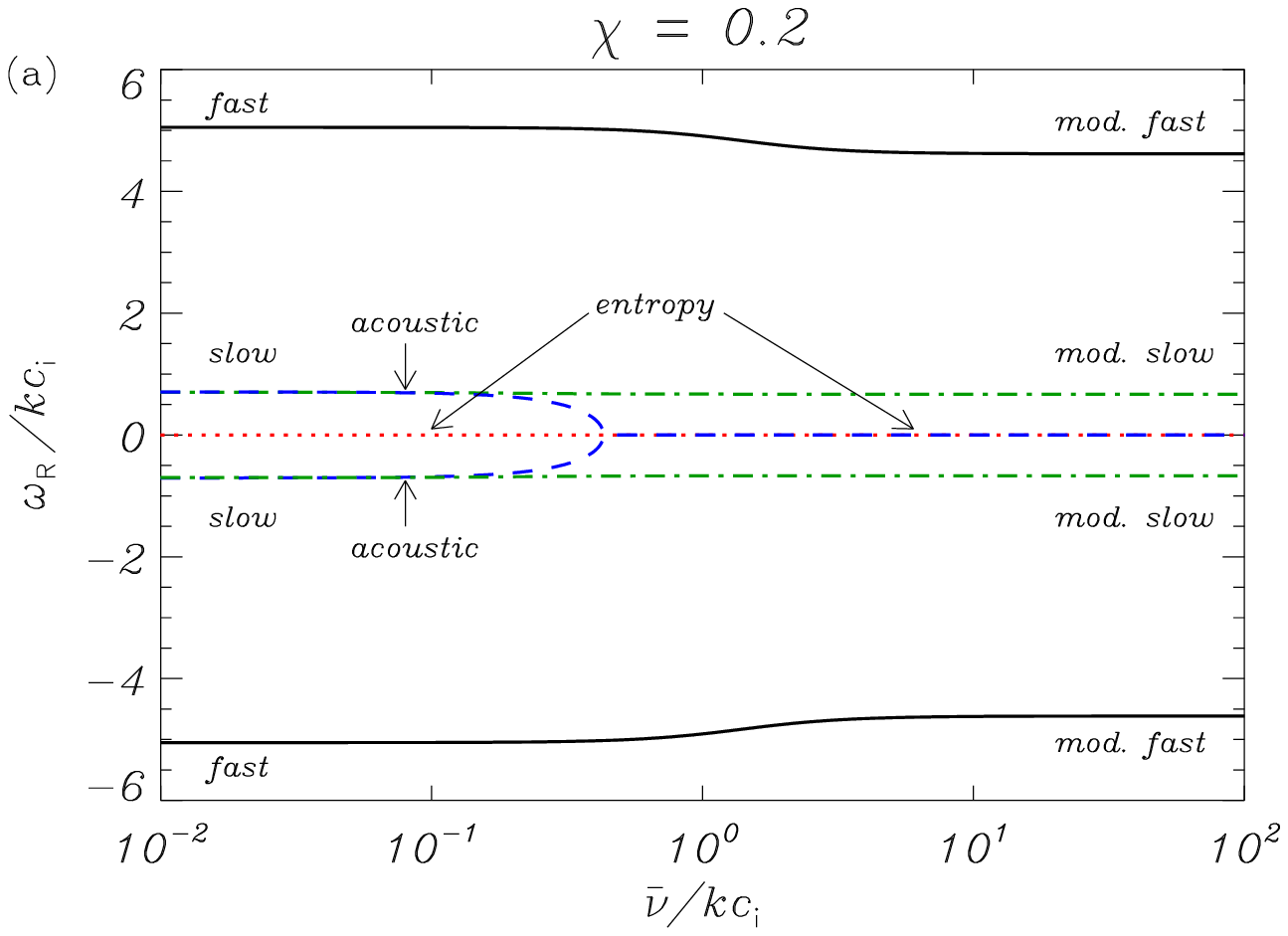}
	\includegraphics[width=.49\columnwidth]{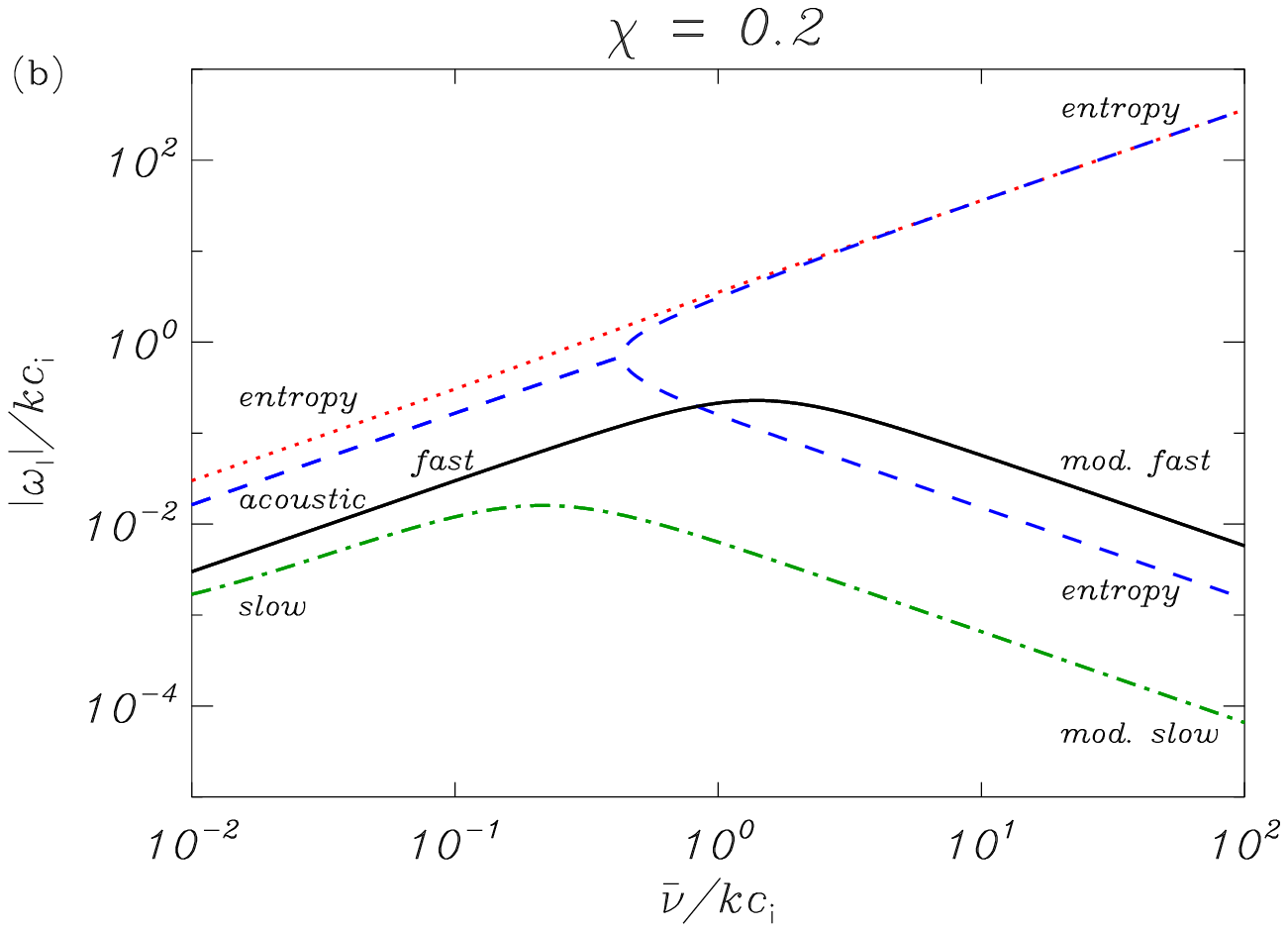}
	\includegraphics[width=.49\columnwidth]{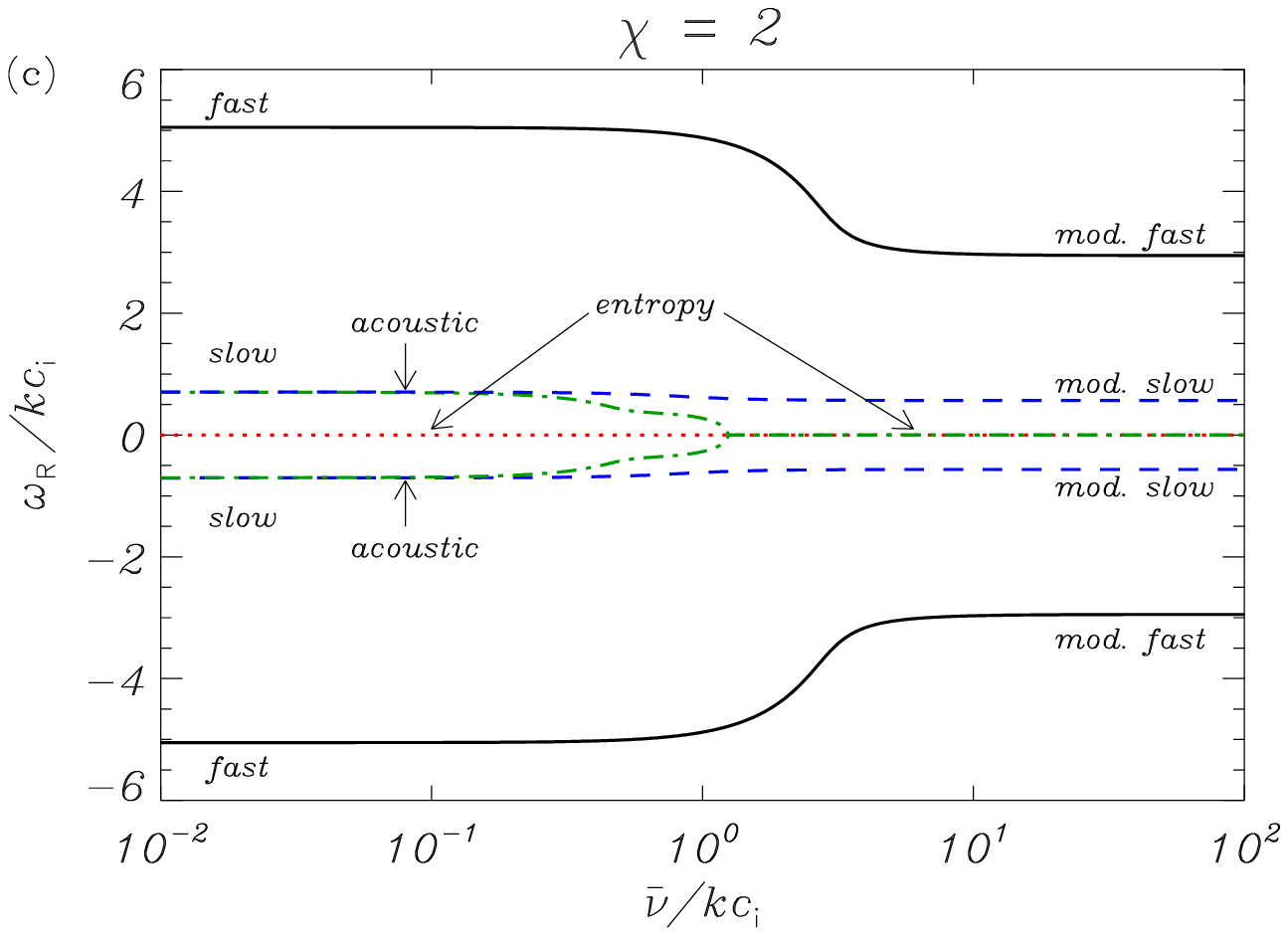}
	\includegraphics[width=.49\columnwidth]{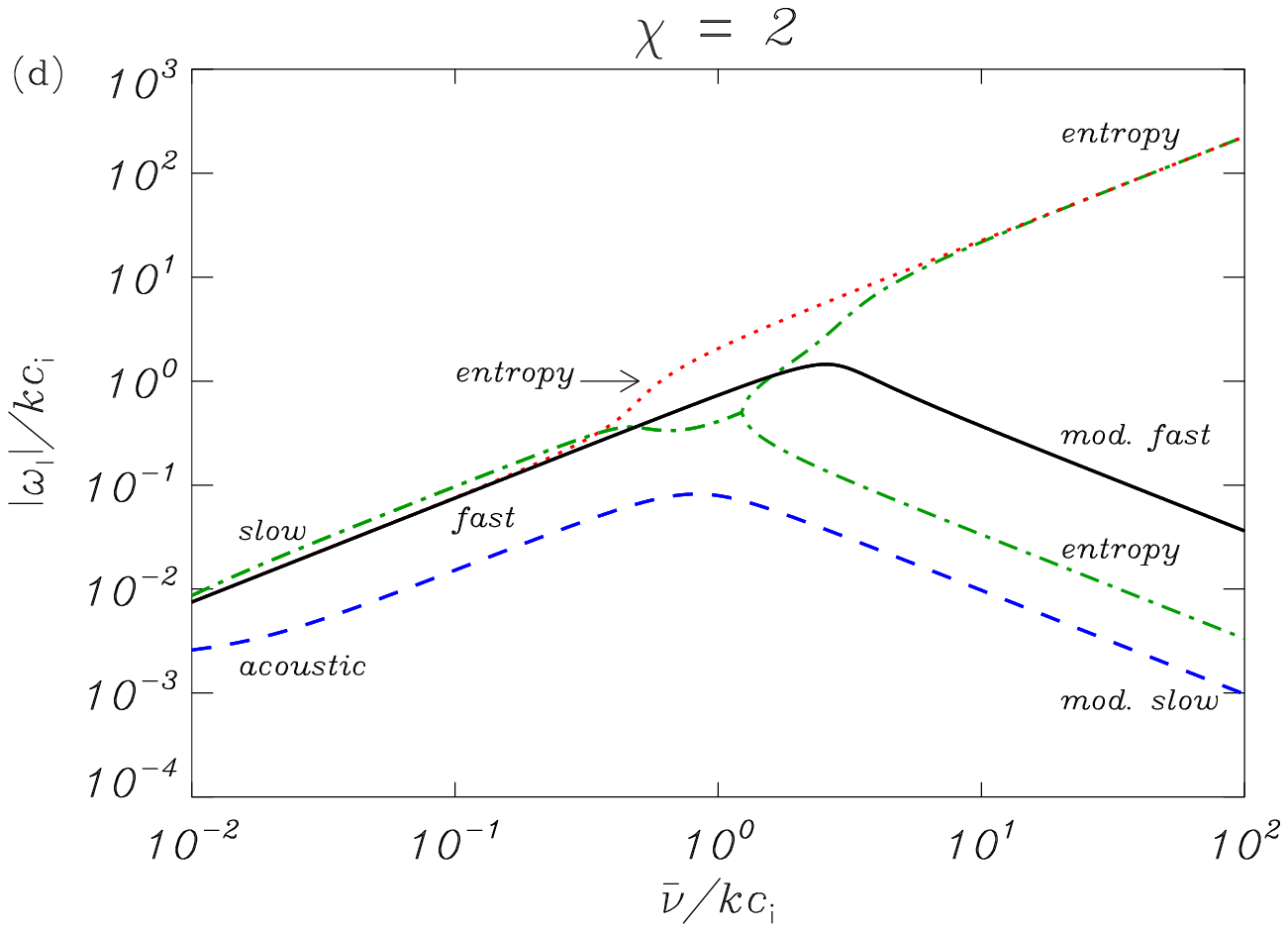}
	\includegraphics[width=.49\columnwidth]{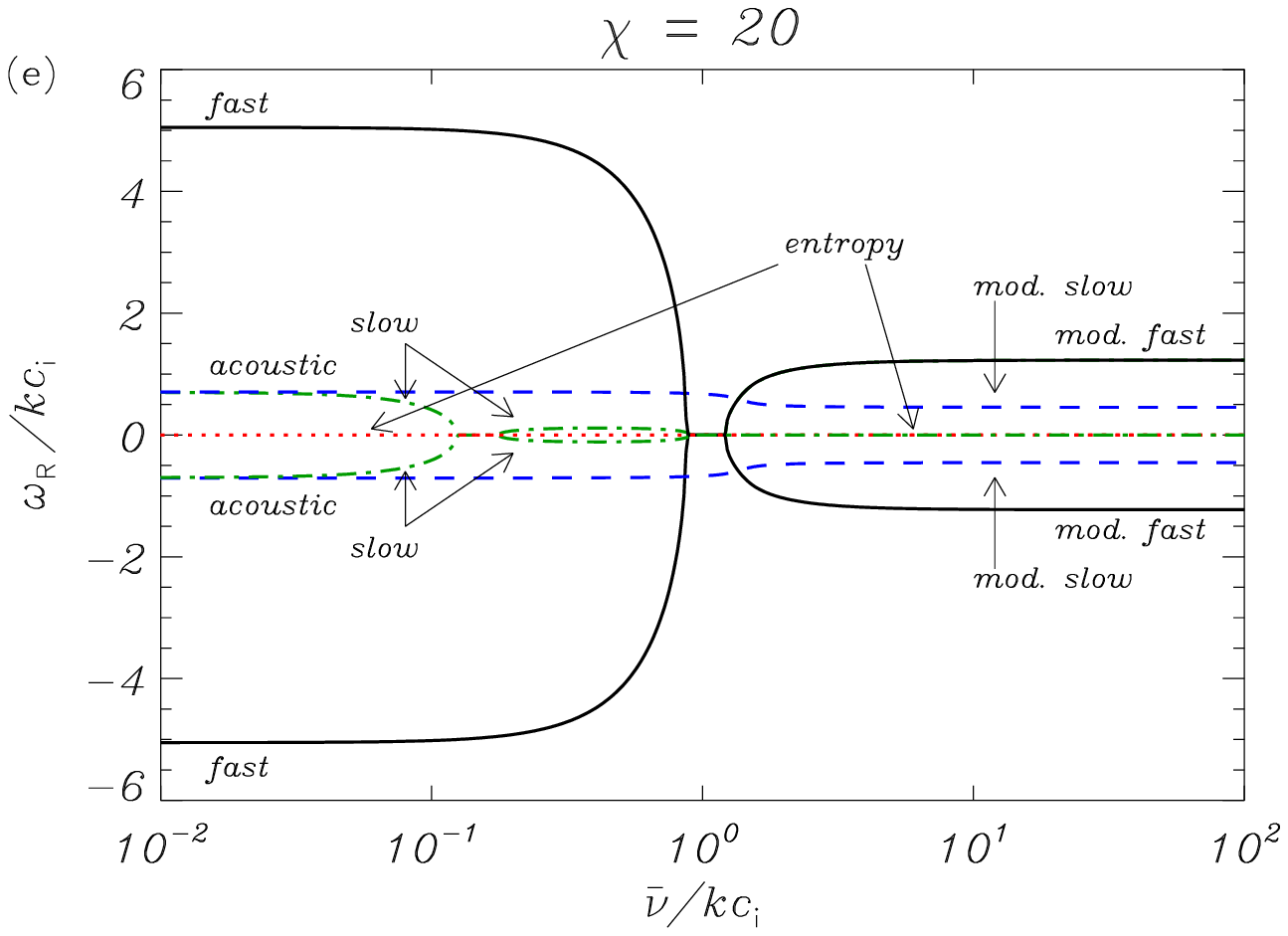}
	\includegraphics[width=.49\columnwidth]{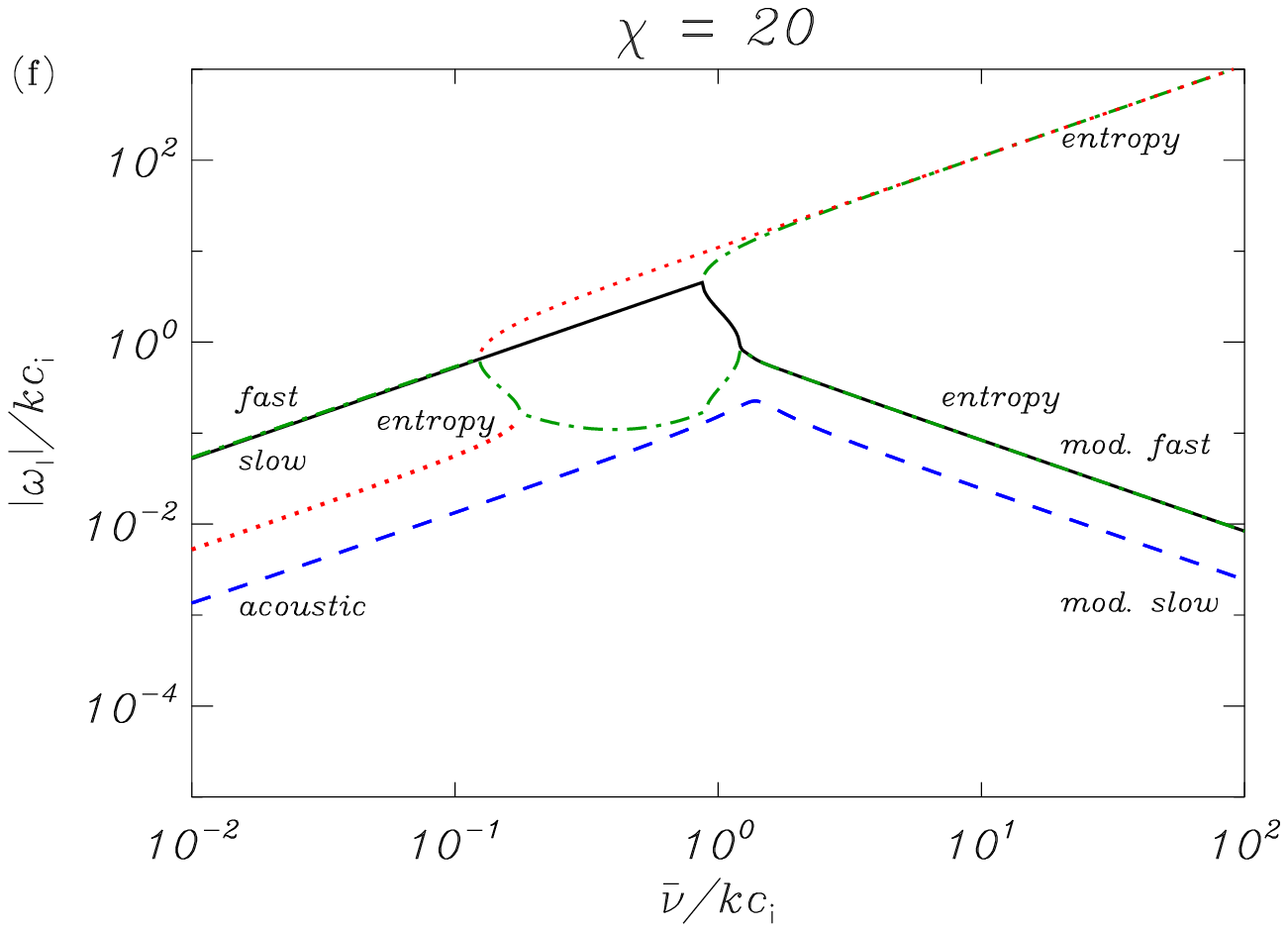}
	\caption{Same as Figure~\ref{fig:perp} but for oblique propagation with $\theta = \pi/4$. We use $\beta_{\rm i}=0.04$.}
	\label{fig:pi4}
\end{figure*}

\begin{figure*}
	\centering
	\includegraphics[width=.49\columnwidth]{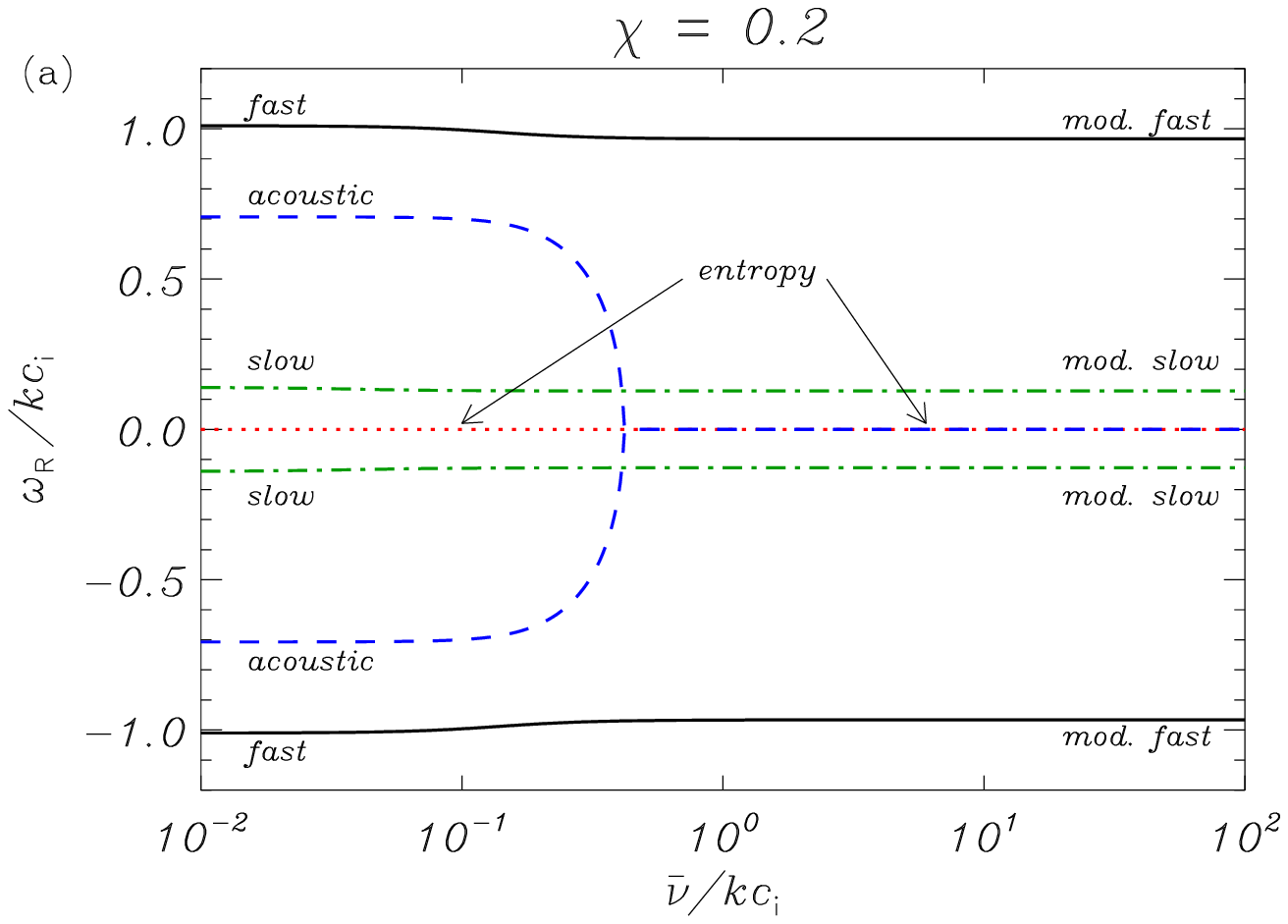}
	\includegraphics[width=.49\columnwidth]{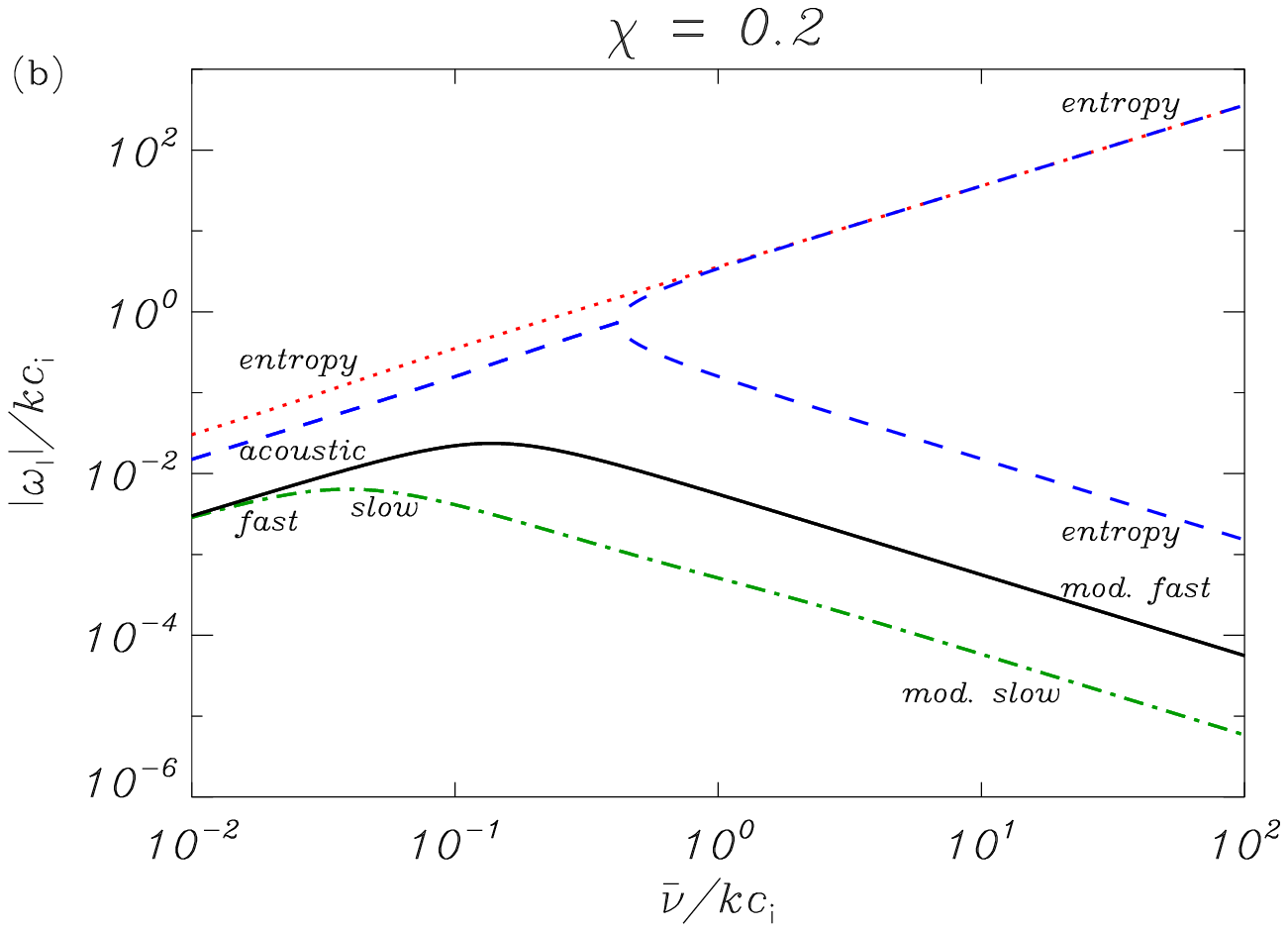}
	\includegraphics[width=.49\columnwidth]{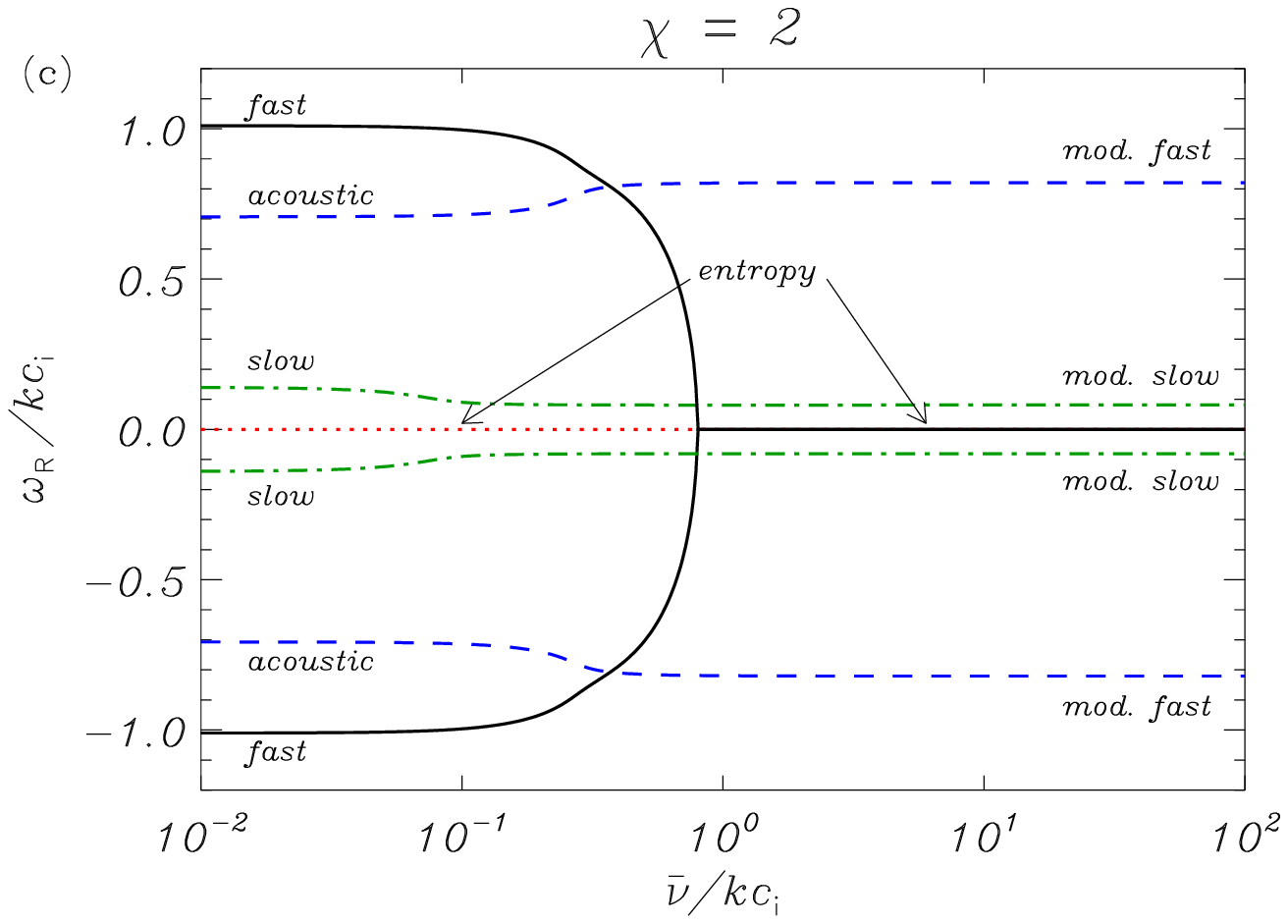}
	\includegraphics[width=.49\columnwidth]{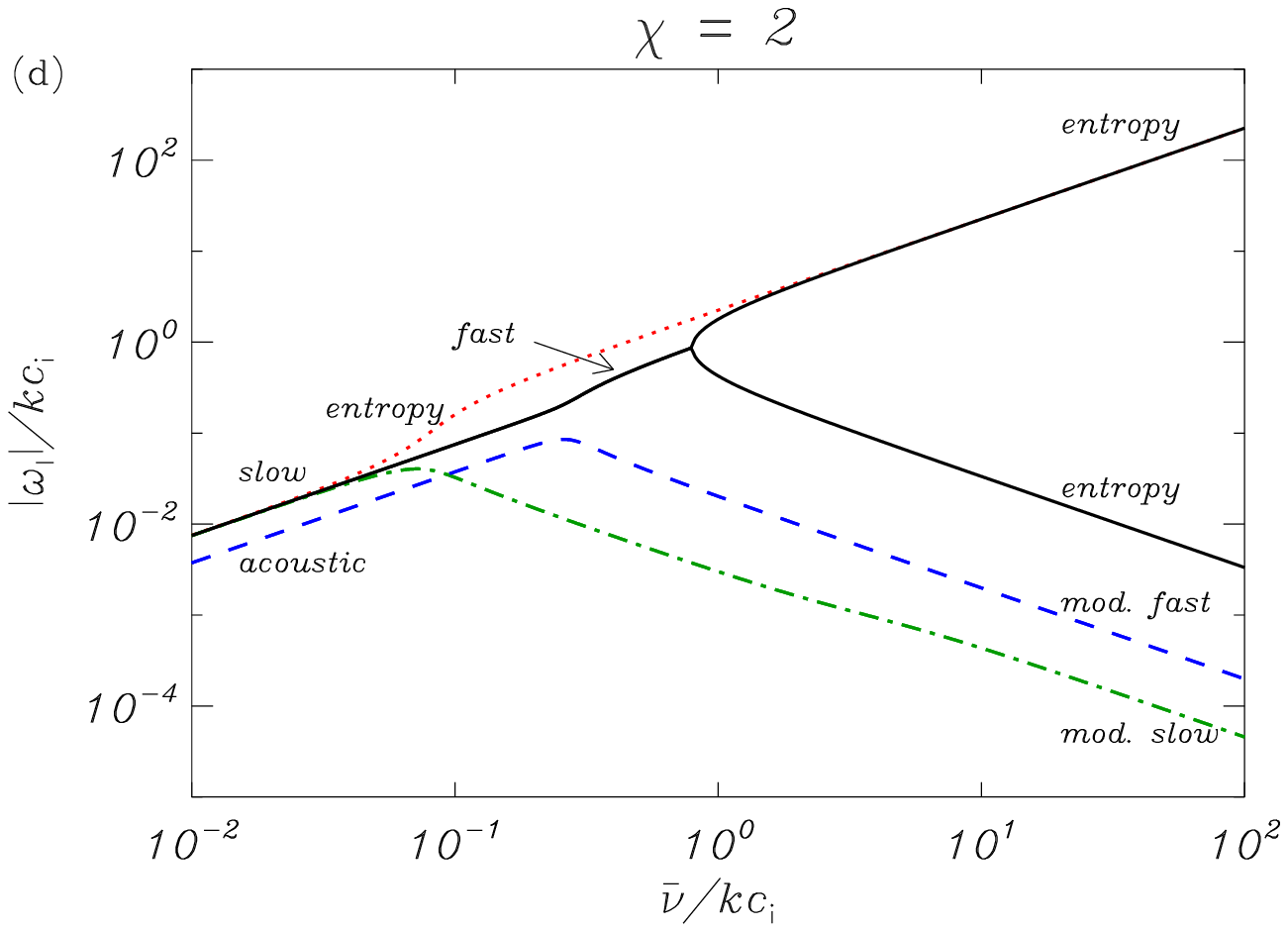}
	\includegraphics[width=.49\columnwidth]{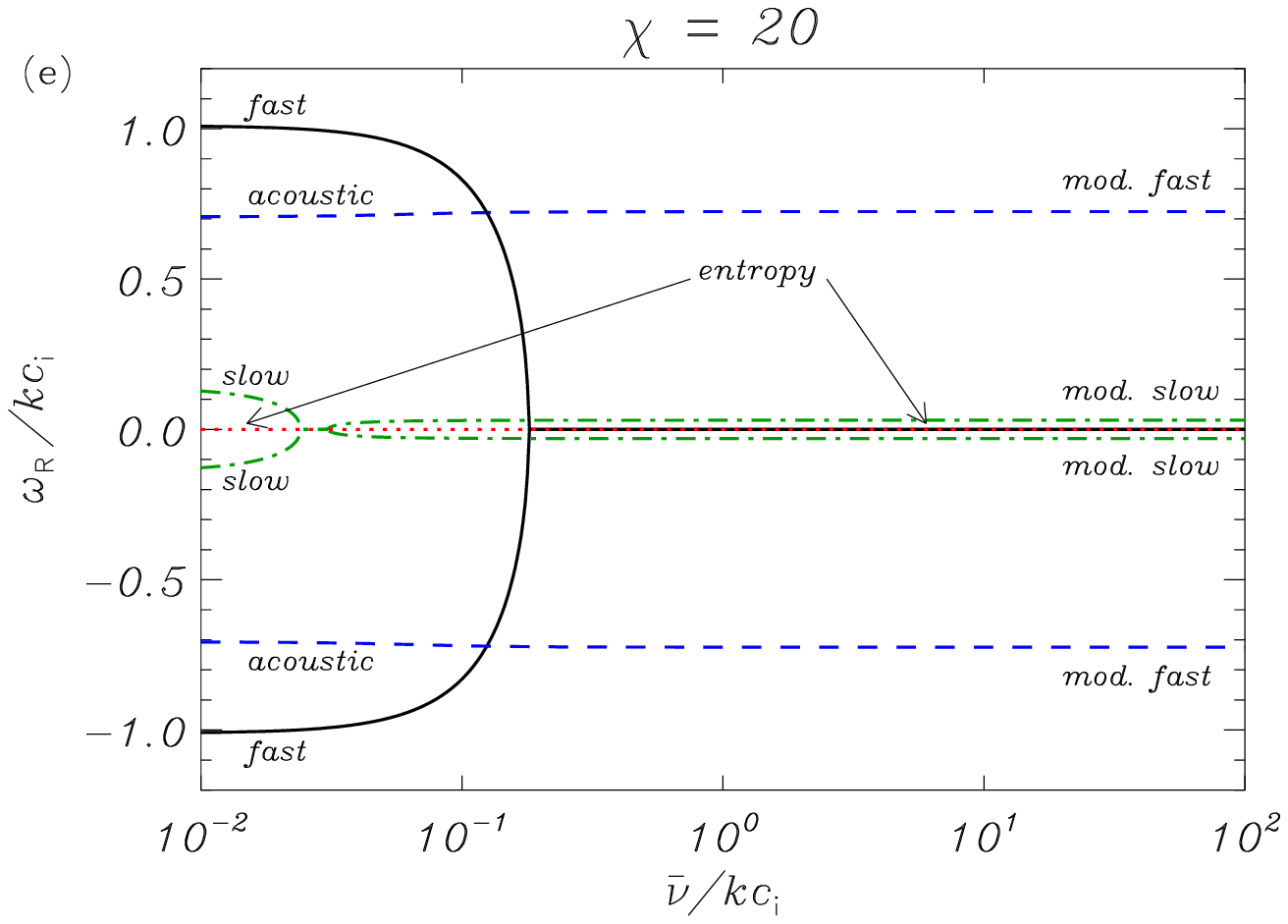}
	\includegraphics[width=.49\columnwidth]{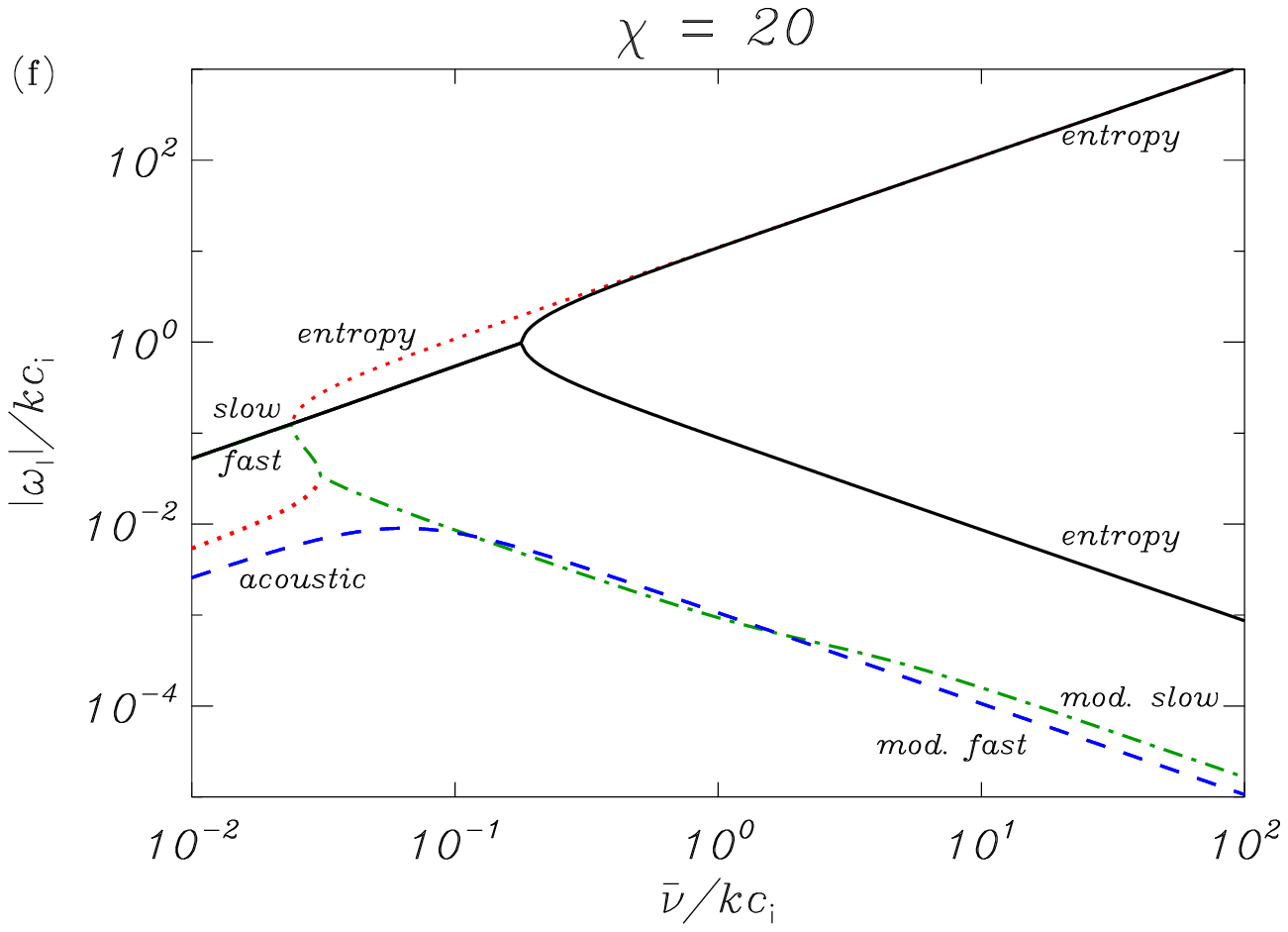}
	\caption{Same as Figure~\ref{fig:pi4} but for $\beta_{\rm i}=25$.}
	\label{fig:pi4h}
\end{figure*}

Finally, here we consider the case in which the direction of wave propagation is in between the  limit cases studied in the previous two Subsections. Hence, we use $\theta = \pi/4$. The results are displayed in Figures~\ref{fig:pi4} and \ref{fig:pi4h}. The behavior of the various solutions is more complex than in the previous limits and can be understood as a combination of the results for perpendicular and parallel propagation. Compared to the case with $\theta=0$,  the acoustic and fast modes display similar behavior, while the slow mode is the solution whose behavior is most altered, specially in the low-$\beta_{\rm i}$ case (compare Figures~\ref{fig:para} and \ref{fig:pi4}). Both slow and fast modes display nonpropagating intervals for oblique propagation, whereas only the fast mode have a nonpropagating interval for parallel propagation. The fast mode nonpropagating interval is also approximately described by Equation~(\ref{eq:nonprop}), which points out that the location of this interval does not depend on $\theta$.

\begin{figure*}
	\centering
	\includegraphics[width=.69\columnwidth]{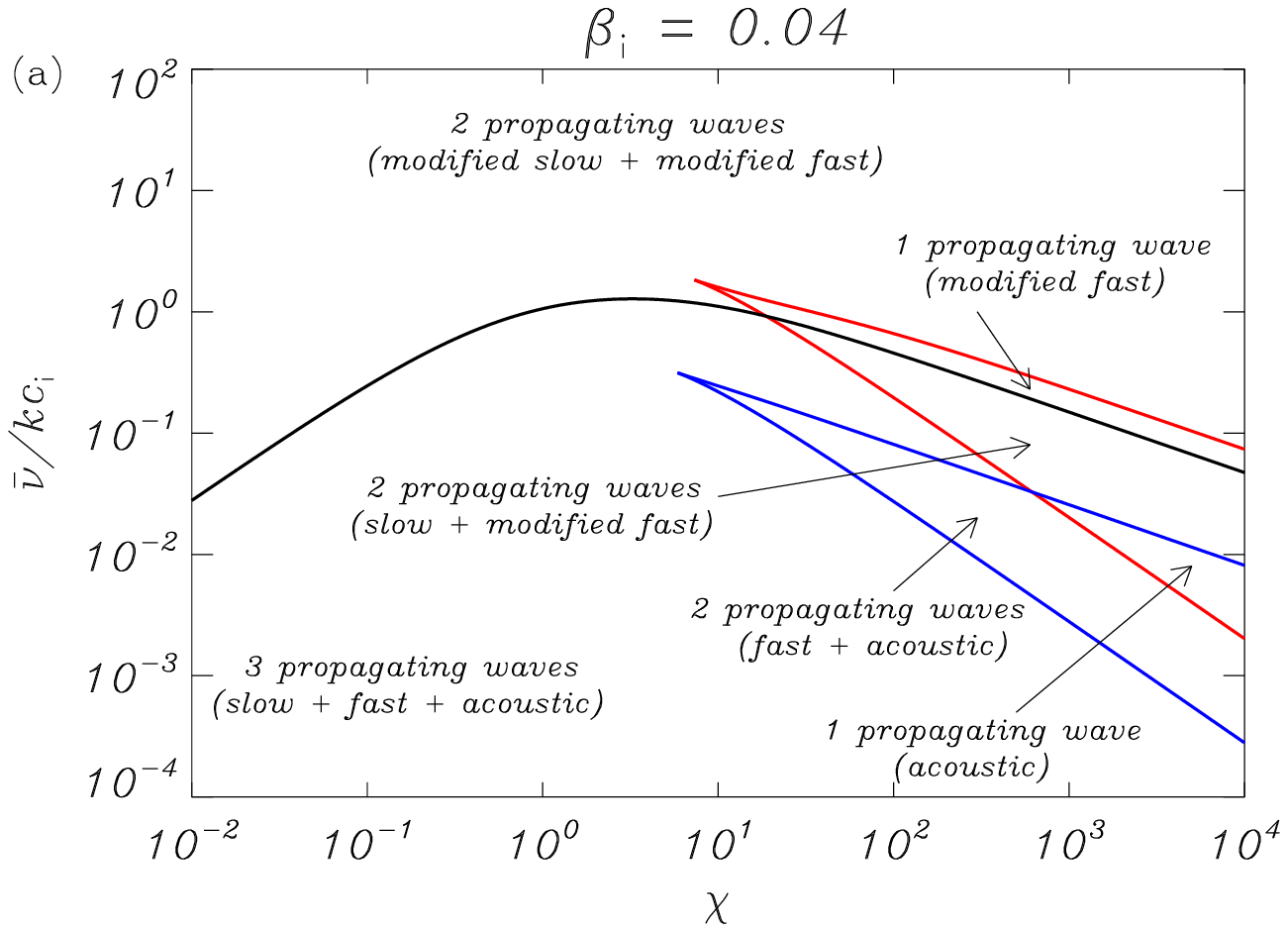}
		\includegraphics[width=.69\columnwidth]{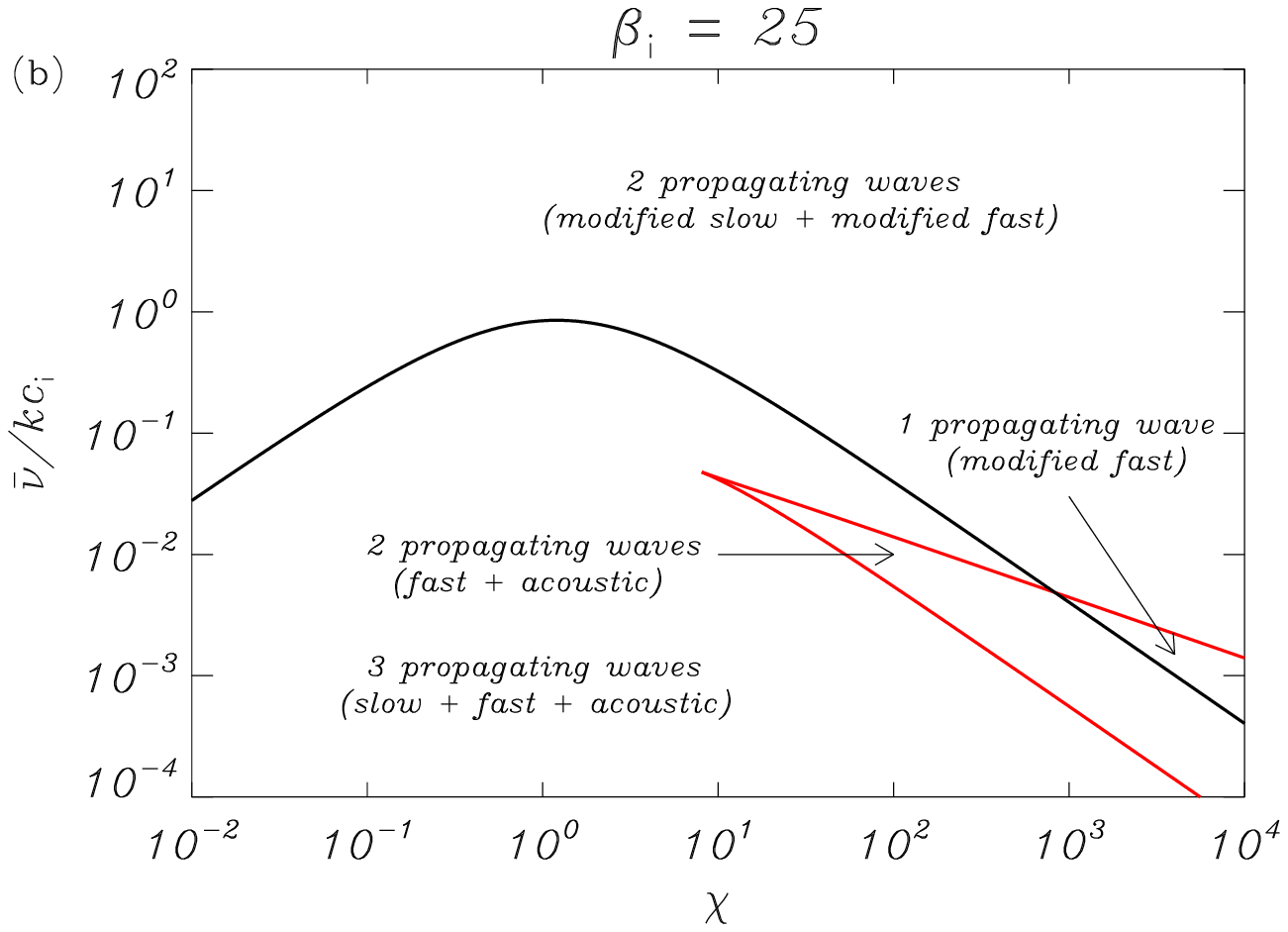}
	\caption{Number and nature of the propagating solutions of the dispersion relation in the $\chi$--$\bar{\nu}/k\cie$ plane  for $\theta = \pi/4$ and (a) $\beta_{\rm i} = 0.04$ and (b) $\beta_{\rm i} = 25$. The red and blue lines correspond to the nonpropagating intervals for slow and fast waves.}
	\label{fig:discpi4}
\end{figure*}

We proceed as before and show in Figure~\ref{fig:discpi4} the various  possible scenarios for wave propagation in the $\chi$--$\bar{\nu}/k\cie$ plane of parameters. The high-$\beta_{\rm i}$ case (Figure~\ref{fig:discpi4}b) is very similar to the equivalent result for $\theta = 0$ (Figure~\ref{fig:discpara}b), so that no additional comments are needed. However, in the low-$\beta_{\rm i}$ case, the presence of nonpropagating intervals for both slow and fast modes and their intersections cause the presence of six possible scenarios (Figure~\ref{fig:discpi4}a). The result  represented in Figure~\ref{fig:discpi4}a clearly points out the high complexity of the interactions between the various waves in a partially ionized plasma, which do not occur in  fully ionized plasmas \citep{mouschovias2011}.

The existence of two nonpropagating intervals in the low-$\beta_{\rm i}$ case is consistent with the results by \citet{soler2013aa} of slow waves propagating along a partially ionized magnetic flux tube. \citet{soler2013aa} also found the presence of two nonpropagating intervals, see their Figure~5. Here we see that this double nonpropagating interval is a result also present for magnetoacoustic waves in a homogeneous, partially ionized medium. Thus, we can now conclude that this result is not caused by the geometry of the waveguide, but it is an intrinsic property of magnetoacoustic waves propagating obliquely to the magnetic field  in a partially ionized plasma. The slow magnetoacoustic waves in a magnetic flux tube studied by \citet{soler2013aa} behave as the waves studied here for oblique propagation.

\section{APPLICATION TO THE SOLAR CHROMOSPHERE}
\label{sec:app}

Here we perform a specific application to magnetoacoustic waves propagating in the partially ionized solar chromosphere. This application is motivated by the recent observations of ubiquitous compressive waves in the low solar atmosphere \citep{morton2011,morton2012}. To represent the chromospheric plasma we use the same simplified model considered in Paper~I. We use the quiet sun model C of \citet{vernazza1981}, hereafter VALC model, to account for the variation of physical parameters with height from the photosphere to the base of the corona. The expression  of  $\ain$ is taken after \citet{brag}, namely
\begin{equation}
\ain = \frac{1}{2} \frac{\rhoi \rhon}{m_{\rm i}}\sqrt{\frac{16 k_{\rm B} T}{\pi m_{\rm i}}}\sigma_{\rm in}, \label{eq:ainhydrogen}
\end{equation}
where $m_{\rm i}$ is the ion (proton) mass, $k_{\rm B}$ is Boltzmann's constant, and $\sigma_{\rm in}$ is the collision cross section. Equation~(\ref{eq:ainhydrogen}) considers only hydrogen and ignores the influence of heavier species. In Equation~(\ref{eq:ainhydrogen}) it is implicitly assumed that the ion and neutral masses are approximately equal. For the collision cross section of protons with neutral hydrogen atoms we consider the typical value of $\sigma_{\rm in}\approx 5\times 10^{-19}$~m$^2$ \citep[used in, e.g.,][among others]{khodachenko2004,leake2005,arber2007,soler2009PI,khomenko2012}. Here we must note that, in the application done in Paper~I for the case of Alfv\'en waves, we took $\sigma_{\rm in}\approx 10^{-20}$~m$^2$ based on the hard sphere collision model \citep[see, e.g.,][]{brag,zaqarashvili2013}. In a recent paper, \citet{vranjes2013} claim that the realistic value of $\sigma_{\rm in}$ in the solar chromosphere is about two orders of magnitude higher than that estimated in the hard sphere  model. The value $\sigma_{\rm in}\approx 5\times 10^{-19}$~m$^2$ used here is closer to the value proposed by \citet{vranjes2013} than to the hard sphere value.

 We adopt the magnetic field strength model by \citet{leake2006}, which aims to represent the field strength in a  chromospheric vertical magnetic flux tube expanding  with height, namely
\begin{equation}
B = B_{\rm ph} \left( \frac{\rho}{\rho_{\rm ph}} \right)^{0.3},
\end{equation}
where $\rho = \rhoi + \rhon$ is the total density, and $B_{\rm ph}$ and $\rho_{\rm ph} = 2.74\times10^{-4}$~kg~m$^{-3}$ are the magnetic field strength and the total density, respectively, at the photospheric level.  The variation of $\rho$ with height is taken from the VALC model. Accordingly, the magnetic field strength decreases with height. Regarding the value of $B_{\rm ph}$, we consider two possible scenarios: an active region and the quiet Sun. For the active region case, we set $B_{\rm ph} = 1.5$~kG, so that $B\approx 100$~G at 1,000~km  and  $B\approx 20$~G at 2,000~km above the photosphere. For the quiet Sun case, we set $B_{\rm ph} = 100$~G, so that $B\approx 7$~G at 1,000~km  and  $B\approx 1$~G at 2,000~km above the photosphere.

As explained in Paper~I, the physical parameters in this simplified model of the low solar atmosphere depend on the vertical direction, while in the previous theoretical analysis all the parameters are taken constant. Here we perform a local analysis and use the dispersion relation derived for a homogeneous plasma. We use  the physical parameters at a given height to locally solve the dispersion relation at that height.  A limitation of the present approach is that it ignores the possible presence of cutoff frequencies due to gravitational stratification \citep[see, e.g.,][]{roberts2006}. The condition for this method to be approximately valid is that the wavelength,  $\lambda = 2\pi/k$, is much shorter than the stratification scale height. This is fulfilled by the wavelengths used in this analysis.

\begin{figure*}
	\centering
	\includegraphics[width=.59\columnwidth]{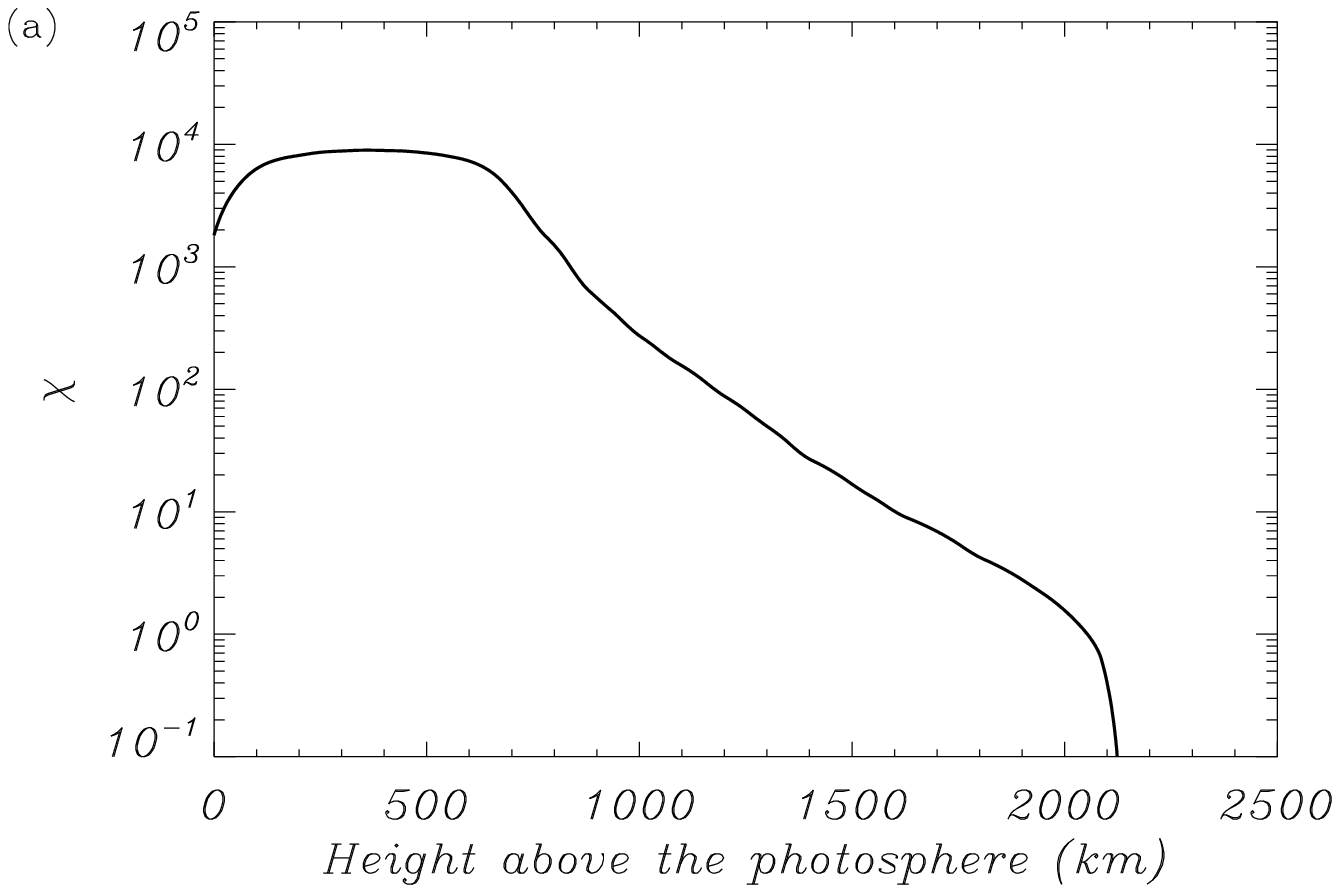}
	\includegraphics[width=.59\columnwidth]{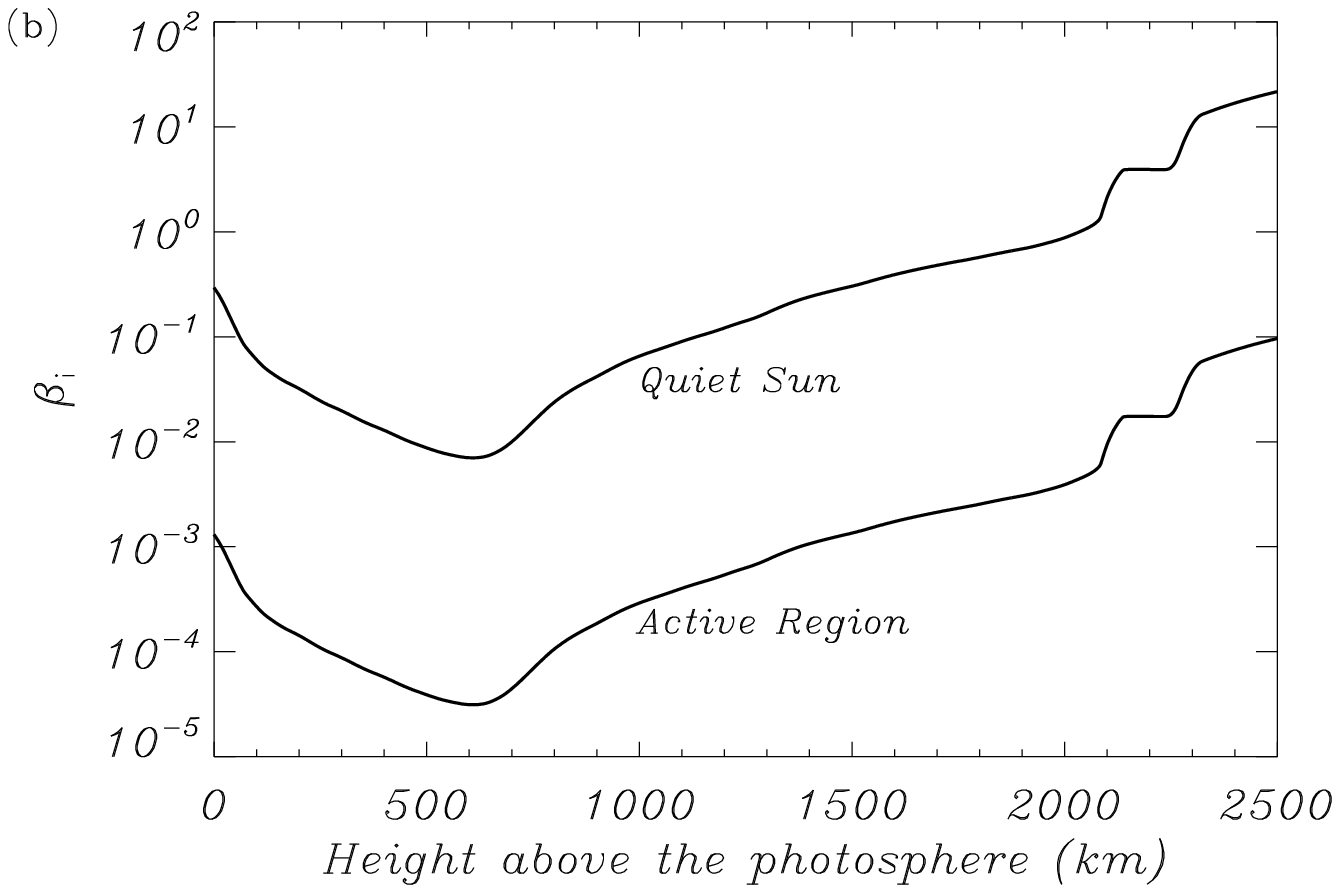}
	\includegraphics[width=.59\columnwidth]{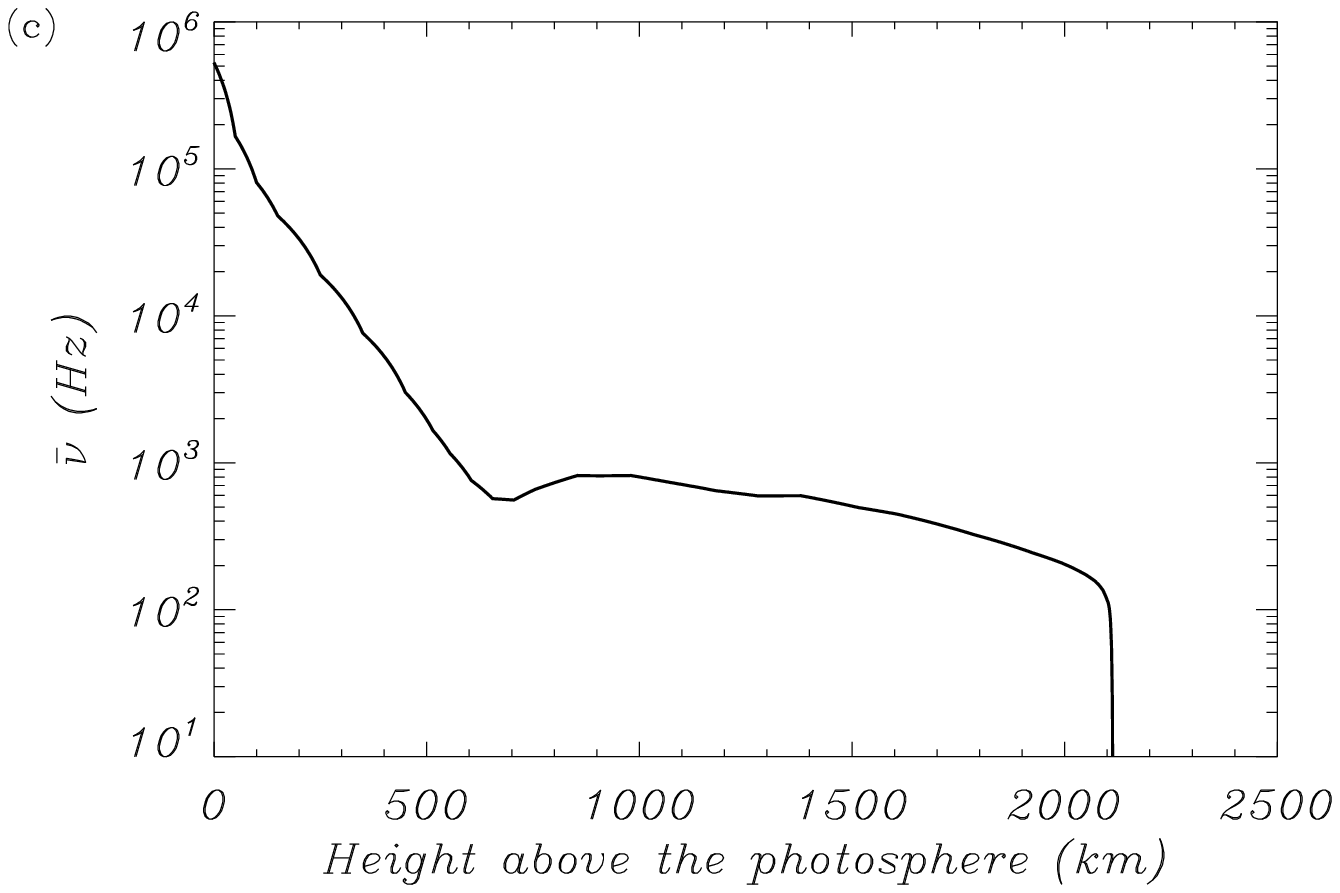}
	\caption{Variation of (a) $\chi$, (b) $\beta_{\rm i}$, and (c) $\bar \nu$ with height in the solar chromosphere model used in Section~\ref{sec:app}.}
	\label{fig:valcmodel}
\end{figure*}

Figure~\ref{fig:valcmodel} displays the variation with height above the photosphere of the three physical quantities relevant for the behavior of the waves: the ionization fraction, $\chi$, the ionized fluid $\beta_{\rm i}$, and the averaged ion-neutral collision frequency, $\bar \nu$. Only $\beta_{\rm i}$ is affected by the magnetic field scenario considered. First of all, we see that $\chi$ ranges several orders of magnitude from very weakly ionized plasma at the low levels of the chromosphere to fully ionized plasma when the transition region to the solar corona is reached. Full ionization of hydrogen takes place at 2,100~km above the photospheric level, approximately. The value of $\beta_{\rm i}$ in the active region case is much smaller than unity throughout the chromosphere, which informs us that we are dealing with a low-$\beta_{\rm i}$ situation. Conversely, in the quiet Sun case the value of $\beta_{\rm i}$ ranges from $\beta_{\rm i} \ll 1$ at low levels to $\beta_{\rm i} \gg 1$ at high levels in the chromosphere. Moreover, $\bar \nu$ also ranges several orders of magnitude, specially in the low chromosphere up to 700~km, approximately. Then, $\bar \nu$ takes values in between $10^2$~Hz and $10^3$~Hz until the full ionization level is reached. 

\begin{figure*}
	\centering
	\includegraphics[width=.49\columnwidth]{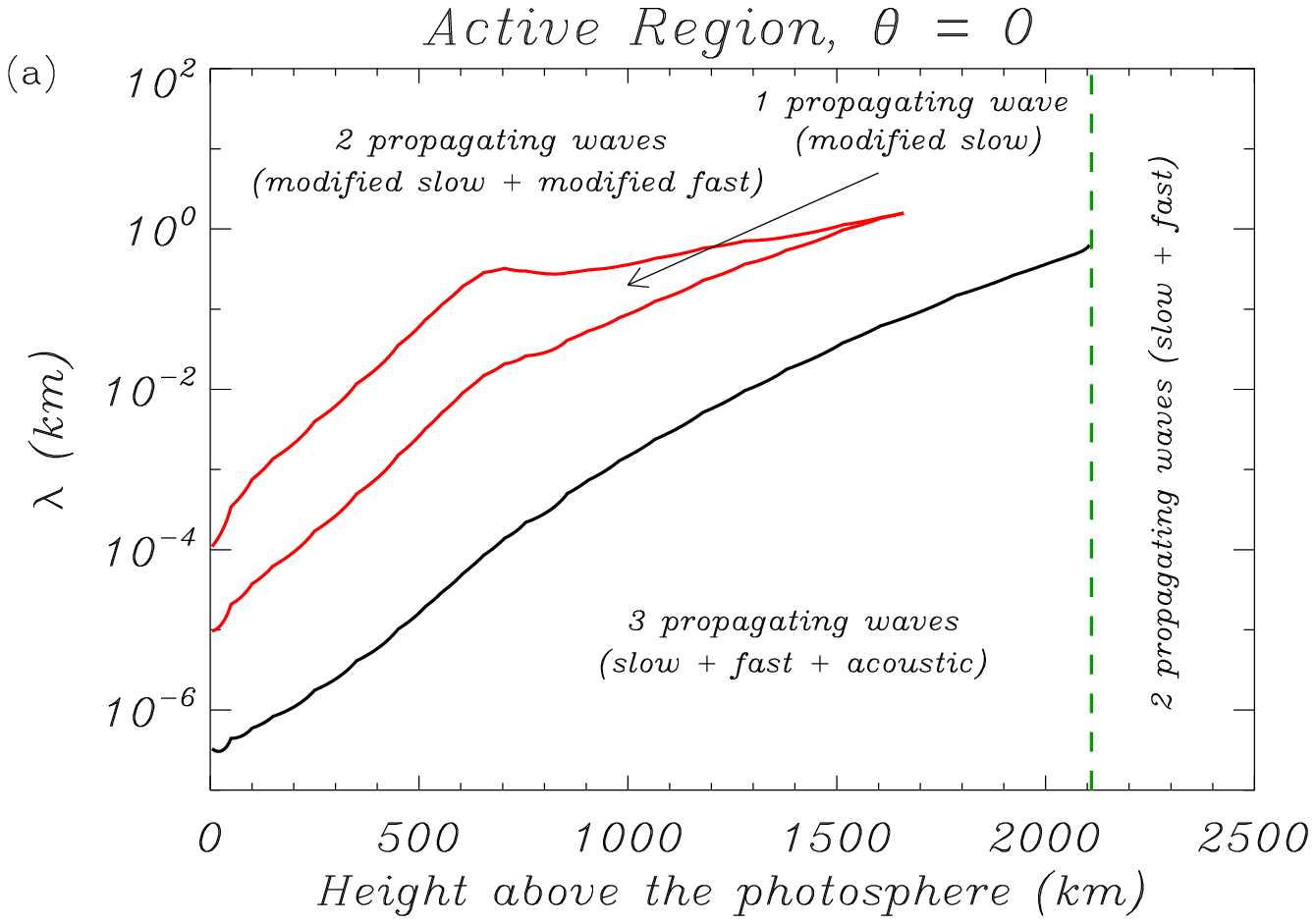}
	\includegraphics[width=.49\columnwidth]{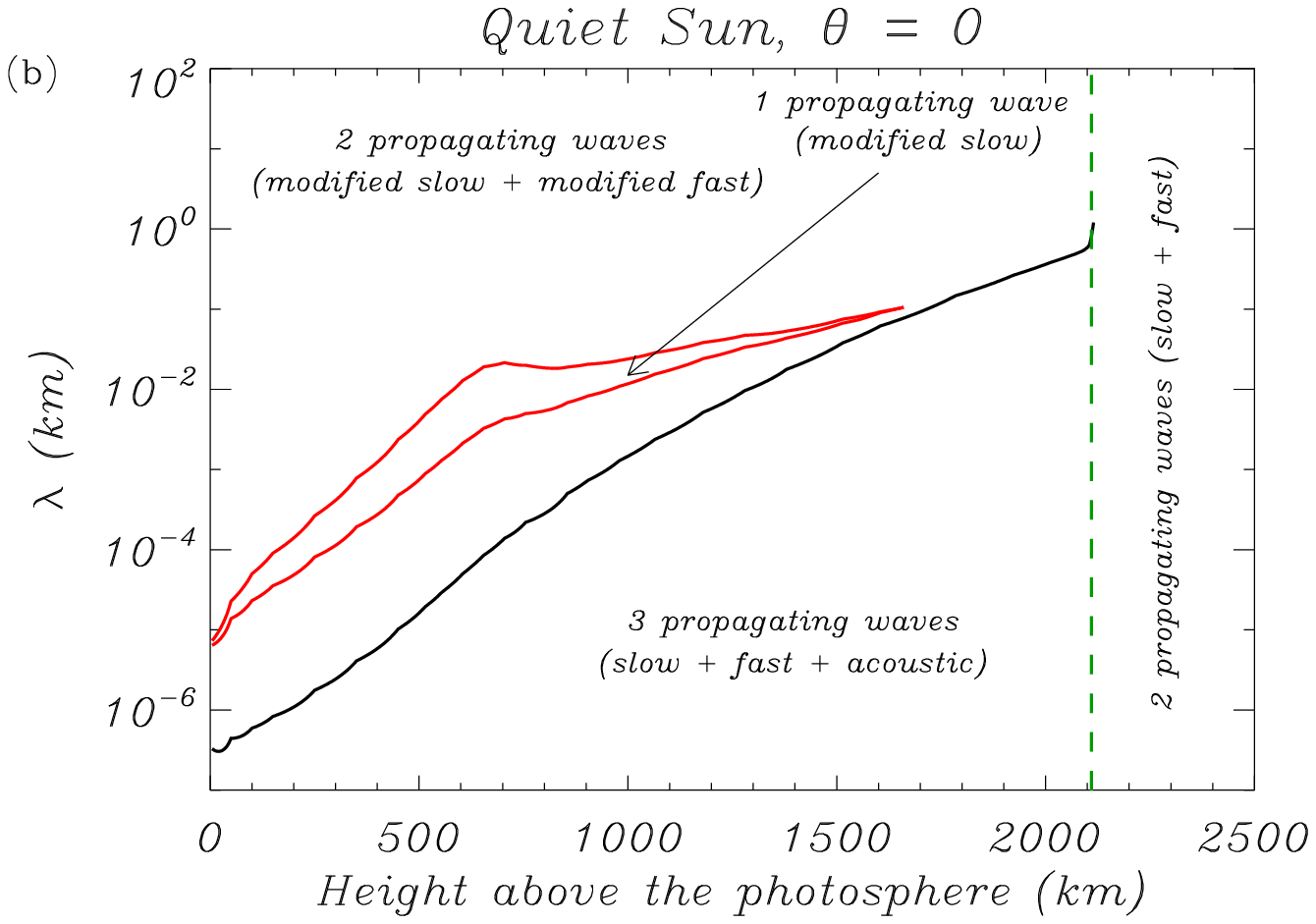}
	\includegraphics[width=.49\columnwidth]{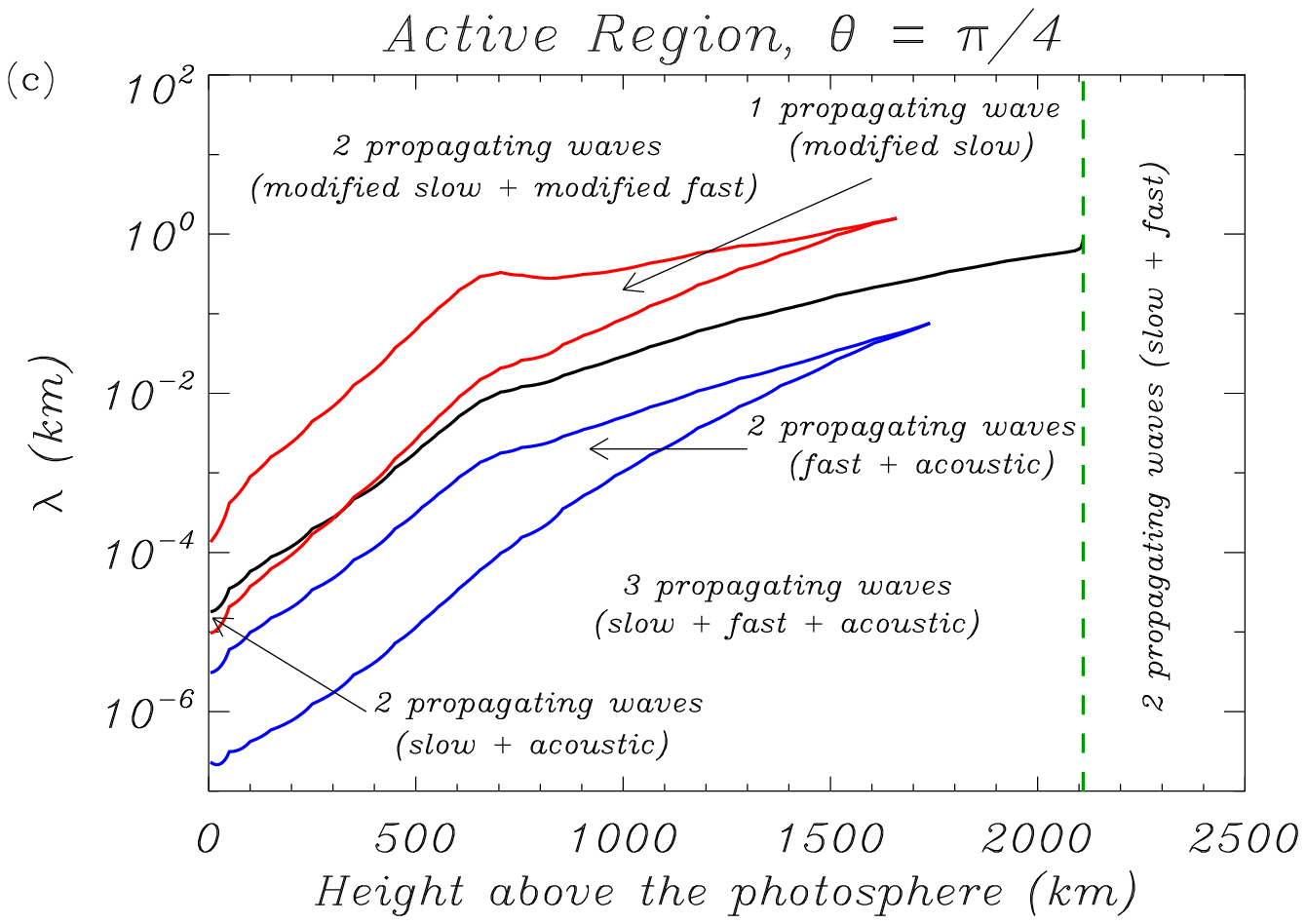}
	\includegraphics[width=.49\columnwidth]{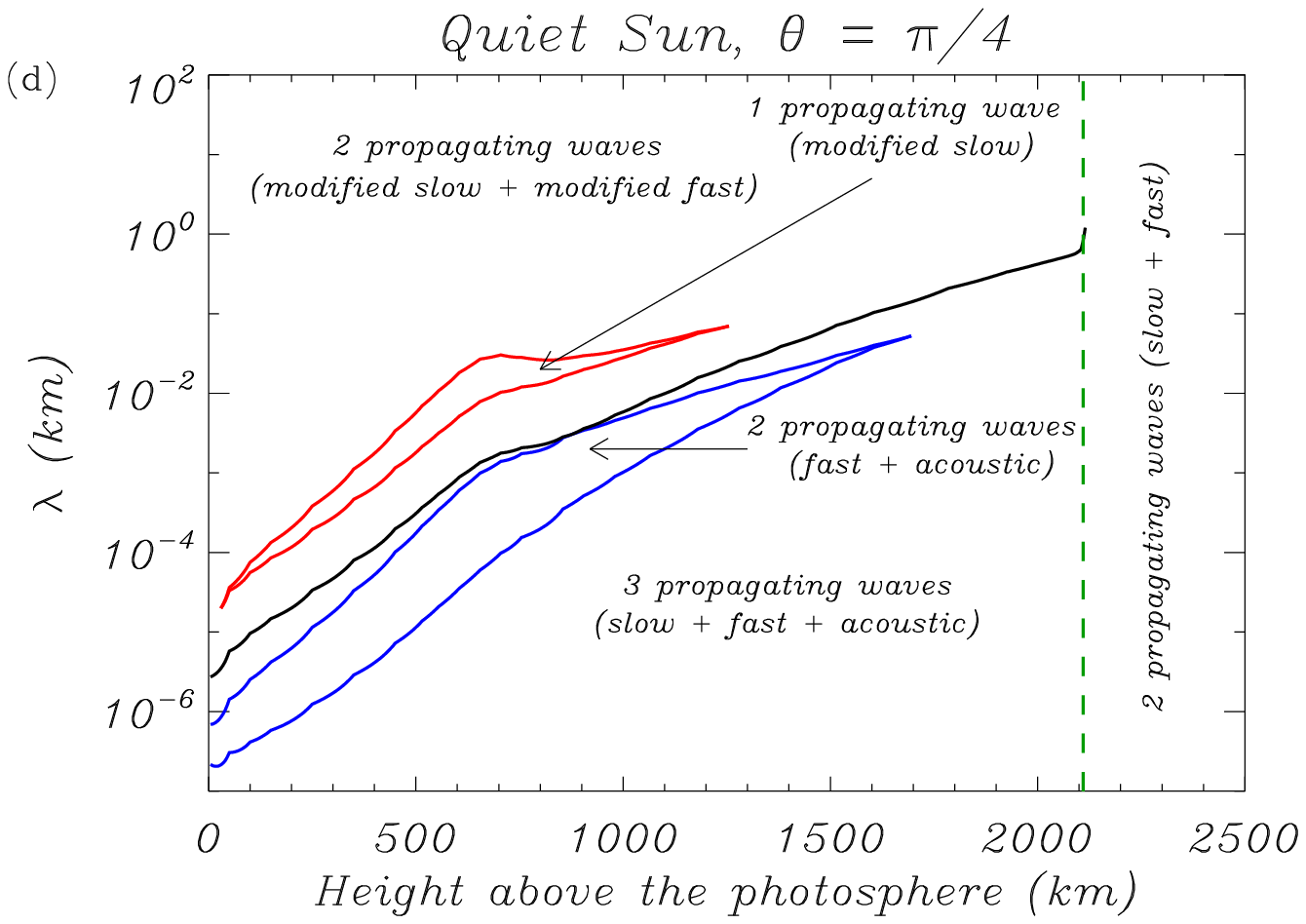}
	\includegraphics[width=.49\columnwidth]{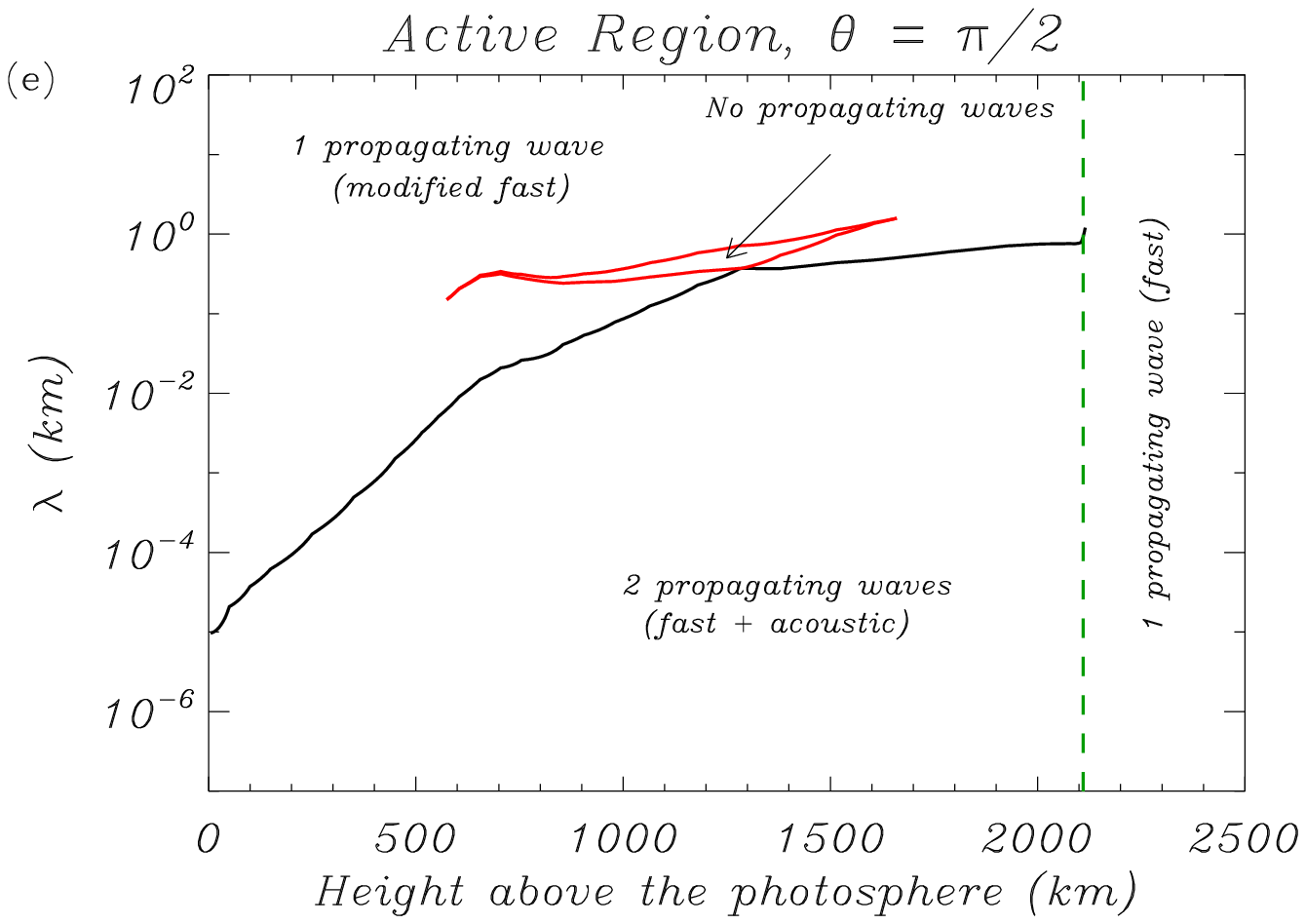}
	\includegraphics[width=.49\columnwidth]{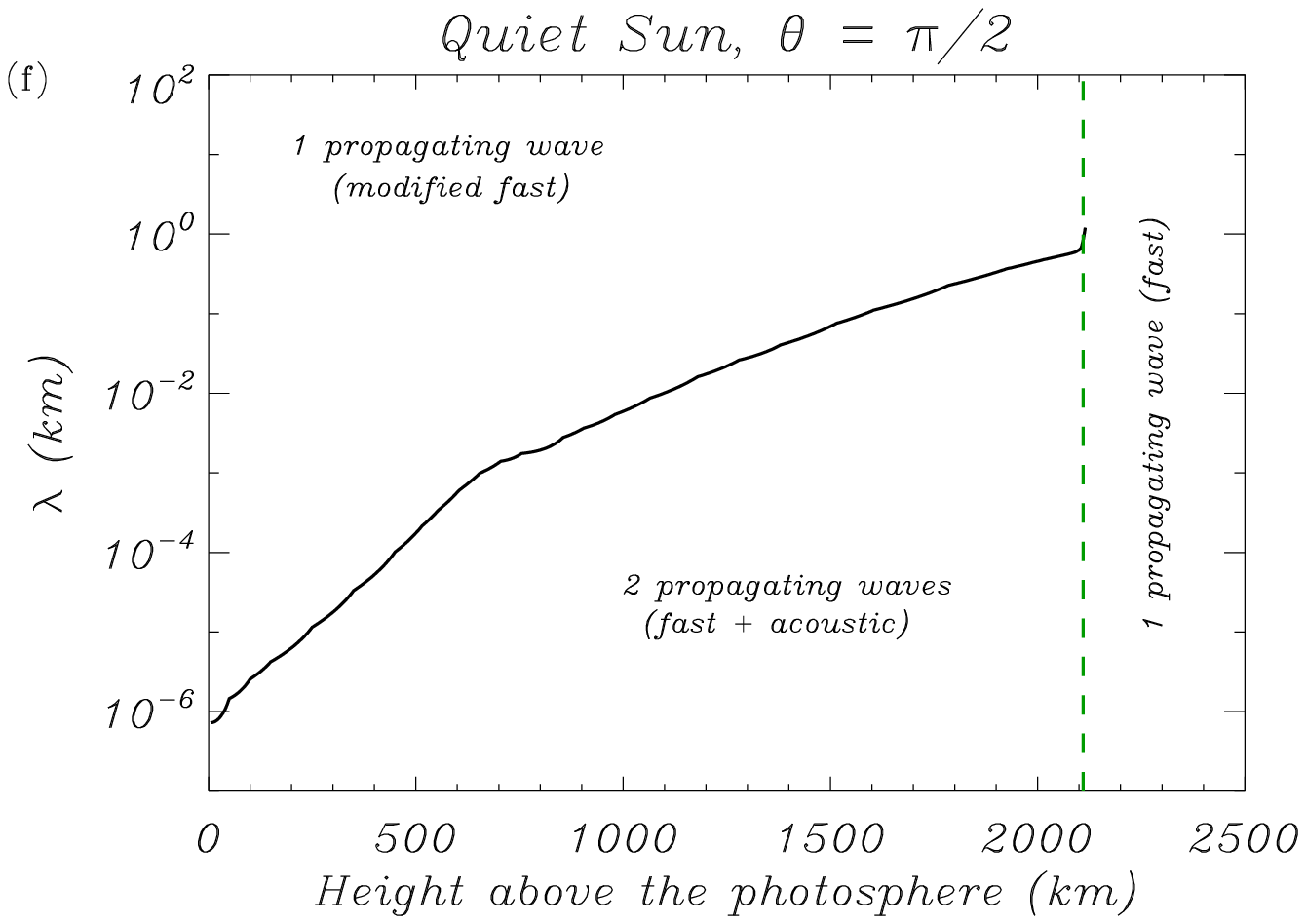}
	\caption{Number and nature of the propagating waves with a given wavelength, $\lambda$, as function of height in the solar chromosphere for the active region case (left) and the quiet Sun case (right) with $\theta = 0$ (top), $\theta = \pi/4$ (middle), and $\theta = \pi/2$ (bottom). The vertical dashed line denotes the height at which hydrogen becomes fully ionized.}
	\label{fig:valcdis}
\end{figure*}

The high values of  $\bar \nu$ displayed in Figure~\ref{fig:valcmodel}(c) indicate that only waves with  short wavelengths, i.e., high frequencies, would be affected by two-fluid effects. To investigate this, we use the discriminant of the dispersion relation to determine the nature of the propagating solutions as function of height for a given value of the wavelength. This is shown in Figure~\ref{fig:valcdis} for three values of $\theta$, namely $\theta=0$, $\pi/4$, and $\pi/2$. In the active region case, we see that two-fluid effects are important for  propagation of magnetoacoustic waves  when $\lambda \lesssim 1$~km. For longer wavelengths, ions and neutrals behave as a single fluid and, consequently, the propagating waves are the  slow and fast waves modified due to the presence of neutrals (see Section~\ref{sec:highly}). For short wavelengths, however, there appear a number of nonpropagating intervals that constrain the propagation of the various modes. Also, due to the strong ion-neutral coupling, the presence of  neutral acoustic waves in the active region chromosphere is only possible for very short wavelengths. In the case of the quiet Sun, the nonpropagating intervals are shifted towards  values of $\lambda$ about an order or magnitude shorter than in the active region case. This result points out that two-fluid effects are of less relevance in those regions of the chromosphere where the magnetic field is weak. For representation purposes,  in Figure~\ref{fig:valcdis} we have varied $\lambda$ in between $10^{-7}$~km and $10^2$~km, although we must warn the reader that the fluid approximation for the various species  may not hold for wavelengths approaching the lower boundary  of this range. Therefore, the results shown  in Figure~\ref{fig:valcdis} for  wavelengths near the lower limit should be taken with extreme caution.  Also, the assumption that gravitational effects are negligible breaks down for very long wavelengths, although this probably happens for  wavelengths longer than those in  Figure~\ref{fig:valcdis}.

\begin{figure*}
	\centering
	\includegraphics[width=.49\columnwidth]{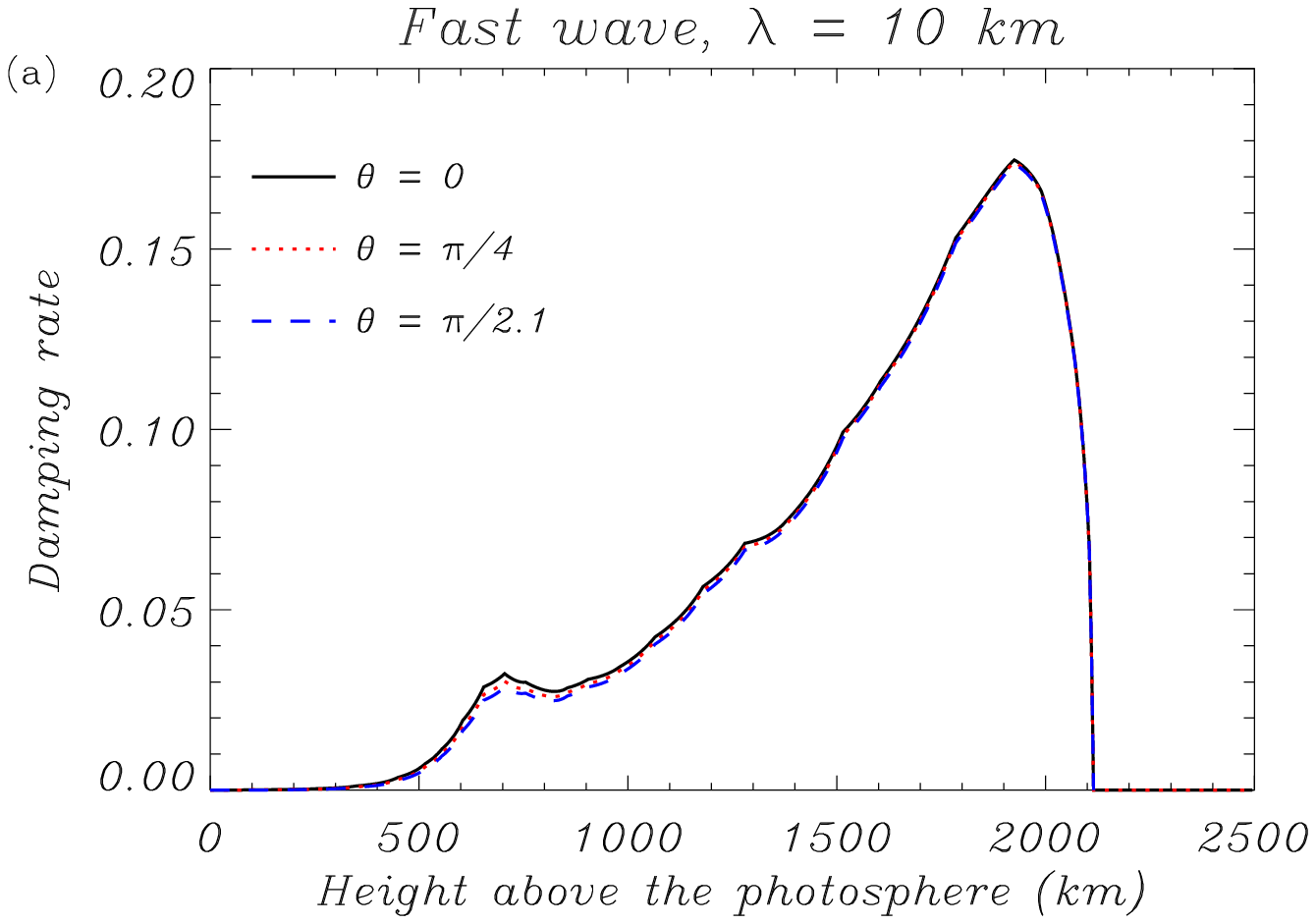}
	\includegraphics[width=.49\columnwidth]{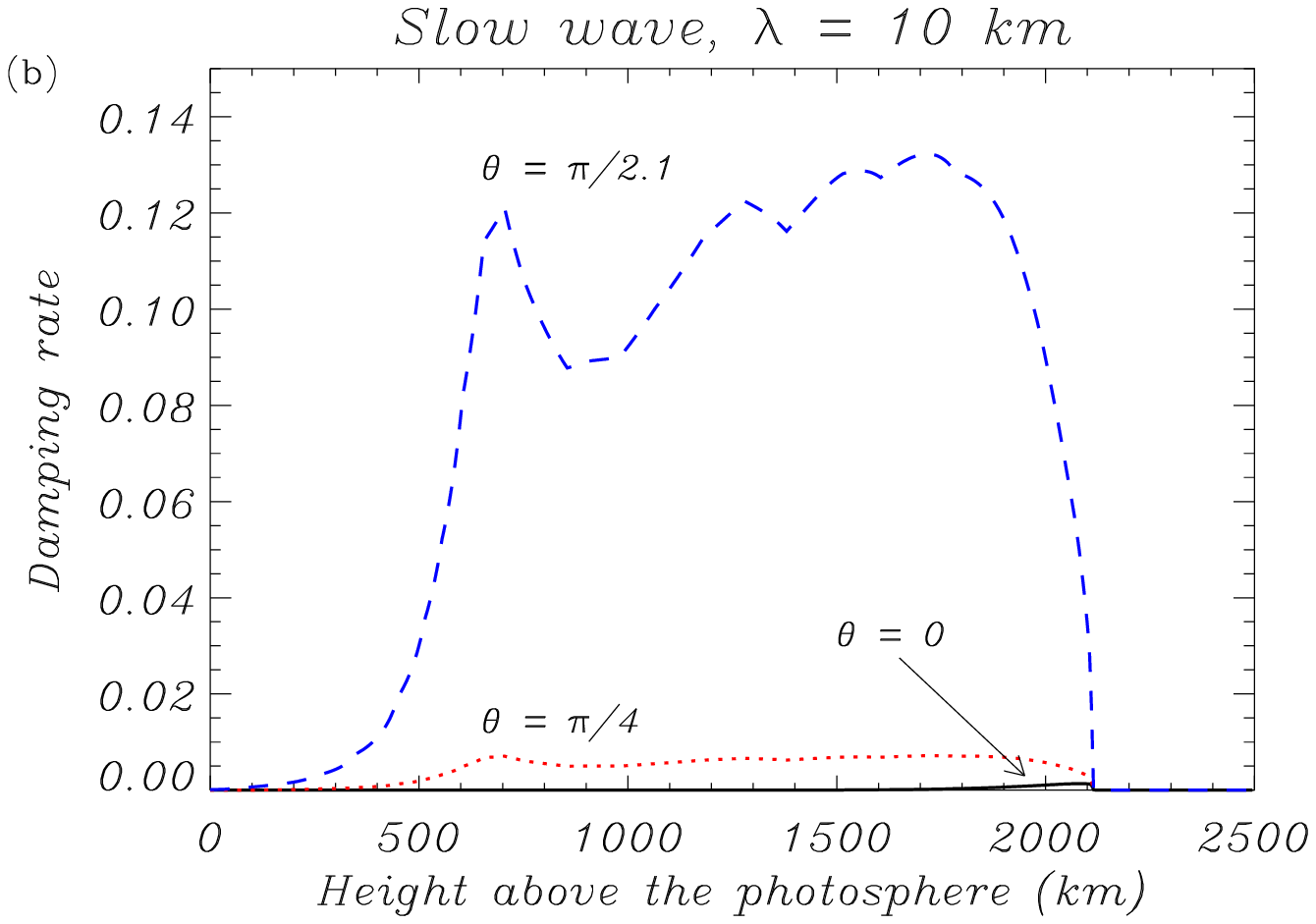}
	\includegraphics[width=.49\columnwidth]{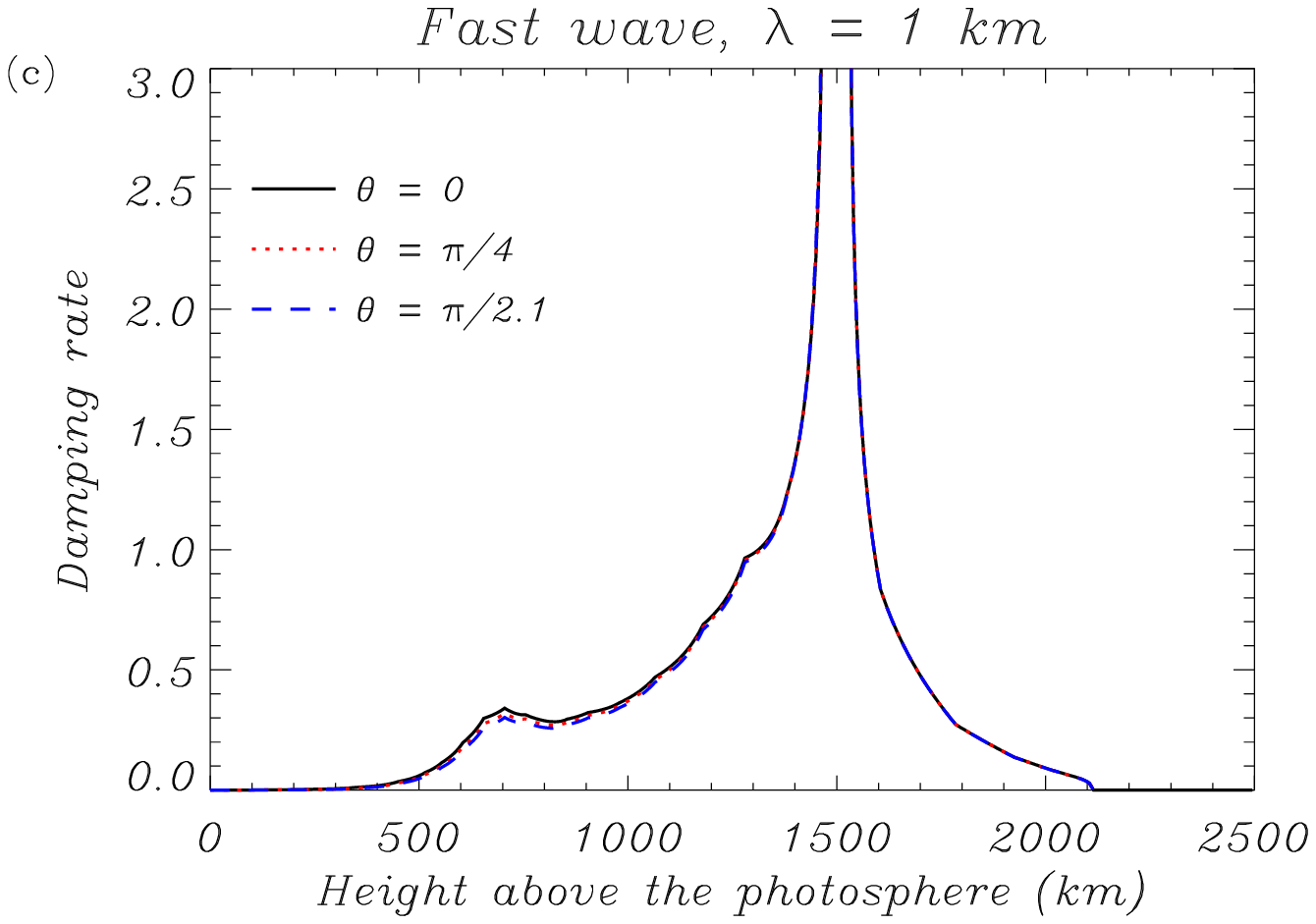}
	\includegraphics[width=.49\columnwidth]{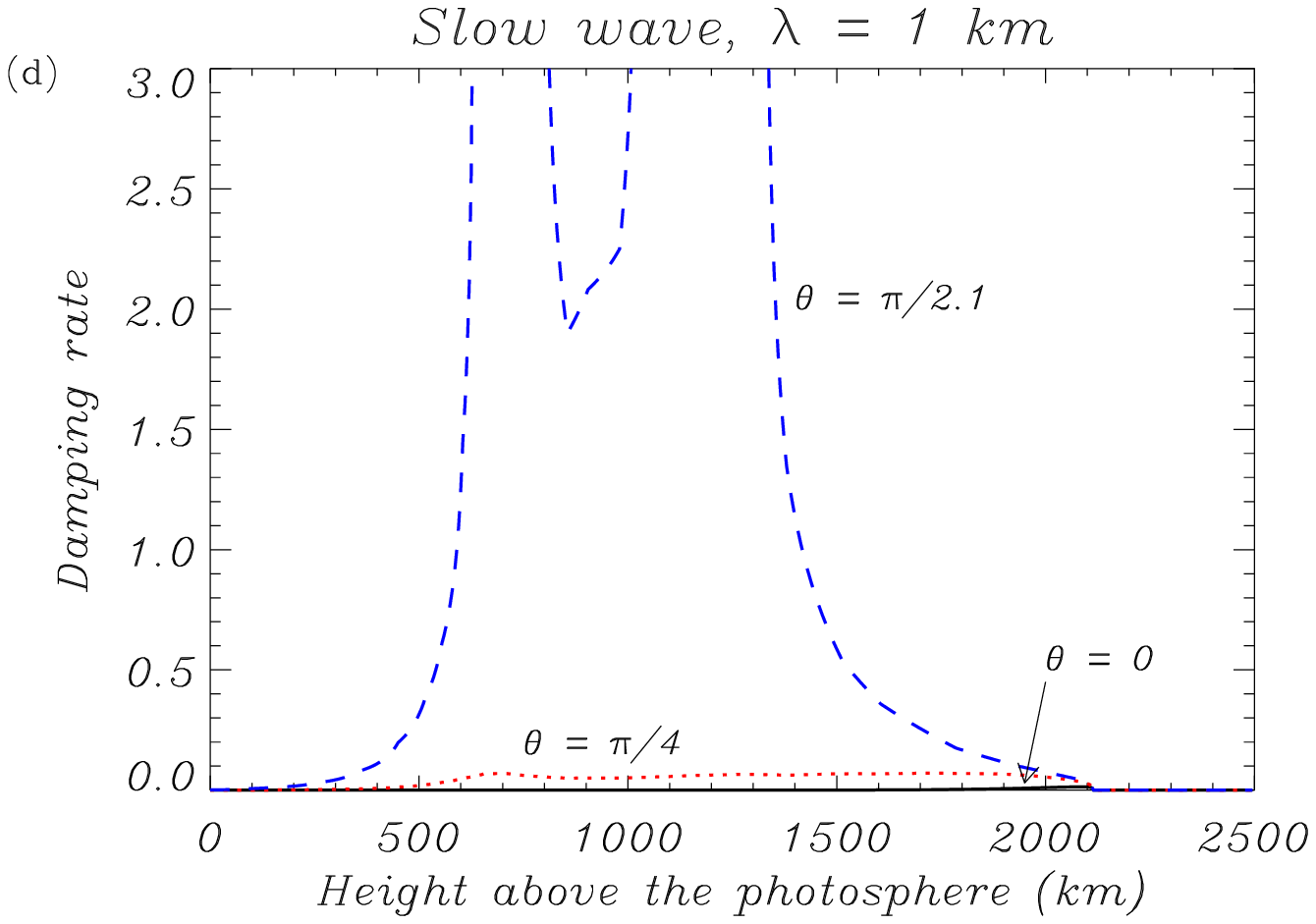}
	\caption{Active region case. Damping rate, $\delta = -\omega_{\rm I} / \omega_{\rm R}$, of (a) fast waves and (b) slow waves as function of height in the solar chromosphere for $\lambda = $~10~km. The various lines are for $\theta =$~0 (solid), $\pi/4$ (dotted), and $\pi/2.1$ (dashed). Panels (c) and (d) are the same as panels (a) and (b) but for $\lambda = $~1~km.}
	\label{fig:ratevalc}
\end{figure*}

Next, we define the damping rate as $\delta = -\omega_{\rm I} / \omega_{\rm R}$. This quantity informs us about the efficiency of  damping due to ion-neutral collisions. Indirectly, $\delta$ is also an indicator of the  ability of ion-neutral collisions to heat the plasma by dissipating magnetoacoustic waves. For the active region case, Figure~\ref{fig:ratevalc} shows the damping rate of the modified slow and fast waves as function of height for $\lambda =$~10~km and $\lambda =$~1~km. When $\lambda =$~10~km (Figure~\ref{fig:ratevalc}(a)-(b)), the waves are unaffected by two-fluid effects and do not have nonpropagating intervals. The only effect of ion-neutral collisions is to produce the damping of the waves. The fast wave damping rate is independent of the propagation angle, $\theta$, and is maximal at 1,900~km, approximately. This suggests that the main contribution of fast waves to plasma heating might take place at the higher levels of the active region chromosphere.  Conversely, the slow wave damping rate strongly depends on $\theta$. The damping of the slow wave is in general very weak. We have to consider propagation almost perpendicular to the magnetic field, i.e., $\theta \to \pi/2$, to obtain a significant slow wave damping. Two-fluid effects play a role when the wavelength is decreased to $\lambda =$~1~km (Figure~\ref{fig:ratevalc}(c)-(d)). Consistent with Figure~\ref{fig:valcdis}, when  $\lambda =$~1~km the fast wave has a nonpropagating zone around 1,500~km above the photosphere. Again, the results for the fast wave are independent of $\theta$. Near the location of the nonpropagating zone, the fast wave damping rate boosts dramatically since $\delta \to \infty$ in the nonpropagating interval. Conversely, the slow wave only has nonpropagating regions when  propagation is nearly perpendicular to the magnetic field. In this case, the slow wave has two nonpropagating intervals at two different heights \citep{soler2013aa}. These forbidden intervals for the nearly perpendicular slow wave cover a significant part of the chromosphere.

\begin{figure*}
	\centering
	\includegraphics[width=.49\columnwidth]{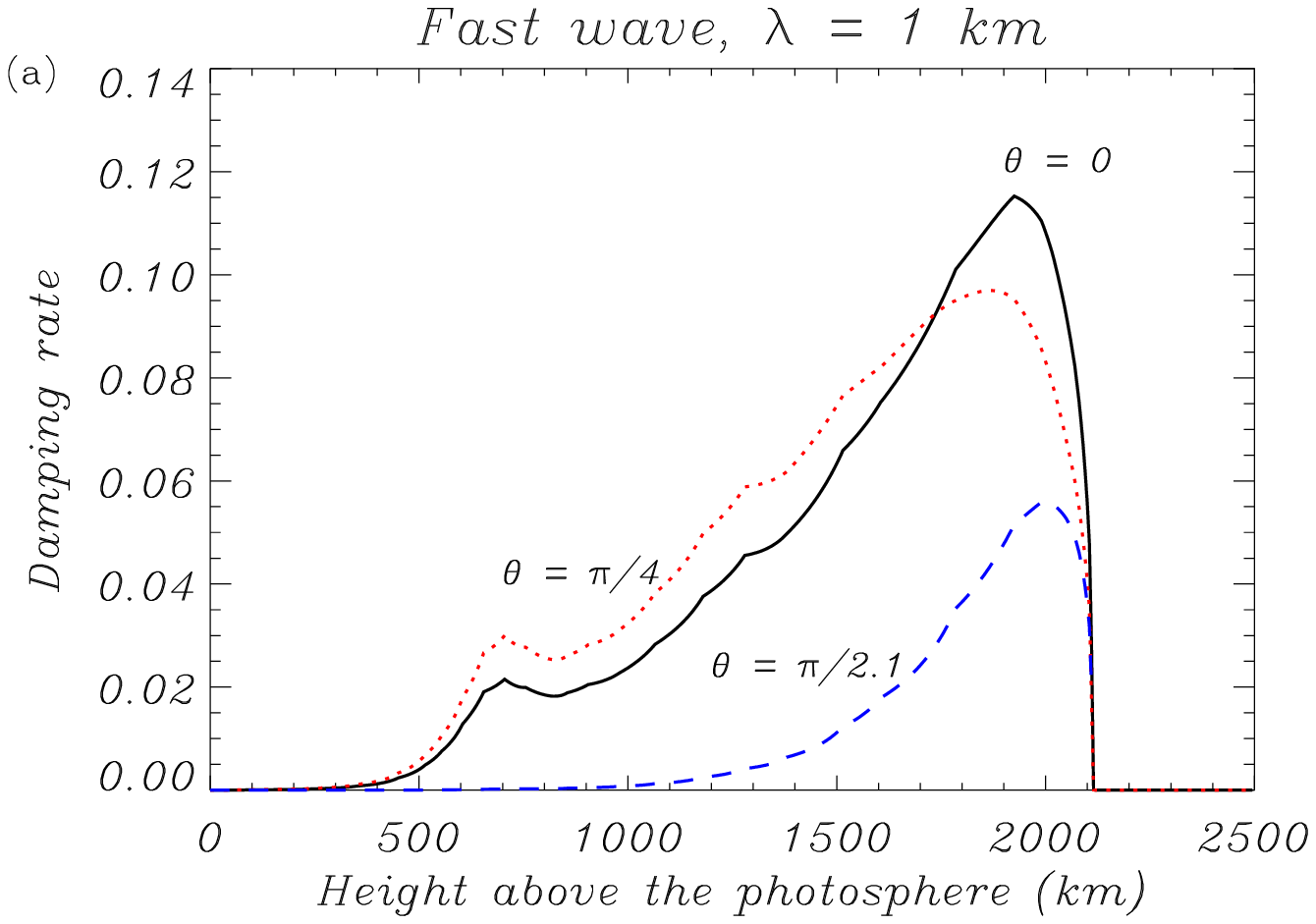}
	\includegraphics[width=.49\columnwidth]{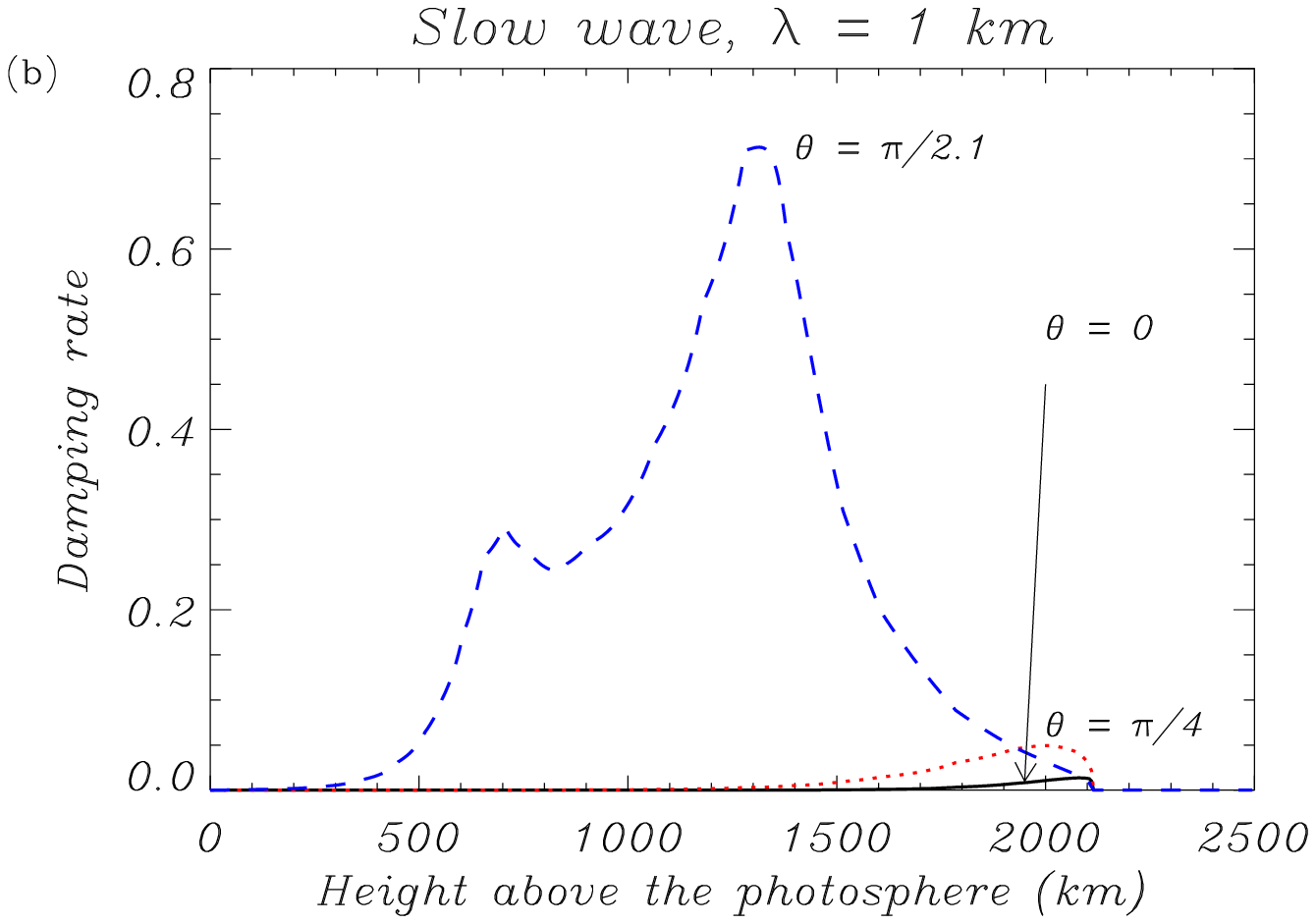}
	\caption{Same as Figure~\ref{fig:ratevalc} but for the quiet Sun case with $\lambda =$~1~km.}
	\label{fig:ratevalcquiet}
\end{figure*}

Figure~\ref{fig:ratevalcquiet} shows the same results as  Figure~\ref{fig:ratevalc} with $\lambda =$~1~km but for the quiet Sun case. Now, the solutions do not display nonpropagating intervals. As indicated by Figure~\ref{fig:valcdis}, in the quiet Sun case we need to consider shorter wavelengths than in the active region case for nonpropagating intervals to be present. In addition, Figure~\ref{fig:ratevalcquiet} shows that both fast and slow waves damping rates depend upon the propagation angle. The reason for this result is that in the quiet Sun case $\beta_{\rm i}$ is higher than in the active region case and, in fact, $\beta_{\rm i} > 1$ at the upper levels. This causes the fast wave behavior to be dependent on the propagation angle too.

In summary, in this application we conclude that  magnetoacoustic waves propagating in  regions of the solar chromosphere with strong magnetic fields are  affected by two-fluid effects when $\lambda \lesssim$~1~km. The presence of nonpropagating intervals heavily constrains the propagation of magnetoacoustic waves of short wavelengths. In addition,   damping due to ion-neutral collisions is very efficient in the vicinity of the nonpropagating regions, which points out that significant wave energy dissipation may take place at those heights in the chromosphere. The fast wave is the propagating solution that may contribute the most to heat the plasma. For $\lambda \gtrsim$~1~km, however, two-fluid effects are of less relevance because ions and neutrals behave as a single fluid. Then, wave damping is appropriately described by the single-fluid approximation \citep[e.g.,][]{khodachenko2004}. In quiet Sun regions, we need to consider much shorter wavelengths for two-fluid effects to be relevant.

At present, we lack of appropriate observations to compare with the theoretical predictions discussed above. Unfortunately, current instruments do not have enough spatial and temporal resolutions to observe the range of wavelengths where two-fluid effects would be important. It is possible that new and future instruments as, e.g., the Atacama Large Millimeter/submillimeter Array (ALMA), may reach sufficiently high resolutions to observe wavelengths of 1 km and shorter. Angular resolution of 0.01 arcseconds or smaller is  needed to observe wavelengths on the order of kilometres in the chromosphere. The ALMA telescope may reach the required resolution.

\section{CONCLUDING REMARKS}
\label{sec:discussion}

In this paper we continued the work started in \citet{soler2013} about the effect of ion-neutral collisions on MHD wave propagation in a two-fluid plasma.  In the previous paper, we focused on incompressible Alfv\'en waves. Here, we investigated compressible magnetoacoustic waves. The present work is also related to recent investigation of  wave propagation in a partially ionized magnetic flux tube \citep{soler2013aa}. The study presented in this paper reveals that ion-neutral coupling strongly affects the well-known behavior and properties of classic magnetoacoustic waves. There is a large number of possible scenarios for wave propagation depending on the plasma physical properties. As pointed out by \citet{mouschovias2011}, the various waves supported by a partially ionized plasma display complex interactions and couplings that are not present in the fully ionized case. Among these complex interactions, the presence of cutoffs and forbidden intervals has an strong impact on waves since the allowed wavelengths of the propagating modes get constrained.

After performing a general study, we considered the particular case of the solar chromosphere and showed that magnetoacoustic waves with $\lambda \lesssim$~1~km are affected by two-fluid effects  in regions with intense magnetic fields, while much shorter wavelengths have to be considered for these effects to be relevant in quiet Sun conditions. In addition, we discussed the possible role of these waves in heating the chromospheric plasma due to dissipation by ion-neutral collisions. Here, we must refer to the comment by \citet{khodachenko2004} about that the correct description of MHD wave damping in the solar atmosphere requires the consideration of all energy dissipation mechanisms. We have focused on the effect of ion-neutral collisions alone and have not taken into account the roles of, e.g., thermal conduction, viscosity, resistivity, etc., that may have an important impact on the dissipation of wave energy. The consideration of these mechanisms using the two-fluid description of a partially ionized plasma is an interesting task to be done in future works.

Finally, a natural extension of the investigation done in \citet{soler2013,soler2013aa} and in the present paper is to abandon the normal mode approach and to study the behavior of impulsively excited disturbances. This requires the solution of the initial-value problem by means of time-dependent numerical simulations. The combination of results from both the initial-value problem and the normal mode analysis would give us a complete picture of wave excitation and propagation in a partially ionized plasma. The investigation of the initial-value problem will be tackled in future studies.

\begin{acknowledgements}{}
We acknowledge the support  from MINECO and FEDER Funds through grant AYA2011-22846 and from CAIB through the `grups competitius' scheme and FEDER Funds. 
\end{acknowledgements}

\bibliographystyle{apj}
\bibliography{refs}

\begin{deluxetable}{cccc}
\tabletypesize{\scriptsize}



\centering

\tablecaption{Behavior of the modified fast and slow waves and their approximate frequencies in the highly collisional limit depending on the ordering of the characteristic velocities. \label{tab:summ}}

\tablenum{1}

\tablehead{\colhead{Case} &\colhead{Ordering of Velocities} & \colhead{Modified Fast Wave} & \colhead{Modified Slow Wave} } 

\startdata
1 & $\ca^2 \gg \cie^2 \gg \chi\cn^2$  &   ion-electron IMW, $\omega^2 \approx k^2 \ca^2$ & ion-electron GAW, $\omega^2 \approx k^2 \cie^2 \cos^2\theta$ \\ 
2 & $\ca^2 \gg \cie^2 \sim \chi\cn^2$  &  effective IMW, $\omega^2 \approx k^2 \frac{\ca^2}{1+\chi}$ & effective GAW, $\omega^2 \approx k^2 \frac{\cie^2 + \chi \cn^2 }{ 1+\chi } \cos^2\theta$  \\ 
3 & $\ca^2 \gg \chi\cn^2  \gg \cie^2 $  & neutral IMW, $\omega^2 \approx k^2 \frac{\ca^2}{\chi}$ &  neutral GAW, $\omega^2 \approx k^2 \cn^2 \cos^2\theta$  \\  
4 & $\chi\cn^2 \gg \ca^2 \gg \cie^2 $  &  neutral IAW, $\omega^2 \approx k^2 \cn^2$ & neutral GMW, $\omega^2 \approx k^2 \frac{\ca^2}{ \chi } \cos^2\theta$  \\ 
5 & $\cie^2 \gg \ca^2 \gg \chi\cn^2 $  &  ion-electron IAW, $\omega^2 \approx k^2 \cie^2$ & ion-electron GMW, $\omega^2 \approx k^2 \ca^2 \cos^2\theta$  \\ 
6 & $\cie^2  \gg \chi\cn^2 \gg \ca^2 $  &  ion-electron IAW, $\omega^2 \approx k^2 \cie^2$ & ion-electron GMW, $\omega^2 \approx k^2 \ca^2 \cos^2\theta$  \\ 
7 & $\cie^2 \sim \chi\cn^2 \gg \ca^2 $  &  effective IAW, $\omega^2 \approx k^2 \frac{\cie^2 + \chi \cn^2 }{ 1+\chi }$ & effective GMW, $\omega^2 \approx k^2 \frac{\ca^2 }{ 1+\chi } \cos^2\theta$  \\ 
8 & $\chi\cn^2 \gg \cie^2 \gg  \ca^2 $  &  neutral IAW, $\omega^2 \approx k^2 \cn^2$ & neutral GMW, $\omega^2 \approx k^2 \frac{\ca^2 }{\chi } \cos^2\theta$  \\
\enddata

\tablecomments{IMW, IAW, GMW, and GAW denote isotropic magnetic wave, isotropic acoustic wave, guided magnetic wave, and guided acoustic wave, respectively.}

\end{deluxetable}


\end{document}